\documentclass[fleqn,usenatbib]{mnras}

\usepackage{newtxtext,newtxmath}

\usepackage[T1]{fontenc}

\DeclareRobustCommand{\VAN}[3]{#2}
\let\VANthebibliography\thebibliography
\def\thebibliography{\DeclareRobustCommand{\VAN}[3]{##3}\VANthebibliography}

\usepackage{graphicx}	
\usepackage{amsmath}	

\usepackage{multirow}
\usepackage{soul}

\title[Formation of ring-like structures in flared $\alpha$-discs with X-ray/FUV photoevaporation]
{Formation of ring-like structures in flared $\alpha$-discs with X-ray/FUV photoevaporation}

\author[J. C. Vallejo et al.]{
Juan C. Vallejo,$^{1}$\thanks{E-mail: juancval@ucm.es}
Ana In\'{e}s G\'{o}mez de Castro,$^{1,2}$
\\
$^{1}$Joint Center for Ultraviolet Astronomy, AEGORA Research Group, Universidad Complutense
de Madrid, Avda Puerta de Hierro s/n, 28040 Madrid, Spain\\
$^{2}$U.D. Astronom\'{i}a y Geodes\'{i}a, Fac. CC Matem\'{a}ticas, Universidad Complutense de Madrid, Spain\\
}

\date{Accepted XXX. Received YYY; in original form ZZZ}

\pubyear{2020}

\begin{document}
\label{firstpage}
\pagerange{\pageref{firstpage}--\pageref{lastpage}}
\maketitle

\begin{abstract}

Protoplanetary disks are complex dynamical systems where several processes may lead to the 
formation of ring-like structures and planets. 
These discs are flared following a profile where the vertical scale height 
increases with radius.
In this work, we investigate the role of this disc 
flaring geometry on the formation of rings and holes.
We combine a flattening law 
change with X-ray and FUV photoevaporative winds.
We have used a semi-analytical $1D$ viscous $\alpha$ approach, presenting
the evolution of the disc mass and mass rate in a grid of representative systems.
Our results show that changing the profile of the flared disc may favour
the formation of ring-like features resembling those observed
in real systems at the proper evolutionary times, with proper disc masses and
accretion rate values. However, these features seem 
to be short-lived and further enhancements are still needed for better matching 
all the features seen in real systems.
\end{abstract}



\section{Introduction}

Planets form in accretion discs around Pre-Main Sequence stars, and 
understanding the evolution of these protoplanetary
discs (PPDs) and the mechanisms and timescales by which these discs are eventually
dispersed is a key issue in planet formation theories.

As the mass accretes onto star, the conservation of angular momentum implies that 
the disc spreads out, the accretion decreases with time and the radius 
increases with time. This transport can be explained in terms of internal 
viscous transport and magnetised winds, even when other possible mechanisms to
consider may be planet formation, dust growth and encounters with other stars
\citep{armitage99, dullemond05}.

Transition discs are PPDs characterised by 
a deficiency of disc material close to the star, exhibiting inner cavities or gaps in the distribution
of the dust and/or gas \citep{andrews11,espaillat14,tazzari17}.
Many of these inner cavities have sizes that 
can vary from sub-AU to many tens of AU, while there is 
still mass accretion onto the central star \citep{francis20,rometsch20}. 

In addition to these internal cavities, ALMA data shows that the dust 
continuum emission of many, if not all bright and large, PPDs consists of rings and 
gaps, and one can find ring-like structures
across the full ranges of spectral type and luminosity, up to a distance 
of around $100$au from their host star, with ages about $0.4-10$ Myr
\citep{andrews18, long18}). 

An inside-out dispersal model produced by photoevaporation can naturally produce these internal 
cavities. When the thermal velocity exceeds the local escape velocity,
the surface layer gets unbound and evaporates. Hence, the thermal 
winds remove mass regardless the angular momentum transport internal process, 
and the features seen in transition  discs can be produced.
The depletion can be produced by external radiation fields, 
when the disc is located in a populated and dense star forming region
\citep{winter18,concha19} or by internal radiation fields, 
generated by the host star itself \citep{alexander06,gorti09, owen12}.
Notably, the mass losses from both mechanisms are comparable, 
typically quoted as $10^{-8}$ $M_{\odot}/yr$ \citep{anderson13}. 

These photoevaporative discs reproduce well some of the
observed transition discs,  mainly those with small cavities and 
low accretion rates, less than $10^{-8}$ $M_{\odot}/yr$. 
\citep{ercolano18, wolfer19}. However, 
they have problems to reproduce all observed discs 
\citep{owen12,marel19}. Therefore, many other mechanisms 
have been raised to explain the presence of ring-like
fatures in young discs barely $1$ Myr old. Among these mechanisms, 
we can cite planets \citep{dong15,long18, huang18}, snow lines \citep{zhang15}, 
sintering \citep{okuzumi16}, instabilities \citep{takahashi16}, 
resonances \citep{boley17}, dust pile-ups \citep{gonzalez17}, dead zones \citep{ruge16},
self-induced reconnection in magnetised discs wind systems \citep{suriano18}
or large scale vortices \citep{barge17}.

Obviously, PPDs are complex systems where different mechanisms leading to 
ring-like structures can be present at time, and one of these 
mechanisms might or not predominate. 
A common explanation for the observed ring-like
features is the presence of planet-disc interactions. Therefore, 
these features are frequently considered tracers of new-born planets. 
However, it still remains how to match these widely found gaps 
at given distances with the small number of 
Jupiters mass planets around main sequence stars at those separations \citep{lodato19}.

The goal of our work is to analyse if simple photoevaporative viscous 
models can explain some of the observed
ring-like features in PPDs while keeping the disc masses and accretion rates
within the observed range of values. 
We will analyse the interplay of a variable flaring geometry
and different photoevaporative winds in a variety of viscous $\alpha$-discs,
exploring in a systematic way the impact of the different control parameters on 
the evolution and the formation of ring-like structures on those discs.

Viscous momentum transport is still of interest in disc modelling
even when other transport alternatives are possible \citep{papaloizou05,heinemann09,hartmann18}.
The  magnetorotational instability, or MRI \citep{balbus91}, 
was for long time a leading candidate for turbulence and 
angular momentum transport. However, 
the MRI can be suppressed in non-ideal MHD \citep{riols18},
and other mechanisms such as outflows, hydrodynamical processes 
and gravitational instability can be considered 
\citep{armitage15, kratter16}.
Moreover, magnetic fields can be of importance in momentum transport mechanisms 
and disc accretion may be primarily wind-driven with magnetised disc winds
\citep{pudritz91,bai16,simon17}.
Nevertheless, the presence of these magnetised winds does not exclude the presence 
of viscous transport.


The EUV flux was the first radiation field taken into consideration
as a mechanism for the internal photoevaporation \citep{hollenbach94}. But EUV radiation leads to limited mass losses, while
the FUV and X-rays radiation fields produce larger mass losses than EUV.
Moreover, the X-ray winds are itself optically thick to EUV photons, and they
prevent them from reaching the disc. Therefore, EUV is usually neglected when one considers the other two fields \citep{owen12,kimura16}.

The luminosities in FUV and X-ray wavelengths of young low-mass stars are 
significant and exceed by a factor $50$-$100$ the luminosities found 
in older stars like our Sun \citep{gomez12}. 
When modelling X-ray and FUV radiation, the hydrodynamic and 
thermal structure of the disc must be solved numerically, taking into account 
the irradiating spectrum, grain physics, chemical processes
and cooling by atomic and molecular lines. This makes the 
modelling of each disc complex and computationally expensive.
Hence, previous numerical studies dealing with these issues focused on analysing  
representative values of the host star mass and the viscosity 
\citep{anderson13, lodato17}.
These studies are still being enhanced, by adding three-layers structures, metallicity dependences, 
improved modelling of gas temperatures and improved thermochemistry processes 
\citep{wang17,nakatani18, picogna19, grassi20}. 

Because of the complexity of the problem, we are going to take a semi-analytical $1D$ viscous 
$\alpha$-disc. Here, we can add a flaring profile that changes with time, 
and a photoevaporative term resulting from the combination 
of X-ray and Far-Ultraviolet (FUV) winds of different
efficiencies.

A $1D$ model is by far much straightforward to implement, but 
it might be considered  only valid while one does not
include magnetic fields, envelopes or any other $2D$ or $3D$ structures. 
However, the complex hydrodynamics, chemical processes, and radiative 
processes can be incorporated in the form of function profiles resulting from
numerical fits to the result of the more realistic simulations.
In this way, semi-analytical $1D$ models are useful nowadays 
because their simplicity in interpreting the results allows to 
complement more complex hydrodynamical simulations.

We are going to numerically explore a grid of representative systems, 
analysing the evolution of the disc mass, the accretion rate and the
creation of ring-like features as the stellar masses and initial disc mass
are varied. Aiming to provide some insight on the role played by each parameter
in better matching a numerical model with a real system, 
we will compare our results with an arbitrary sample of real systems in the Taurus star forming region,
the same we selected in \cite{vallejo18}. This will allow to 
probe the usefulness and limitations of this semi-analytical $1D$ approach.

The structure of the paper is as follows.
The section 2 introduces the grid of $\alpha$-disc models. 
The section 3 shows the evolution of these models as the different
control parameters, such as the viscosity, the flaring profile 
and the efficiency of a X-ray dominated wind, are varied. 
The section 4 presents the changes in the evolution 
of the discs when we combine both X-ray and FUV ionising fields. 
Section 5 is devoted to the analysis of the 
accretion rates in both cases.
Finally, the last section summarises the results and makes some concluding remarks.

\section{The numerical $\alpha$-discs}

\begin{figure}
\begin{center}
\includegraphics[width=\columnwidth]{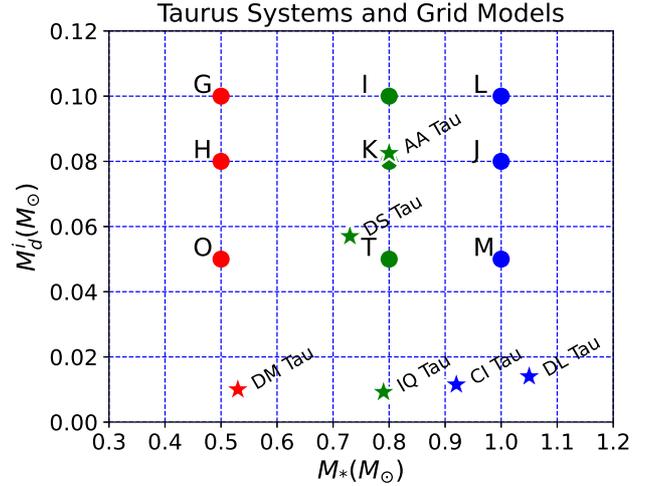} 
\end{center}
\caption{
The selected grid of discs models in the $M_{*} - M_d^i$ plane.
The initial disc masses $M_d^i$ decrease as the discs evolve. 
Hence, these initial points move downwards in the diagram, approaching 
to the Taurus systems from Table~\protect\ref{listofstars} that are also overlaid to the grid. 
}
\label{figmodels}
\end{figure}

The Figure~\ref{figmodels} 
shows the grid of systems that we will study. 
Every model of the grid is labeled with a given capital letter,
keeping the same labels already used in \cite{vallejo18}.
This grid covers typical values for the host star mass $M_{*}$ and the
initial mass of the disc $M_d^i$, meaning $M_{d}(0)$.
This initial mass decreases with time as the disc is eroded, at a 
rate driven by the remaining control parameters.
Conversely, the mass of the star $M_{*}$
can be considered to be constant as the accreted mass term does 
not add enough amount of mass to modify it.

As the initial disc mass of each numerical model decreases, their discs masses
decrease, reaching the masses observed in real system at given ages. In this way,  
we can see how a real observed system can be best matched by a given 
model (or models) when these control parameters
are varied in a given way. For doing so, 
we have overlaid to this figure the arbitrary selection of real discs
listed in Table~\ref{listofstars}. 
These representative systems will be our reference sample.
We have selected Taurus because it is one of the nearest and best-studied large star-forming regions. In addition, 
our work focuses on discs subject to radiation fields coming from the host star, and, 
in Taurus, the molecular cores can be considered isolated and the influence of outflows, jets, 
or gravitational effects is minimised \citep{hartmann00}. 
Other regions such as $\rho$-Oph or Orion are a good place 
for testing external photoevaporation models, but they are not the best environments for the analyses 
of disc sizes as diagnostic for isolated viscous evolution.

\begin{figure}
\begin{center}
\includegraphics[width=\columnwidth]{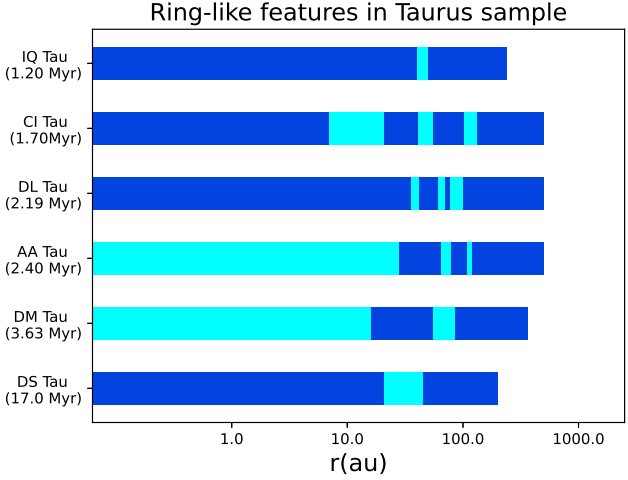} 
\end{center}
\caption{
A schematised view of the position of the inner cavities and gaps of different widths 
that are observed in the Taurus reference sample 
(there is no detailed indication of the real density gradients). The discs are plotted as 
dark blue, and the cavities and gaps in cyan.
For reference, ALMA spatial resolution in Band 6 is around $16$au at Taurus distance.
See references given in Table~\ref{listofstars} for details.}
\label{models}
\end{figure}

The Table~\ref{listofstars} also includes data about the 
ring-like structures (rings and gaps) detected in those discs. 
A schematised view of the position of these structures can be seen in Figure~\ref{models}. 
These data are listed just for reference. Some slight variations can be found along the 
literature because observational uncertainties, but we just aim to gain insights on the different 
mechanisms that may produce similar ring-like features at given ages on PPDs.
For reference, the median age of Taurus stars is just a few Myr \citep{bertout07},
and a recent study \citep{luhman18} shows that the older population of stars $> 10$ Myr
which was proposed to be associated with this region in other studies
has no physical relationship with Taurus.


\begin{table*}
\caption{Representative systems from the Taurus star forming region. 
The values for the star masses are those found in \protect\cite{simon17}. 
The values for disc mass, accretion rate 
and ages are those found in \protect\cite{jones12}. 
The X-ray Luminosity $L_X$ is computed from $M_*$ following \protect\cite{telleschi07}. 
Additional viscosity data can be found in \protect\cite{vallejo18}.}
\label{listofstars}
\begin{tabular}{@{}l|cccccccl}
\hline
Star      & $M_{*}$ & $M_d(t_{*})$ & $M_d(t_{*})/M_{*}$ & $L_X$ & $log_{10} t_{*}$ & $\dot{M}(t_{*})$  & Rings-like  \\
           & ($M_{\odot}$) & ($M_{\odot}$) & (adim) & ($erg \quad s^{-1}$) & ($t_{*}$ yrs) & ($M_{\odot}$ $yr^{-1}$) & brightness profile estructures \\
\hline
\hline
DL Tau & $1.05$ & $0.0140$ & $0.01$ & $1.46 \cdot 10^{30}$ & $6.34$ & $2.18 \cdot 10^{-8}$ & Gaps at $39$au, $66$au, $89$au. \citep{lodato19,long20} \\
CI Tau & $0.92$ & $0.0115$ & $0.01$ & $1.17 \cdot 10^{30}$ & $6.23$ & $1.12 \cdot 10^{-8}$  & Gaps at $14$au, $48$au, $118$au. \citep{long18,lodato19} \\
AA Tau & $0.80$ & $0.0825$ & $0.10$ & $9.23 \cdot 10^{29}$ & $6.38$ & $2.75 \cdot 10^{-9}$  & Cavity up to $28$au, gaps at $72$au, $115$au. \citep{marel19} \\
IQ Tau & $0.79$ & $0.0092$ & $0.01$ & $9.03 \cdot 10^{29}$ & $6.08$ & $5.75 \cdot 10^{-9}$  & Gap at $41$au. \citep{long18,lodato19} &  \\
DS Tau & $0.73$ & $0.0570$ & $0.08$ & $7.90 \cdot 10^{29}$ & $7.23$ & $1.00 \cdot 10^{-8}$  & Gap at $33$au. \citep{lodato19,long20} \\
DM Tau & $0.53$ & $0.0100$ & $0.02$ & $4.58 \cdot 10^{29}$ & $6.56$ & $1.02 \cdot 10^{-8}$  & Cavity up to $16$, gap at $70$au. \citep{marel19} \\
\hline
\end{tabular}
\medskip
\end{table*}

The calculation of the mass accreted by every host star of the grid 
can be done through the analysis of the evolution in time of the surface 
mass density $\Sigma (r,t)$ using the basic laws of mass and momentum conservation,

\begin{equation}
\label{diffusion}
\frac{\partial \Sigma }{\partial t} = \frac{3}{r} \frac{\partial}{\partial r} \lbrack r^{1/2} \frac{\partial }{\partial r} ( \nu \Sigma r^{1/2} ) \rbrack - \dot {\Sigma}_{w},
\end{equation}
where $\dot {\Sigma}_{w}$ denotes the mass loss by a given photoevaporative wind, functionally equivalent to have a sink in this diffusion equation.

The viscous evolution of the disc is computed using 
a first-order explicit FTCS scheme. The computational mesh covers uniformly 
from $0.0025$ au to $2500$ au using the scaled variable $x$, where $r = x^{2}/4$. 
Zero-torque boundary conditions are applied in the inner and outer boundaries, meaning 
$\Sigma = 0.0$ at both edges of the mesh. When solving Eq.~\ref{diffusion}, we use 
an initial profile $\Sigma (r,0)$ that follows the self-similar solution firstly 
used in \cite{lynden74}.

The kinematic viscosity $\nu (r)$ is the coefficient that regulates the diffusion. 
When considering that the disc surface density 
follows a $r^{- \gamma}$ dependency, this kinematic viscosity can be modeled by,

\begin{equation}
\label{viscosity}
\nu (r) = \nu_1 (\frac {r}{R_1})^{\gamma}.
\end{equation}

Different $\gamma$ values can be found in the literature.
Some references use $\gamma = 3/2$ \citep{lodato17}, but 
it is more frequent to use a linear dependency with radius, $\gamma = 1$ \citep{clarke01,alexander06,rosotti17}. 
Therefore, we have selected this value for comparison
purposes. 

As a consequence of above, $R_1$ is the scale radius at which the viscosity 
has the value $\nu_1$. This $\nu_1$ just depends
on the initial mass of the disc $M_d^i=M_d(0)$ and the selection of the initial 
accretion rate $\dot {M_0}$,

\begin{equation}
\label{initialaccretion}
\dot {M_0} = \frac {3 M_d(0) \nu_1}{2 {R_1}^2}.
\end{equation}

Once we have fixed the initial profile $\Sigma (r,0)$ 
and the functional form of Eq.~\ref{viscosity}, 
one can obtain $\nu_1$ from Eq.~\ref{initialaccretion} and compute the 
viscous evolution as per Eq.~\ref{diffusion}.
%
However, the initial value for the accretion rate $\dot {M_0}$ is not a good 
control parameter of the model because it is not an observable quantity. 
By other hand, the Shakura prescription, or $\alpha$-disc prescription, allow us to select the
$\nu_1$ coefficient without presuming $\dot {M_0}$.

This parametrisation was firstly used in \cite{shakura73}
and later refined in \cite{shakura76}. It
hides the details of the specific viscous transport 
mechanism while reflecting the impact of the transport in the disc evolution.
Therefore, as this parametrisation simplifies the implementation of the viscosity
in numerical models, it has been widely used for many years and 
it is still in common use today \citep{kimura16, lodato17, ercolano17}.

This model relies on an optically thick accretion disc 
and a turbulent fluid described by a viscous stress tensor, with this stress tensor
being proportional to the total pressure. 
When the eddy size (mean free-path) is less than the disc height and the turbulent velocity smaller
than the sound speed $c_s$, one can write the viscous profile as,
$\nu (r) = \alpha c_s H(r)$,
with  $H(r)$ the scale height profile that models the disc thickness,
presuming $H(r) << r$ for a thin disc. 

Considering hydrostatic equilibrium perpendicular to the disc plane and a simple relationship between the disc and surface temperatures, the sound speed can be written as $c_s = H(r) \Omega$,
with $\Omega$ is the circular keplerian velocity \citep{jones12}.
Hence, the $\nu$ viscosity at a given distance $r$ depends on rotation, with  
a linear dependency on the mass of the star $M_*$ and on the flaring of the disc,

\begin{equation}
\label{shakura}
\nu (r) = \alpha \frac{c_s^2}{\Omega} = \alpha  \Omega H^2(r).
\end{equation}

\begin{figure}
\begin{center}
\includegraphics[width=\columnwidth]{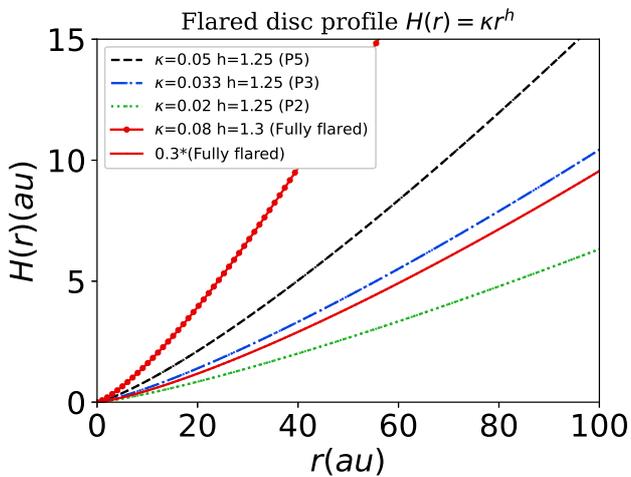} 
\end{center}
\caption{
Different profiles for the considered flared discs. 
The parameters of the power law define the scale height $H(r)$ of the disc. 
The P5 profile means $\upkappa=0.05$, the P3 profile means $\upkappa=0.033$, the P2 profile means $\upkappa=0.02$,
all using $h=1.25$. Hence, when using progressive flattenings, 
the label P52 means a profile changing from P5 to P2.}
\label{flaring_fig}
\end{figure}

Therefore, the flaring profile is a key parameter that defines the 
 $c_s$ profile, and, in turn, the temperature profile of the disc.
The scale height $H(r)$ depends on the competition between thermal pressure and gravity. 
That is, between the temperature and surface density profile. 
And the temperature depends on the amount of stellar radiation impacting the disk, which
also rests on its geometry again. 
By solving these coupled  equations, \cite{chiang97} found 
the scale height $H(r)$ can follow an approximate power law dependence,

\begin{equation}
\label{flaringprofile}
H(r) = \upkappa \ r^{h},
\end{equation}
with $h \sim 1.3-1.5$. However, fully-flared models based on a two-layer dust model 
do not properly reproduce the SED in the far-IR and a flattened disc flaring is required,
typically by a factor $3$ \citep{tazzari17}. This flattened profile is then
similar to those that use $h = 5/4$ \citep{alexander06,owen12,rosotti17} . 
This fullfils the condition for a geometrically thin disk and lead to a viscosity 
through Eq.~\ref{viscosity} that scales linearly with 
radius \citep{hartmann98, alexander07}. Hence, we have used this value of $h$
as baseline profile. The Figure~\ref{flaring_fig} shows these different flaring profiles for reference.

Even when it is customary to keep constant the flaring profile of the disc, it seems
reasonable to consider that this flaring may change with time, with a progressive
flattening of the disc as the age increases. 
This flattening can be due due to settling of dust grains toward the midplane 
\citep{alessio99,chiang01}, and to occur very early in the 
evolution of a disc \citep{dullemond04}. 
Obviously, discs with cavities are very diverse, not forming a single population 
of objects, and they could present very different dust growth and sedimentation
(dust settling) laws \citep{williams11}. 
Finally, the flaring geometry
of the disc may differ when one considers if the photoevaporation
is dominated or not by the X-ray \citep{owen12}. 
Therefore, we will include in our simulations
a changing profile for the flared disc, by decreasing the $\upkappa$ parameter 
in Equation~\ref{flaringprofile} at given times (see Section~\ref{rxwindflatten}).   

\section{Photoevaporation from X-ray radiation dominated winds}
\label{secrxwinds} 

The selection of the photoevaporation mechanism providing the mass sink term $ {\dot {\Sigma}}_{wind}$  in Eq.~\ref{diffusion} has important implications for 
the evolution of the discs. 
The models from \cite{owen12} suggested that the X-ray component dominates the
photoevaporative mass-loss rates from the inner disc unless one may have
high FUV-to-X-ray luminosity ratios ($L_{FUV}/L_X > 100$), even 
when there is not a clear consensus about how X-ray and FUV (and even EUV) must be put together \citep{armitage15}. 
Because we aim to systematically analyse the role of every control parameter, 
our work starts with simulations based on winds presumed to be  
mainly driven by X-ray radiation.

In general, one can consider three sources for the X-rays: emission coming 
from the disc host star,
 emission coming from jets shocks \citep{ustamujic16}, and external emission in clusters \citep{adams12}. 
Here, we will focus on the internal radiation field coming from the host star.
This X-ray heating rely in X-ray photons in the $0.1 - 10$ KeV range that can tear off the electrons from the internal shells of metals. These electrons
 produce further ionisations until the energy is thermalised.
Higher energies photons can penetrate further, but do no provide significant heating \citep{owen12}. 


\begin{figure}
\begin{center}
\includegraphics[width=\columnwidth]{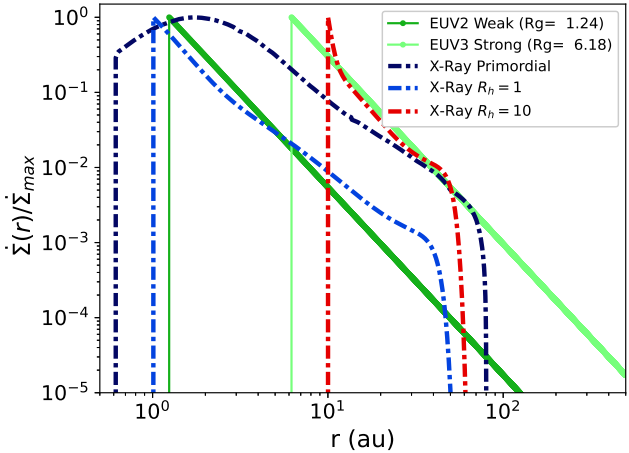} 
\end{center}
\caption{
Photoevaporative wind mass losses for diferent X-ray cases
and $L_x=1 \cdot 10^{30}$ ($ergs$ $s^{-1}$). For comparison, one strong and one weak
EUV wind mass profile are also plotted.
The integrated mass rates along all radii have been set to unity and the
resulting profiles have been divided by their maximun in order to compare the different shapes.}
\label{winds}
\end{figure}

This heating is not straightforward to model because 
the hydrodynamic and thermal structure of the disc must be solved numerically
taking into account the irradiating spectrum, grain physics, chemical processes 
and cooling by atomic and molecular lines. 
As we want to add these winds in our $1D$ semi-analytical model, 
we have used the same approach taken by \cite{rosotti13}, \cite{ercolano17} and
\cite{kimura16}. Hence, we have used a photoevaporative X-wind mass sink term 
$ \dot{\Sigma}_{wind}$ resulting from 
the numerical fit to the results of the simulations done by \cite{owen12}. 

This X-ray wind model has two different wind profiles, that are activated 
by the presence or not of a hole in the disc. 
The Figure~\ref{winds} shows the photoevaporative wind mass losses $\dot {\Sigma}$ corresponding to each phase.
The primordial phase profile corresponds to the initial stage that lasts until the wind eroding the disc
opens a cavity on it. Then, the second stage begins, that 
lasts while this gap gets wider,
 as the inner edge of the outer disc is exposed directly to stellar irradiation. 
The figure also shows the second stage profiles corresponding to different sizes of this hole.
This phase ends when the disc can be considered almost fully depleted as the floor density
has been reached at all radii.



Just for reference, the Figure~\ref{winds} also shows EUV winds: one weak EUV wind (labeled EUV2) 
and one strong EUV wind (labeled EUV3), as described in \cite{vallejo18}.
Meanwhile the EUV winds have more localised wind losses, and the disc is hardly
depleted at large radii, the X-ray winds are stronger, with 
much shallower profiles and significant contributions at large distances. 


One key parameter of this X-ray model is the computation of the instant when 
it switches from the primordial phase to the second phase. 
This can be done by monitoring
the $\Sigma (r,t)$ at every $t$ in two different inner selected locations, $0.1$ au and $1.0$ au \citep{ercolano15}.
The switch will take place when $\Sigma < 10^{-5} g cm^{-2}$. 
As alternative, one can compute the column density $N_H$ at the inner edge of outer disc, and 
the switch will starts when $N_H < 10^{22}$ \citep{ercolano17, kimura16}.
Both approaches lead to similar results, and, for simplicity, we have selected the first option.

The expressions of the total integrated mass loss rate due to the X-ray wind are detailed in \cite{owen12}, and they depend on the stellar mass $M_*$ and X-ray
luminosity $L_X$. For $M_* = 1$ $M_{\odot}$ and $L_X = 10^{30}$ $ergs \quad s^{-1}$, the
 total integrated mass loss rate in the primordial phase is 
 $\dot{M}_X = 6.25 \cdot 10^{-9} $  $M_{\odot} yr^{-1}$, meanwhile the 
 total integrated mass loss rate in the second phase is 
$\dot{M}_X = 4.8 \cdot 10^{-9} $ $M_{\odot} yr^{-1}$.
We will take these values as the baseline option for defining our fiducial X-ray wind.

\subsection{Viscosity and timescales}
\label{sec:rxwindsvisco} 

The $\alpha$ parameter was initially introduced as a convenient scaling factor 
of the friction between adjacent rings. Later on, it turned into a standard
dimensionless measure of viscosity, 
as a convenient way to hide the real viscosity mechanism.
This dimensionless constant can take 
values between $0$ (no accretion) and close to one, presuming a subsonic regime 
for the turbulence. A recent discussion of the $\alpha$ parametrisation in other
regimes can be seen in \cite{martin19}. 

The $\alpha$ prescription can be seen as mere re-parametrisation of 
the viscosity. But, one can also extend this approach, and to use a unique $\alpha$ value for 
modelling the viscosity of the whole disc. 
The variety of values for $\alpha$ then reflects the different effectiveness 
of the hidden viscosity processes found in different discs. 
Indeed, this effectiveness can take different 
values within the same disc at different locations \citep{bai16}.
These issues may explain the variety of values returned from observations,
sometimes being as high as $0.2-0.3$ in fully ionised discs, and sometimes
being much less than that, lowering up to values $< 0.01$ or even $0.001$ for not fully ionised discs \citep{martin19}.

Therefore, $\alpha$ is a main parameter controlling the disc evolution that 
can not be set arbitrarily. The observed 
disc masses and accretion rates in star forming regions such as Taurus, 
Chamaeleon I or Lupus, put constraints on the amount of viscosity to consider, 
see \cite{martin19} and references therein. 
However, there is no general consensus in the most suitable value for  $\alpha$ even in the simplest scenario, when one neglects that the viscosity can evolve with time.
Different $\alpha$ values may mean different metallicity values for the disc \citep{ercolano18}, and one can find a variety of
values. The $\alpha=0.01$ was used in early studies because 
a viscosity sourced to MRI seemed to lead to this value \citep{hawley95}, and also because
it provides evolutionary timescales in line with known properties of discs \citep{hartmann98}.
Later on, a wider rage of values was in used, with values 
in the literature spanning typically from $0.1$ to $0.001$ \citep{andrews09, gorti09, jones12}. 
Finally, lower values like $10^{-4}$ are nowadays also found \citep{anderson13,ercolano17}, and our previous
work with EUV winds points to these small viscosity values \citep{vallejo18}.

\begin{figure*}
\begin{center}
\begin{tabular}{cc}
\includegraphics[width=0.5\linewidth]{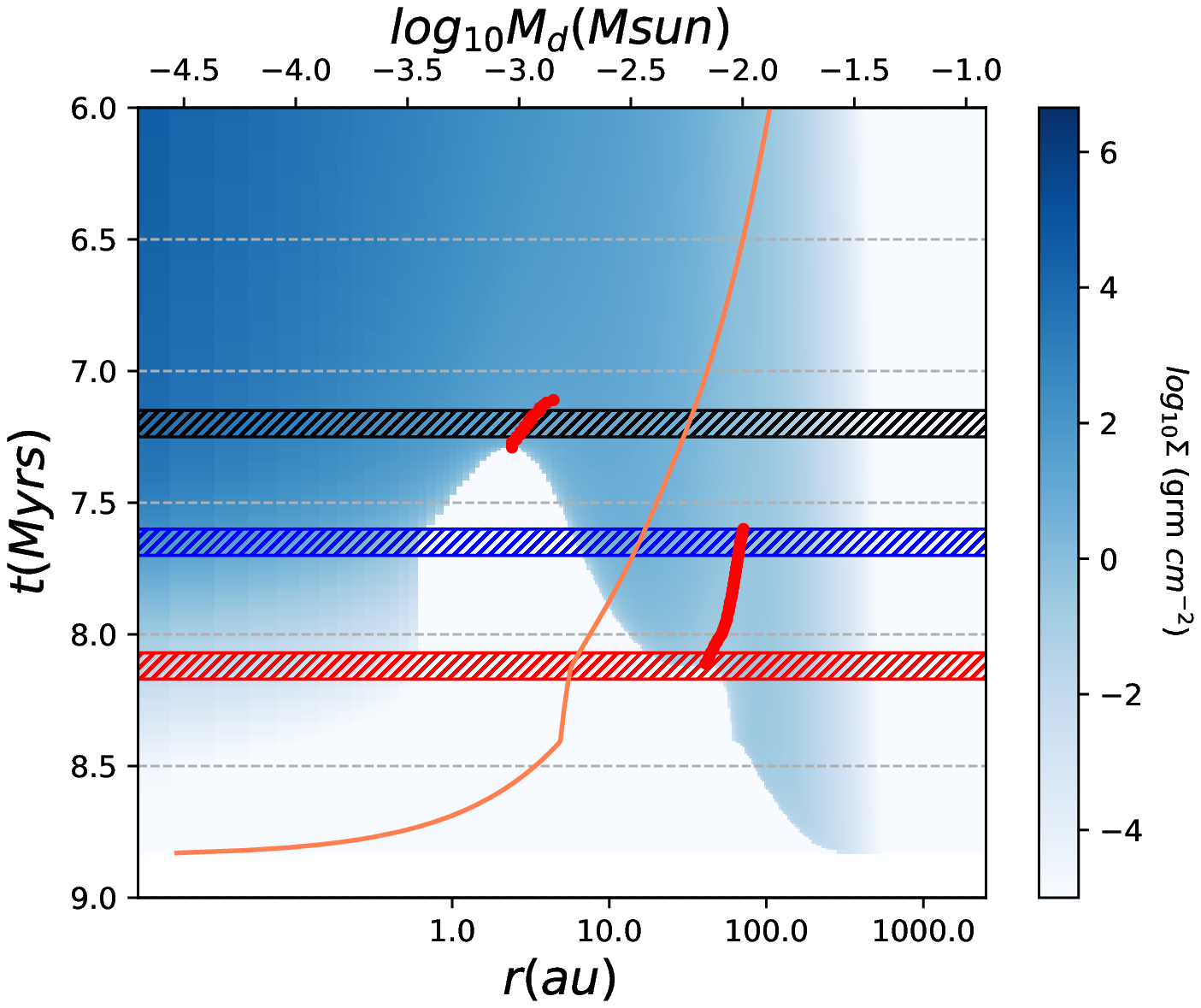} &
\includegraphics[width=0.5\linewidth]{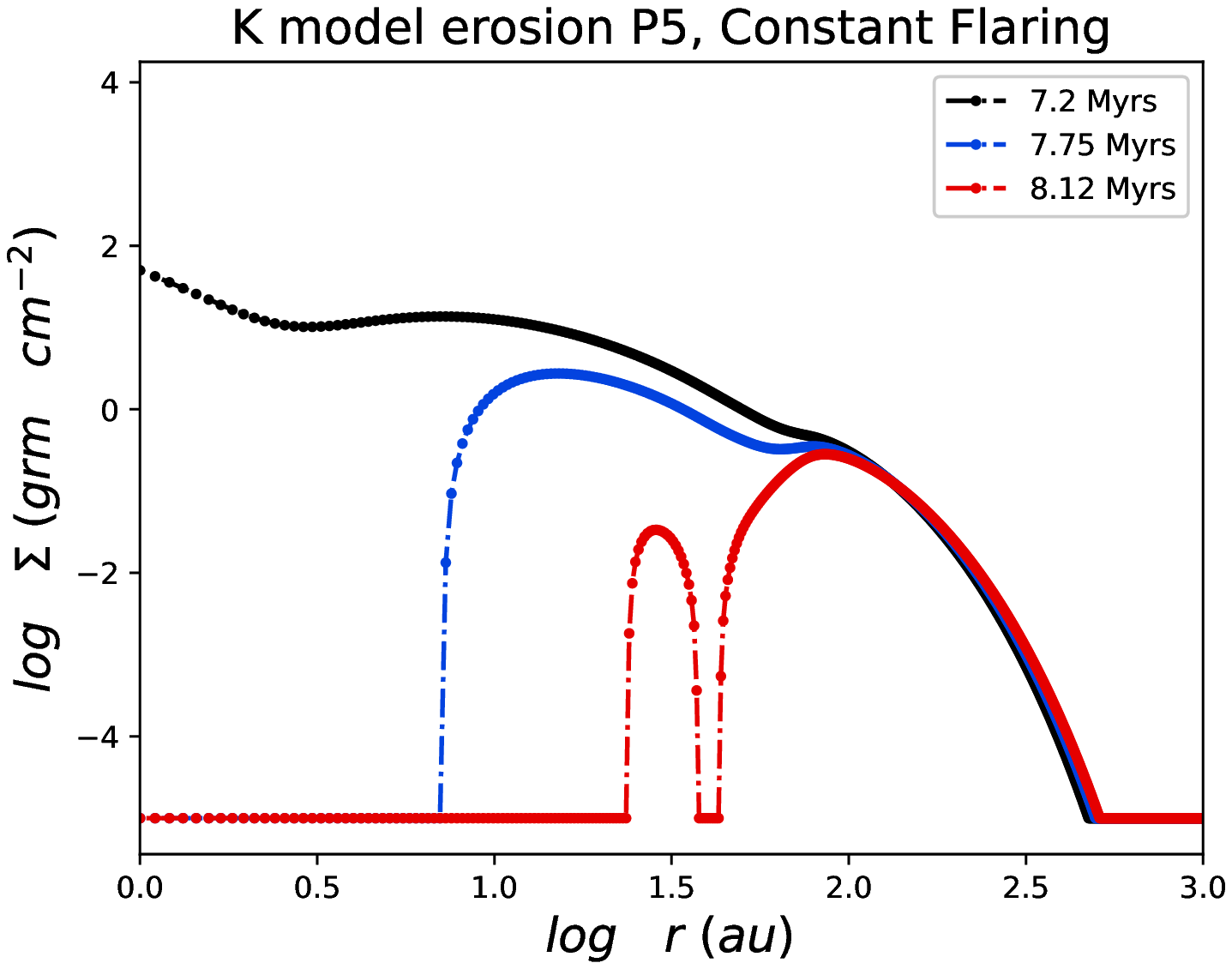}\\
\end{tabular}
\end{center}
\caption{
(Left panel) Density diagram showing the evolution of model $K$ under photoevaporation 
produced by X-ray dominated heating and a viscosity value of $\alpha=10^{-4}$,
with a constant flaring profile $P5$.
Earlier times are at the top of the figure, and the $y$-axis shows the time evolution 
of the surface density from top to down. 
The lighter the colour, the lower the density, with white colour indicating that the density
floor level has been reached. The red dots indicate the local minima in the density profile.
The upper horizontal axis shows the evolution with time of the disc mass as an orange curve.
The black, blue and red horizontal patterns mark the times corresponding to the three density profiles on the right panel.
(Right panel) Density profiles at three given times produced when slicing the density diagram.
The initial dent can grow until it produces an internal cavity. 
If an additional dent in the density is produced, further gaps and outer ring-like features can be observed.}
\label{rx_erosion}
\end{figure*}

Hence, as starting point one can take
the X-ray wind mass losses described in the previous section,
a viscosity value of $\alpha=10^{-4}$, and a flared profile with a 
constant value $\upkappa=0.033$ (the P3 profile).

The wind flow first
produces a dent in the density profile.
This dent can will typically grow with time, 
becoming deeper and deeper until a gap in the density profile is produced. 
As the inner disc density decreases, 
it eventually reaches the floor level and an internal cavity is created.

The creation of this internal cavity flags the start of 
the second phase, when there is a direct flux wind coming from the
host star. The wind profile in this second phase also has a peak at given location
that moves farther away as the time goes, and will erode the outer portions
of the disc.
The larger the mass of the disc, the longer it takes 
to disappear. Conversely, the larger the stellar mass, the stronger
the wind and the shorter the disc lifetime.
However, the internal cavity typically seems to develop at around $10$Myrs, and the 
synthetic disc lifetimes can reach up to $20$Myrs for the largest initial discs and smallest host stellar masses.

One can make these lifetimes shorter by increasing the initial flaring,
considering a flared profile with a constant value $\upkappa=0.05$.
The left panel of Figure~\ref{rx_erosion} corresponds to the evolution of the surface density  $\Sigma(r,t)$ with time
of the central model of the grid, the $K$ model, using the fiducial X-ray wind mass losses,
a viscosity value of $\alpha=10^{-4}$, and the P5 profile. The bottom horizontal axis scale provides the density at given radii.
The lighter the colour, the lower the density, with white colour indicating that the density
floor level has been reached. These diagrams also provide information about when a dent in the density 
profile is produced, if the erosion is slow enough. These dents
might eventually produce further gaps and outer ring-like features. 
The red dots indicate the dents as local minima in the density profiles.
The right panel shows how slicing these diagrams at given times one can visualise these features in the density profiles.

Earlier times are at the top of the figure, and the $y$-axis shows the time evolution 
of the surface density from top to down. The internal gap begins to develop around $7$Myrs
and the disc is completely eroded around $9$Myrs.
In terms of temperature, the larger the $\upkappa$, the higher the temperature at a given radius. 
Hence, one gets a faster eroding in this case, and a shorter lifetime.
As summary, the cavities seen in these simulations are very short lived, lasting just a few Myrs 
but for the smallest stellar masses and largest initial disc masses.

This Figure~\ref{rx_erosion} also shows the evolution with time of the disc mass as an orange curve.
The upper horizontal axis shows how fast the disc mass decreases with time.


Simulations using a higher viscosity value, $\alpha=0.001$, 
lead to the surface density to decrease faster. 
The synthetic models produce inner cavities, but have difficulties on creating ring-like structures at any age,
unless we use very flattened discs.
This is because the peak belonging to the wind front moves fast
inside out, quickly eroding the external disc.

Dents and ring-like features can be also formed, but they are short-lived and created at very late times.
Therefore, these simple photoevaporative models 
with constant flaring profiles can explain the generation of inner cavities and
some ring-like features, but in a limited way.
They do not fully account for creating the observed structures 
in transition disks of ages of Myr or less.
Moreover, the internal cavities are only formed when the disc masses are much lower
than the observed values, as reflected by the orange curves in the plots.

Hence, the age of the disc when the 
cavities and ring-like features appear is a key issue. One might consider that some of the parameters 
of the Taurus sample may be subject to observational uncertainties, mainly the ages.
One may also take into account that the generation of a dent is not always followed by the creation 
of a gap in our models.
However, the possibility of creating dents 
with this simple approach is important because once the dents
are present, gas removal processes may trigger 
collective mechanisms for planetesimal formation such 
as the streaming instability or some increases in the  dust-to-gas ratio,
that may act as dust traps \citep{ercolano17}. Hence, in the following sections 
we will explore how to produce earlier in time these dents, keeping at the same time the proper 
disc mass.

\subsection{The impact of a progressive flattening in the profile of flared discs}
\label{rxwindflatten} 

Discs are flared, as seen by spatially-resolved mid-IR imaging in nearby 
PPDs \citep{doucet07,maaskant13}. 
In these discs, the outer parts of the disc intercepts more radiation from the central star. 
Moreover, according to Equation~\ref{shakura}, the selection of the flaring profile directly impacts the viscosity, thus the evolution of the disc.
We want to analyse the impact of a changing profile of the flared disc
as the disc evolves, with a progressive
flattening of the disc as the age increases
and the dust may settle at different rates. 
Hence, this section aims to analyse the role that these changes may have in the generation of ring-like features
at outer radii.


We will model a decreasing flattening of the disc by 
decreasing the $\upkappa$ parameter in Equation~\ref{flaringprofile}.
A decrease in the flaring of the disc means a decrease in the temperature profile.
Hence, the accretion rate is made stronger, and one gets
a faster starvation of the disc and the ring-like features might appear before.

As a first approach, we can take a model where the disc starts 
with a given flaring profile, keeping $\upkappa$ constant during the primordial phase. 
When the direct-flux phase begins, let us say, at $t_{direct}$, the disc is flattened by lowering the $\upkappa$ value.
This new value remains fixed for the whole remaining disc lifetime.

This one step piecewise flattening is a very rough modelling of the
expected continuous flattening to happen in real systems. Hence, the baseline for our 
synthetic models will consider a second run for each system. This second run 
is based on a $P52$ profile. This means 
that we start with an initial flaring value of $\upkappa=0.05$ that gradually decreases
up to $\upkappa=0.02$ in five steps.
The center point with the fiducial value $\upkappa=0.03$ is set at time $t_{direct}$,
when the direct flux started in the first simulation run. Of course, one should note
that the direct phase can start in the second run at a different time with respect the first simulation,
because we have obviously changed the flaring profile of the disc.

\begin{figure*}
\begin{center}
\begin{tabular}{ccc}
\includegraphics[width=0.3\linewidth]{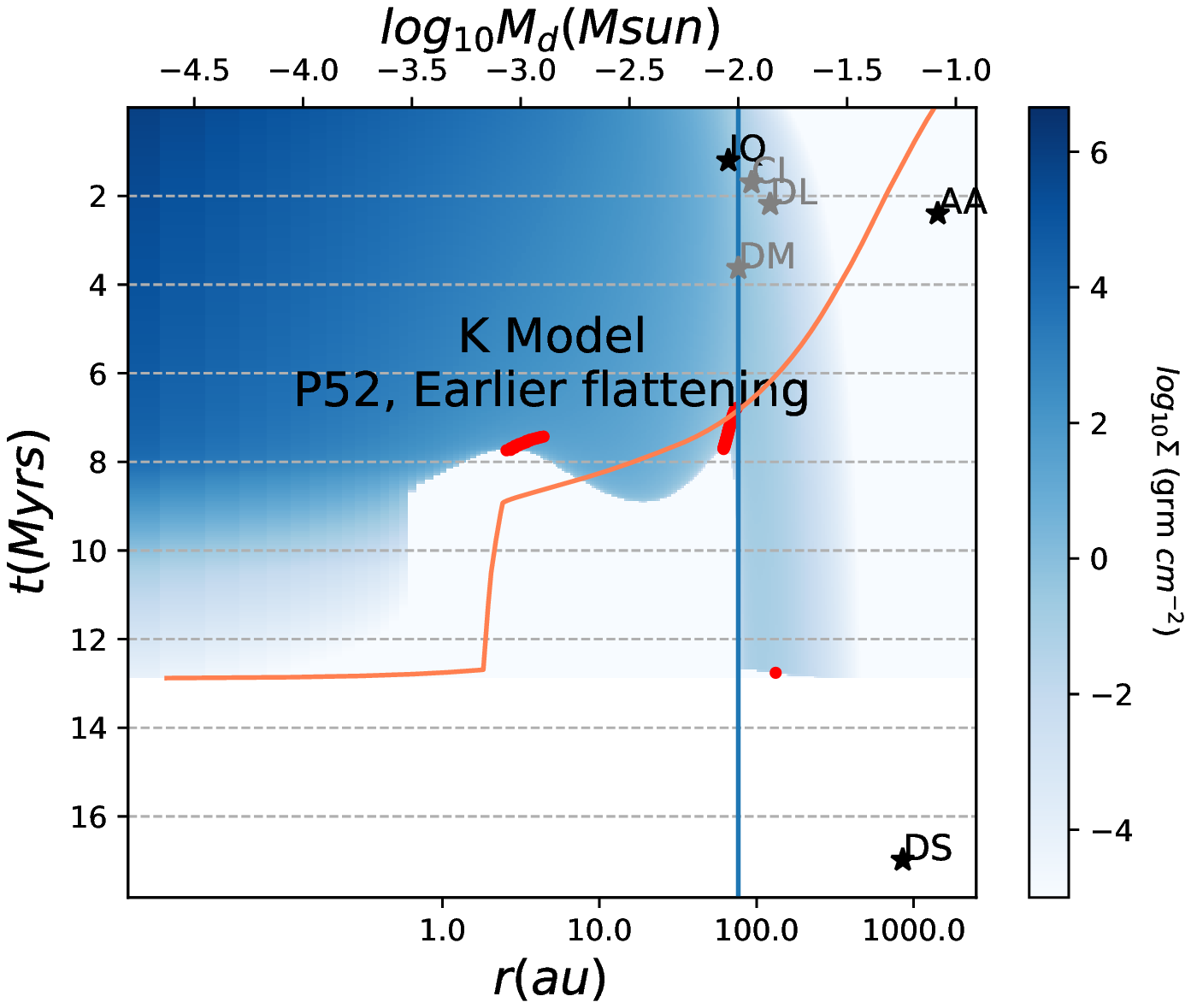} &
\includegraphics[width=0.3\linewidth]{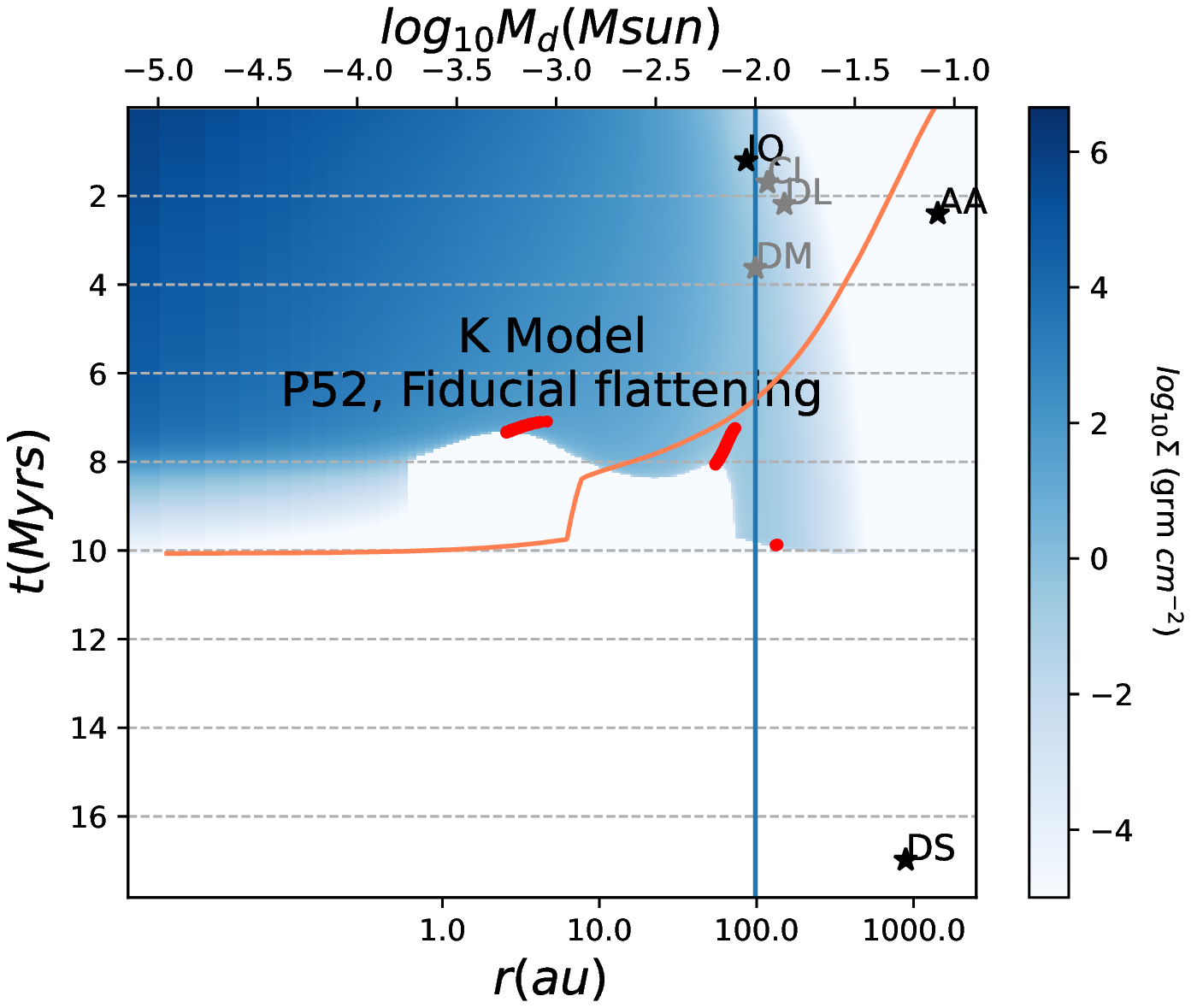} &
\includegraphics[width=0.3\linewidth]{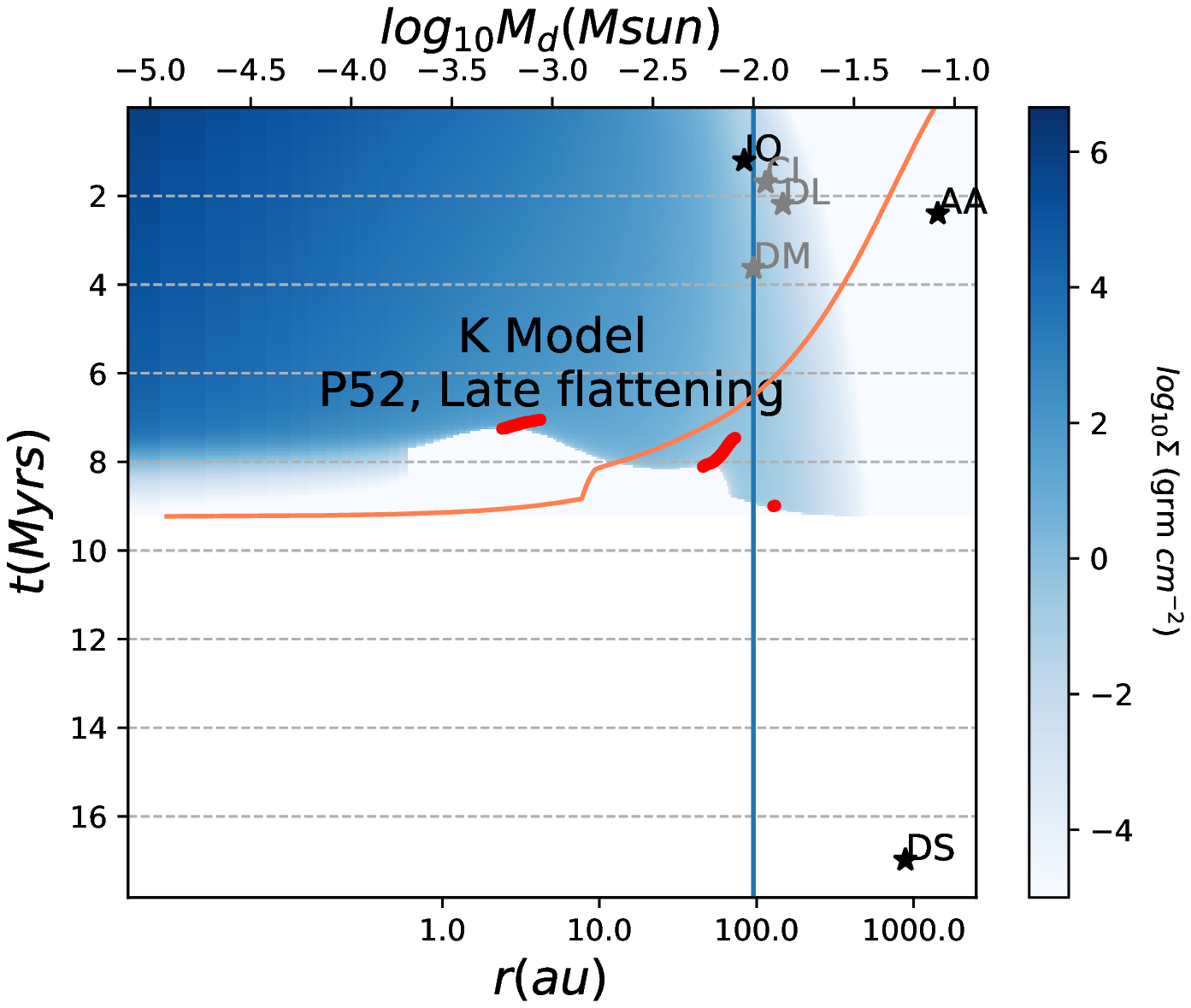} \\
\end{tabular}
\end{center}
\caption{Density diagrams showing the impact of an earlier or late flattening process.
They show the evolution of the central model of the grid (model $K$) under photoevaporation 
produced by X-ray dominated heating and a viscosity value of $\alpha=10^{-4}$,
subject to a progressive $P52$ flattening of the disc.
The center column is the fiducial flattening, that sets the value $\upkappa=0.033$ at $t_{direct}$ time
(see text for details).
The earlier the time when the progressive flattening starts,
the longer the cavity lasts and the longer is the phase when ring-like features 
can exist.}
\label{rx_erosion2}
\end{figure*}

The Figure~\ref{rx_erosion2} shows the impact of this progressive flattening in the evolution of the
the central model of the grid (model $K$).
The center column is the fiducial flattening, that sets the value $\upkappa=0.033$ at $t_{direct}$ time.
As visual reference, the blue vertical line marks an arbitrary value of $0.01M_{\odot}$, that 
indicates when disc mass is very small.

One can consider that the actual flattening may start earlier or later than $t_{direct}$. 
Hence, this figure also shows the results when we move forward and backwards the time when flattening steps
take place. The earlier the time when the progressive flattening starts,
the longer the cavity lasts and the longer is the phase when ring-like features 
can exist. The lifetime of the disc seems to grow, as reflected by the orange curve.
Conversely, when the flattening takes place later, it seems that the lifetimes are similar or slightly shorter.

When one compares these diagrams with those resulting from a constant flaring profile, 
one can see that internal eroding phase takes slightly longer. The orange curve shows that the disc mass will 
slow down in this phase. This allows the wind to slowly erode the outer parts of the disc before
starting the direct phase, which can ease the creation of a third dent in the outer part of the disc.

\begin{figure*}
\begin{center}
\begin{tabular}{cccc}
\multirow{3}{*}{\rotatebox{90}{$\xrightarrow{\makebox[6cm]{$M_{d}$}}$}} &
\includegraphics[width=0.3\linewidth]{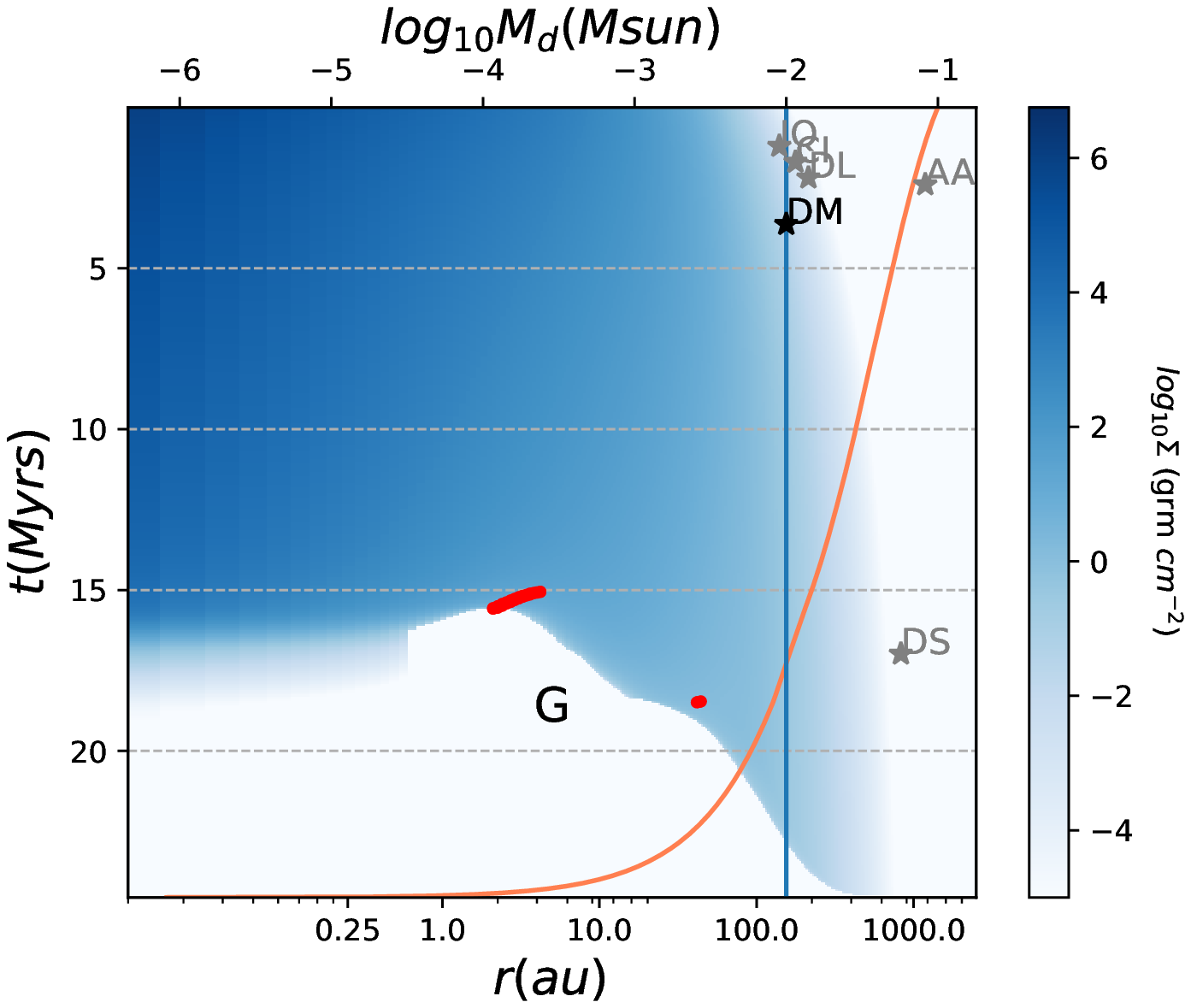} &
\includegraphics[width=0.3\linewidth]{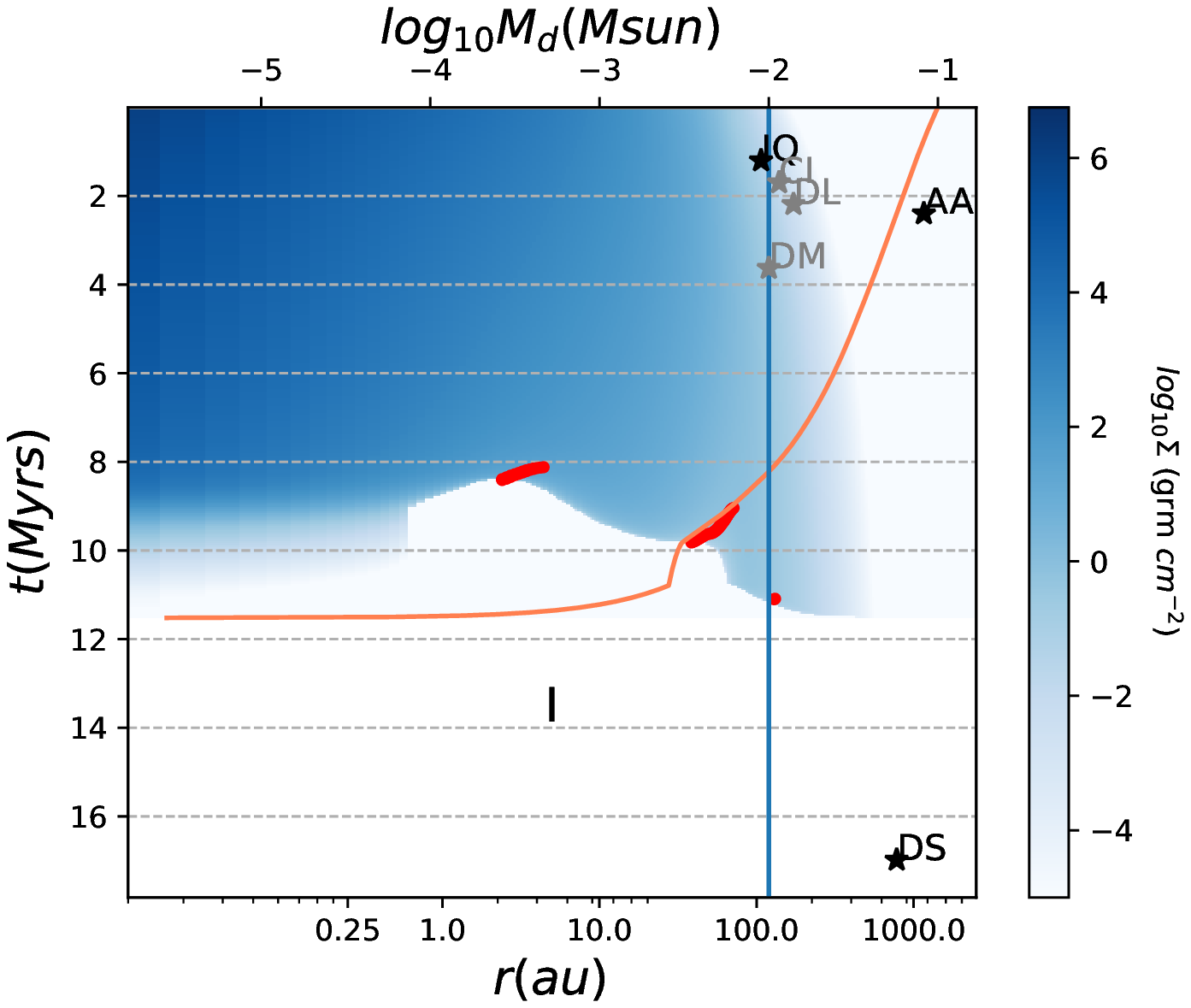} &
\includegraphics[width=0.3\linewidth]{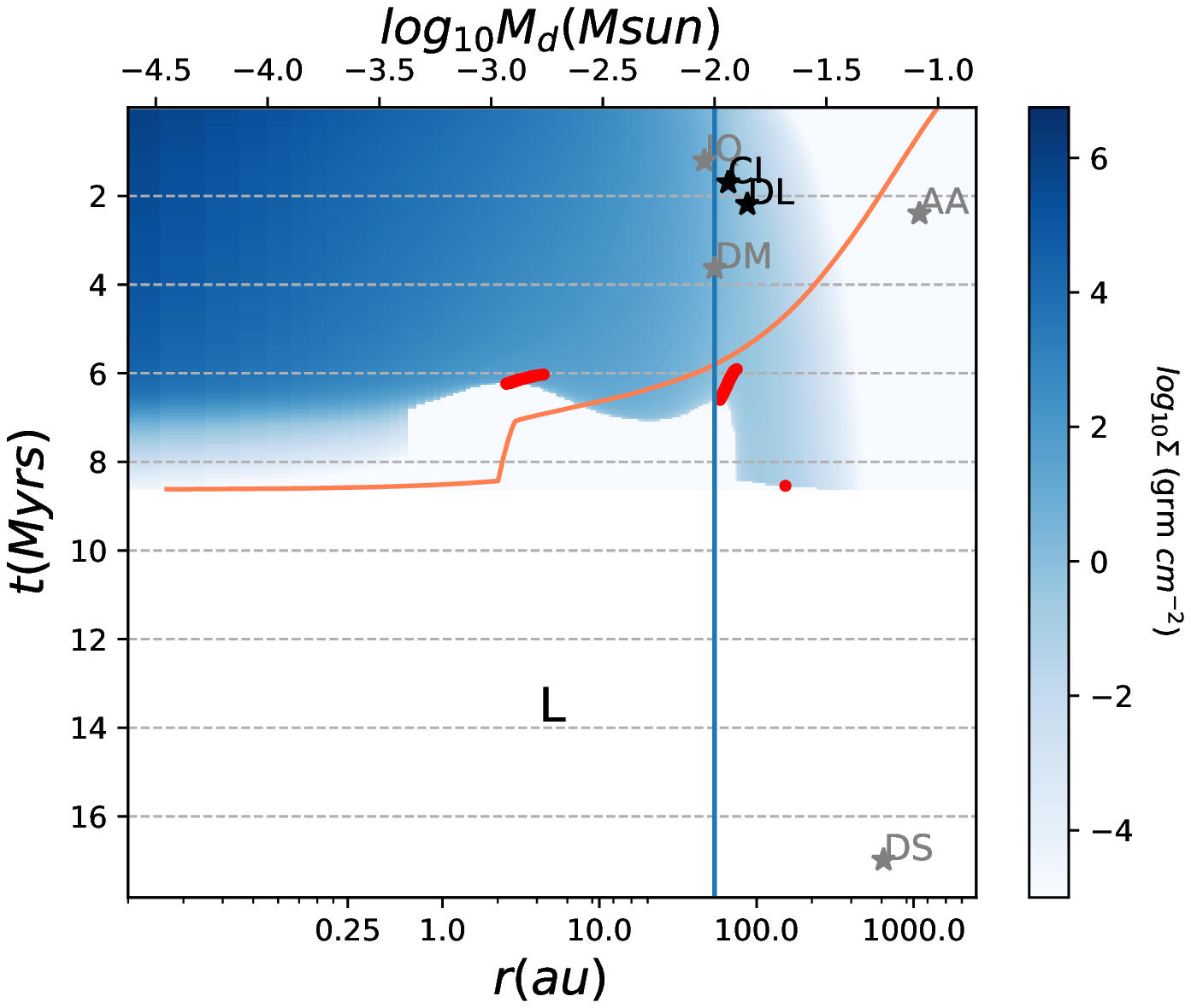} \\
&\includegraphics[width=0.3\linewidth]{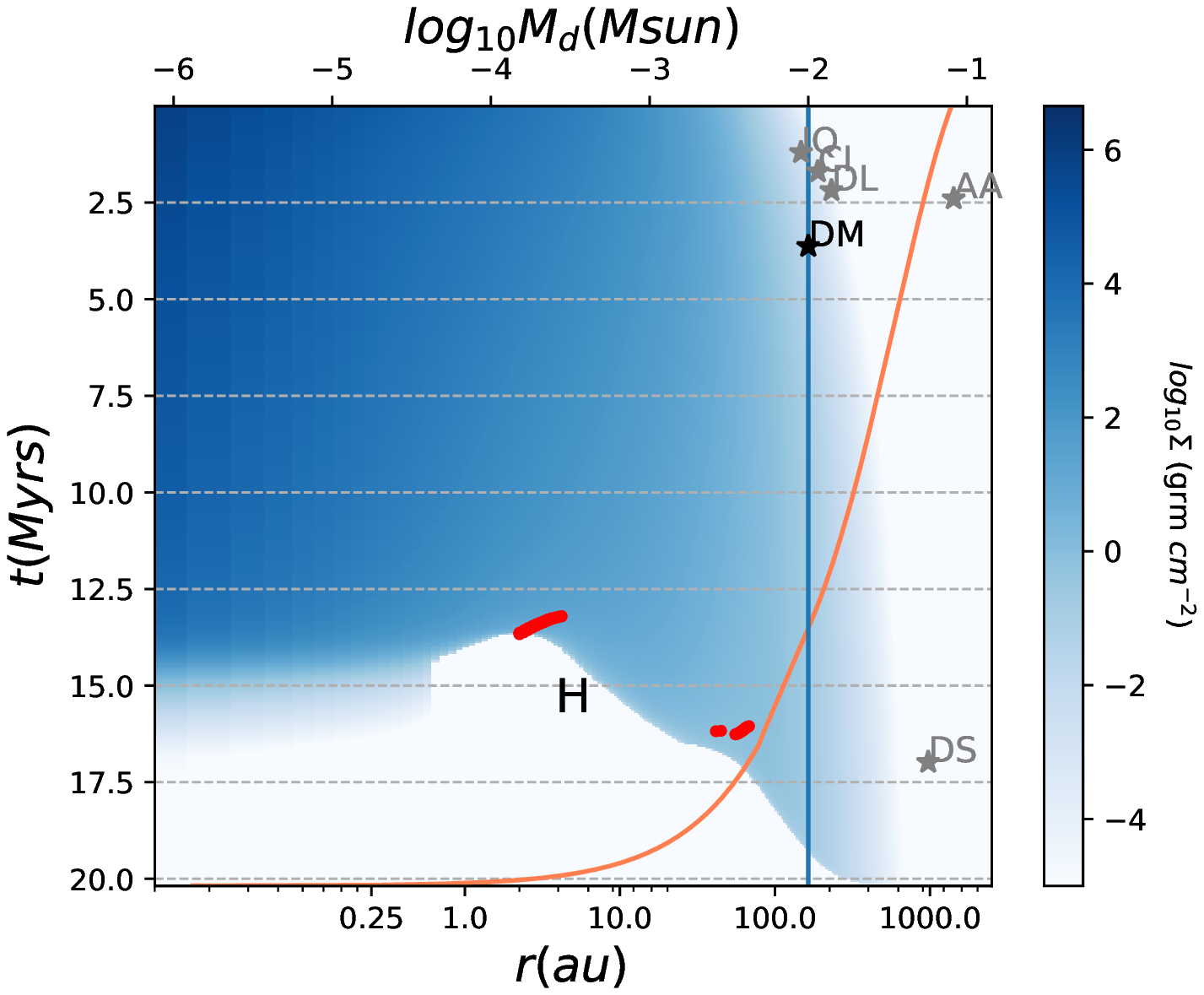} &
\includegraphics[width=0.3\linewidth]{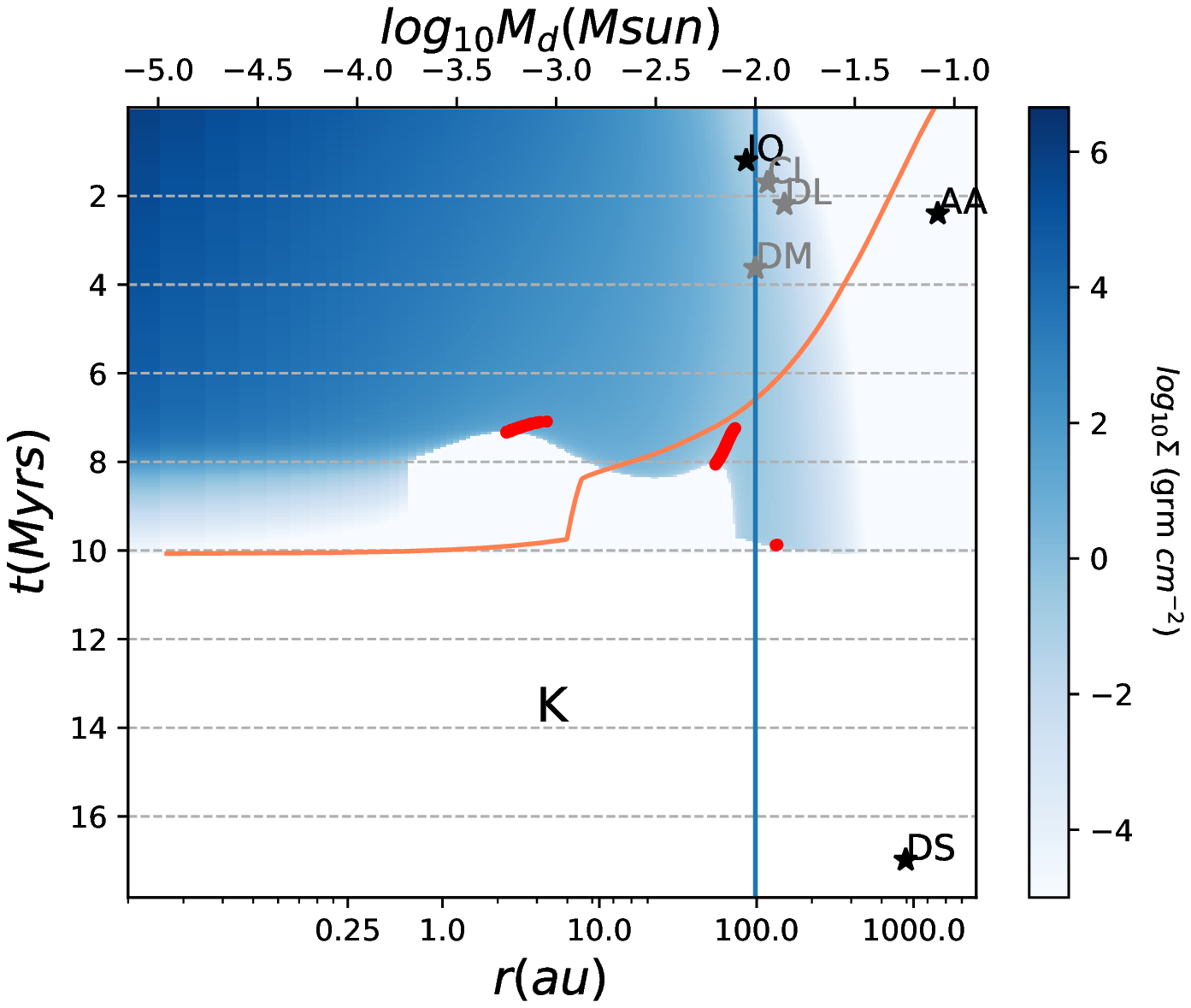} &
\includegraphics[width=0.3\linewidth]{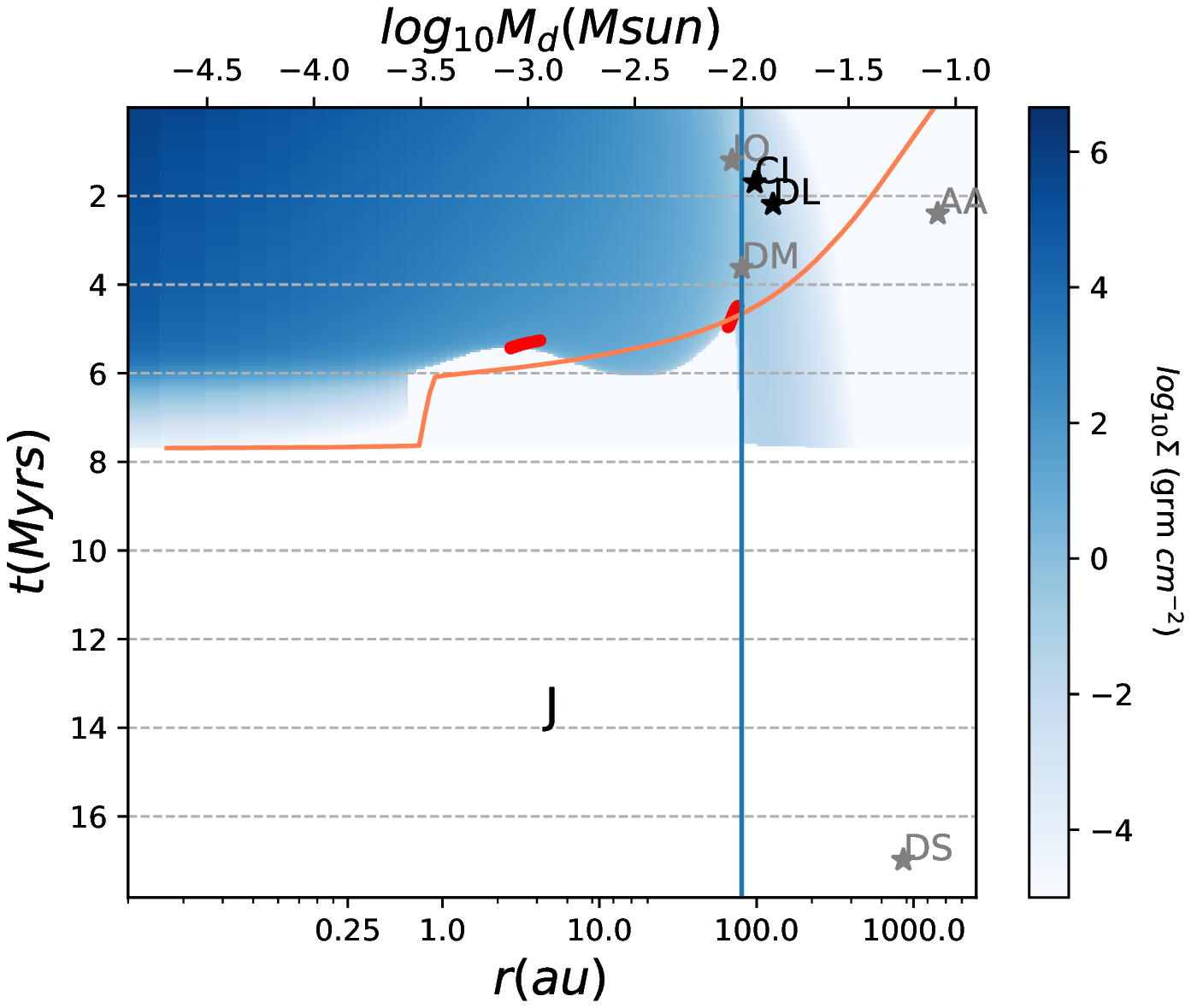} \\
&\includegraphics[width=0.3\linewidth]{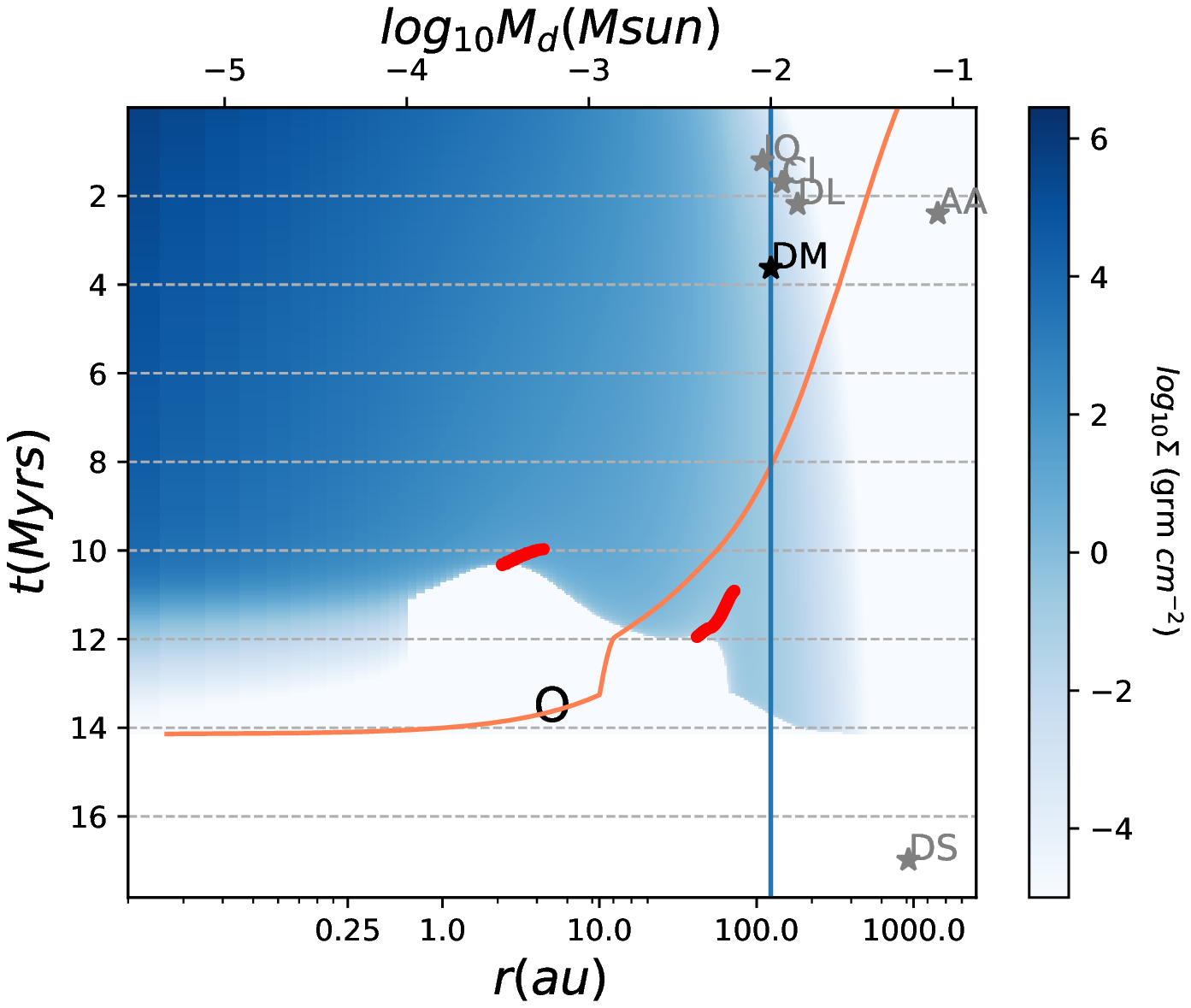} &
\includegraphics[width=0.3\linewidth]{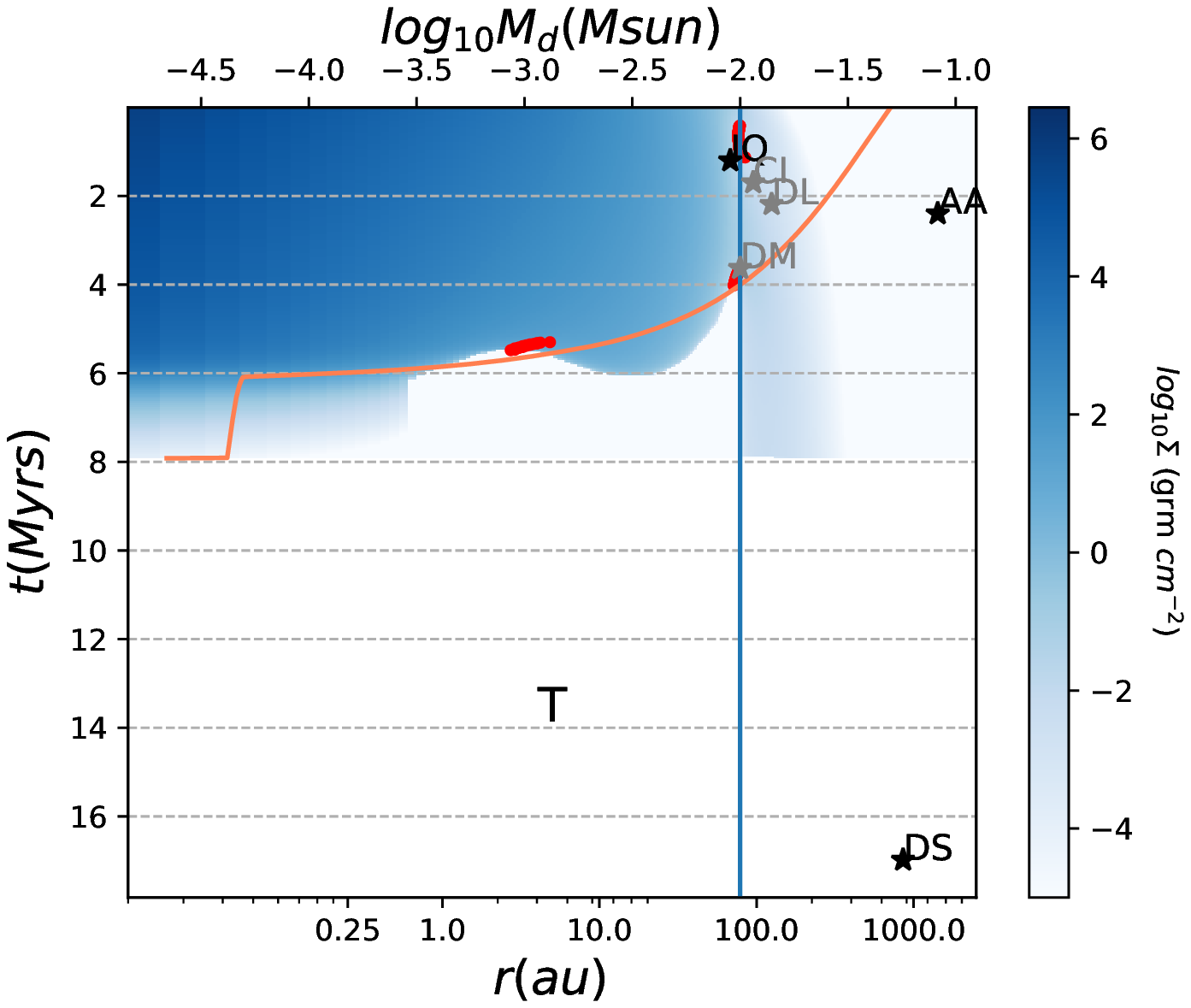} &
\includegraphics[width=0.3\linewidth]{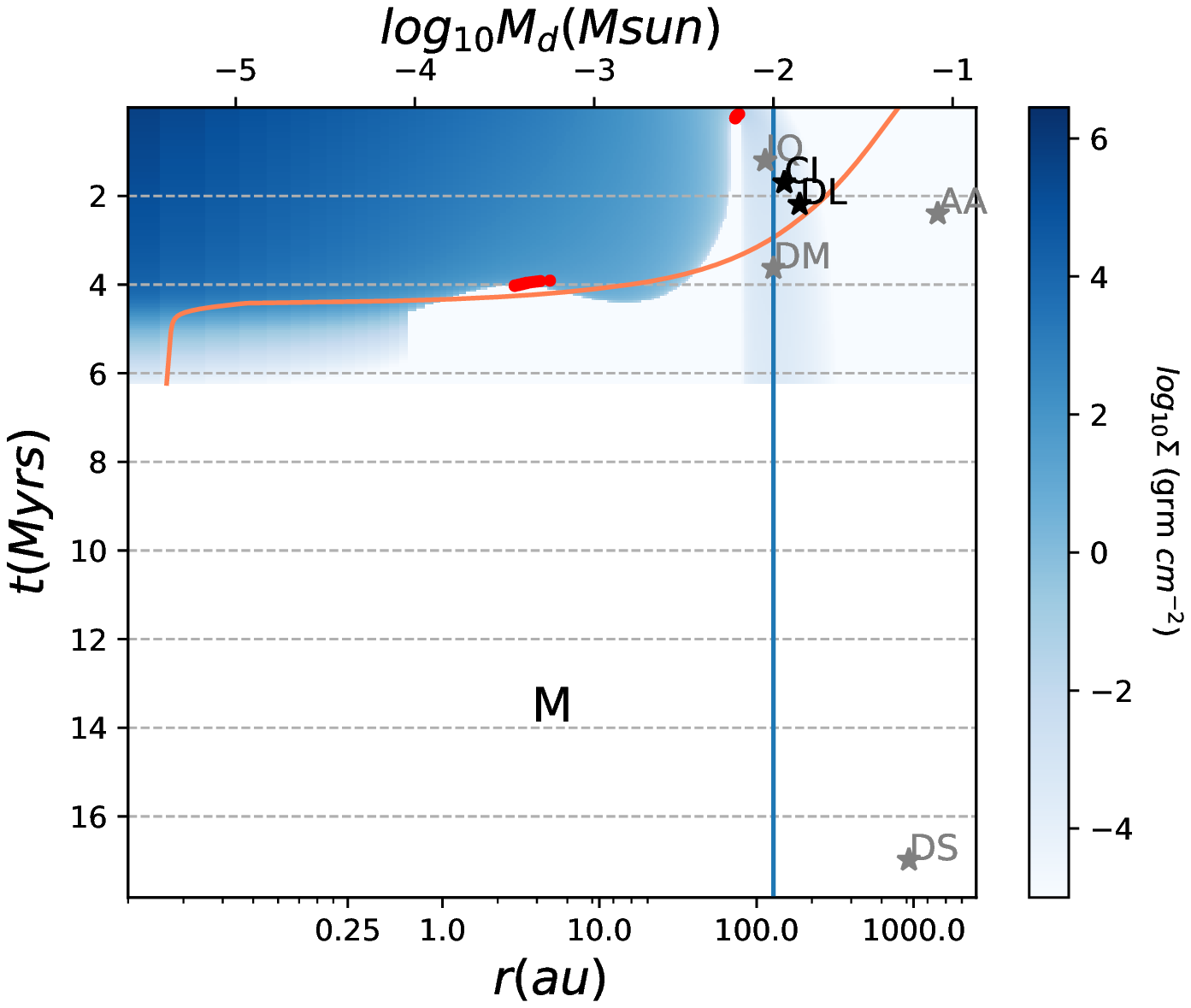} \\
&\multicolumn{3}{c}{  $\xrightarrow{\makebox[6cm]{$M_{*}$}}$  } \\
\end{tabular}
\end{center}
\caption{Evolution of the grid of models for a fiducial photoevaporation 
controlled by X-ray heating, $e_x=1.0$, a viscosity value of $\alpha=10^{-4}$
and a progressively flattened flaring profile $P52$.
The surface density in blue gets lighter as the disc is eroded.
The red points indicate local minima in the density profiles.
The second dent can be carved at the same time or even at earlier times than the 
the inner dent. 
These two dents can produce ring-like features
when exist at the same time.
Sometimes, these two dents can be accompanied by a third outer dent.
The Taurus discs are also plotted for reference. They  
appear in black when their stellar masses are similar to the mass of the analysed synthetic model. Otherwise, they appear in grey.
}
\label{rx_3a410}
\end{figure*}

The Figure~\ref{rx_3a410} shows the density diagrams corresponding to the evolution of the grid of models
for a fiducial photoevaporation controlled by X-ray heating, a viscosity value of $\alpha=10^{-4}$
and this progressively flattened flaring profile $P52$.
Because the time when the direct flux stage gets active is different in each model, 
the time when the flattening occurs is also different in each case.

The larger the stellar mass, the faster the creation
of the inner cavity, and faster the decrease of the disc mass.
Changes in the flaring profile means changes in the eroding rate.
As a consequece, the systems with smaller host stellar masses and largest initial disc masses 
($G$ and $H$) present somehow long-lived cavities, while 
the remaining systems show very short cavity lifetimes.

The red points indicating local
minima (dents) show that as the stellar masses and initial disc masses increase,
the second dents appear earlier. These dents (sometimes eventually converted into gaps) can produce ring-like features
when exist at the same time that the first dent. Moreover, 
an additional third dent can be found in the systems with large enough disc masses ($K$,$I$ and $L$).


Following our simulations, 
the first close-to-star and the second external dents 
are not created at the same time.
First, the inner dents are created. Then, a few Myr later, 
the outer structures appear. 
Notably, the second dent can be produced earlier than the first one
when the initial disc mass is small enough, as best seen 
in models $T$ and $M$.

These dents may or not convert into gaps and lead to true rings. 
The dents sometimes do not evolve into real gaps, because the disc in the surroundings 
reaches the floor density before the dent fully develops. However, 
we will still refer to them as ring-like features, because 
such dents may be considered seeds for true rings. 
As we are dealing with very simple models, one can consider
that such dents, linked to density gradients, may trigger
additional mechanisms for forming traps that 
may enhance the efficiency of the gap generation process beyond our simple semi-analytical model.

In real systems, gaps and ring-like features are found
at essentially any radial location, from close-to-star to outer structures
\citep{andrews20}. This can be seen in Figure~\ref{models},
even when it is important to remark that neither the 
measured positions of the gaps nor the ages are fully 
consolidated.

Some of the features produced in our simulations resemble those found
in real systems. However, one also must consider the ages when the 
ring-like features are produced and the value of the mass of the disc at that time.
This evolution of the disc mass is given by the orange curves. Taking into 
account these issues, the model $M$, the system with the largest stellar mass and smaller initial disc mass,
is the only model roughly closer to some of the real systems of our sample ($CI$ Tau and $DL$ Tau).

Therefore, a key issue is the comparison of the ages when the features are produced 
in the synthetic models and when they are observed in real systems. Note that 
the observed discs sometimes seems to be relatively short-lived, with
transition times from primordial to discsless of roughly $10^5$ yr \citep{owen12,anderson13}. 
The mass depletion mechanism, once there is a hole in the inner regions of the disc, 
seems to lead to a quicker evolution of the discs, making these discs to be typically observed at ages 
of $1$Myr, but to have disappeared at ages around $10$Myr.
Therefore, one may need to modify another parameter for getting these short-lived discs.

\subsection{The impact of the wind efficiency}

Up to now, we have considered fiducial values those given by the profiles from \cite{owen12}
as the baseline for the integrated mass loss rates. 
These results come from a numerical fit to a population 
synthesis study, based on the dependency on $L_X$ and $M_*$.
Following those models, the mass-loss rates scale linearly with
X-ray luminosity, having values from $10^{-11}$ to $10^{-7}$ $M_{\odot}/yr$. 
The mass losses in real systems can deviate from those 
values, and one can observe a variety of mass rates, 
averaged as $(7.5 \pm 2.6) 10^{-9} M_{\odot}/yr$, 
that may span from $10^{-10}$ to $10^{-7}$  \citep{jennings18},
with these values coming actually from observations of Taurus
cluster \citep{gudel07}. 

We have included these variations in the photoevaporative 
wind mass loss rates by adding an arbitrary 
\emph{efficiency factor} $e_x$. This factor will model the possible increases 
(or decreases) in the wind mass loss rates. 
When $e_x=1.0$, we have the fiducial value $10^{-8}$ $M_{\odot}/yr$. A factor $e_x=0.5$ 
lower the mass rates to be half of the fiducial mass rate, meanwhile a factor of 
$5.0$ indicate a mass loss rate five times the baseline value.
We note that this strong efficiency is far in the tail of
the X-ray luminosity distribution in \cite{gudel07}. However, we will 
use such a large value aiming to observe the consequences of shortening 
the disc lifetimes when increasing the efficiencies. 

\begin{figure*}
\begin{center}
\begin{tabular}{cccc}
\multirow{3}{*}{\rotatebox{90}{$\xrightarrow{\makebox[6cm]{$M_{d}$}}$}} &
\includegraphics[width=0.3\linewidth]{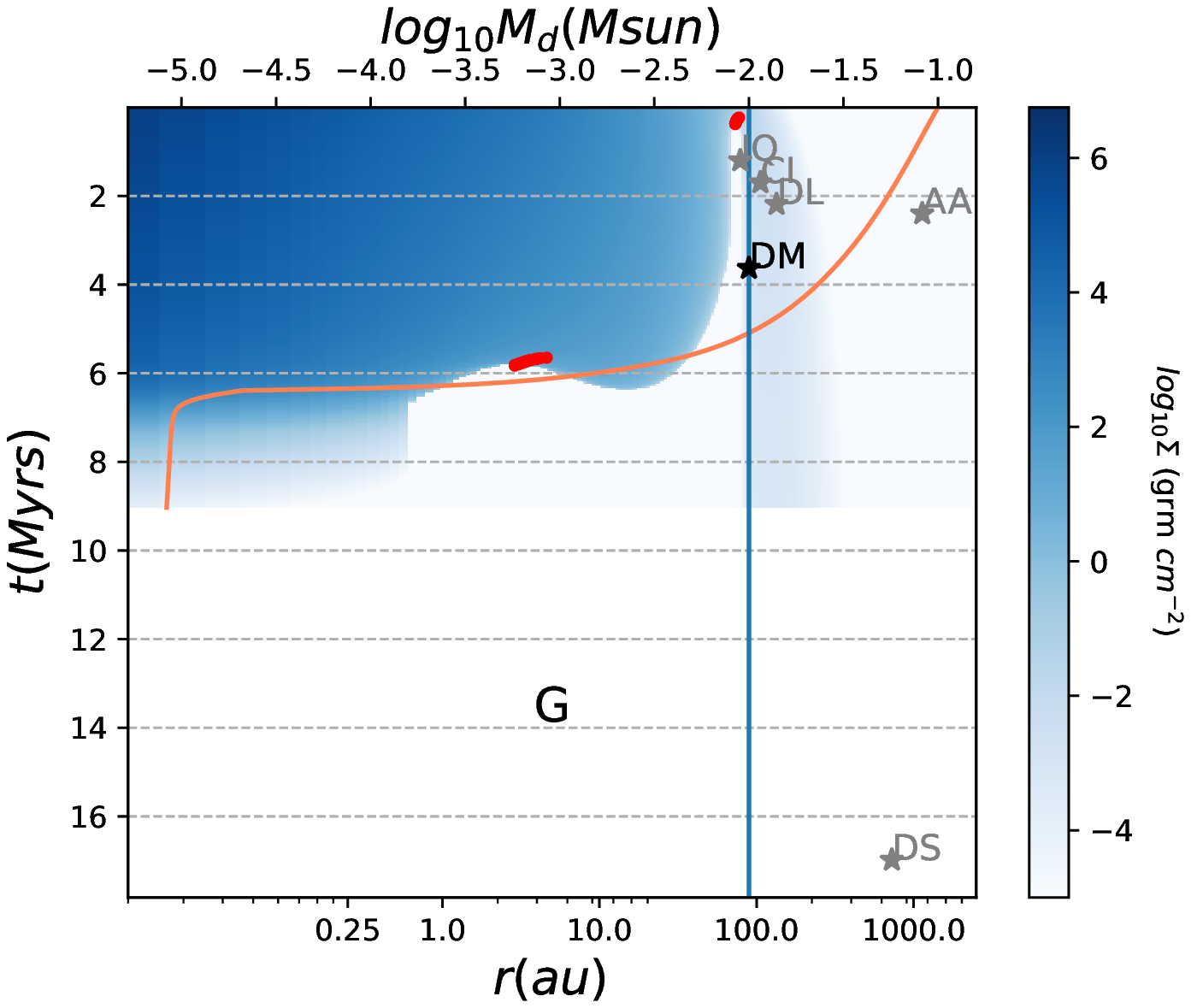} &
\includegraphics[width=0.3\linewidth]{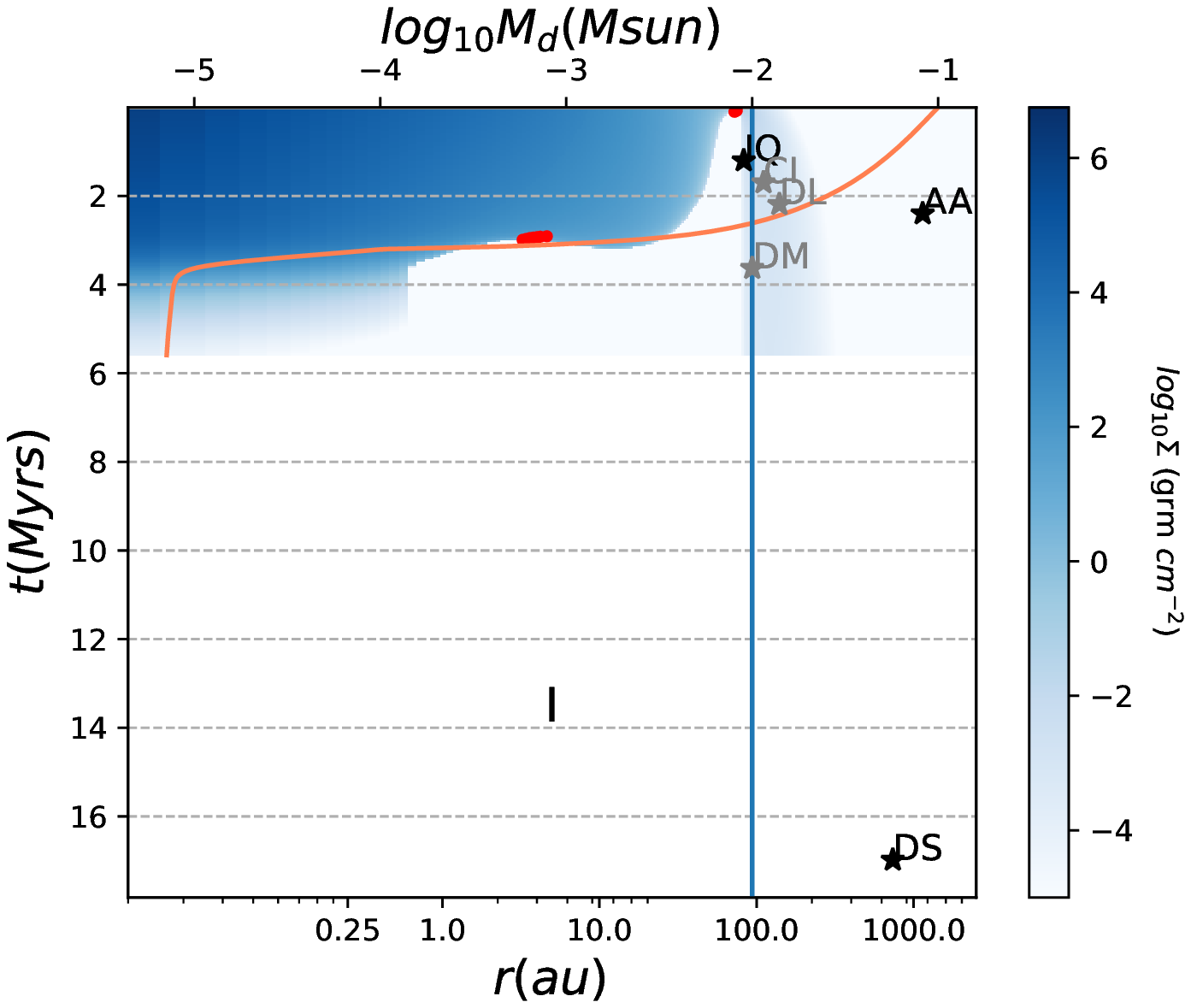} &
\includegraphics[width=0.3\linewidth]{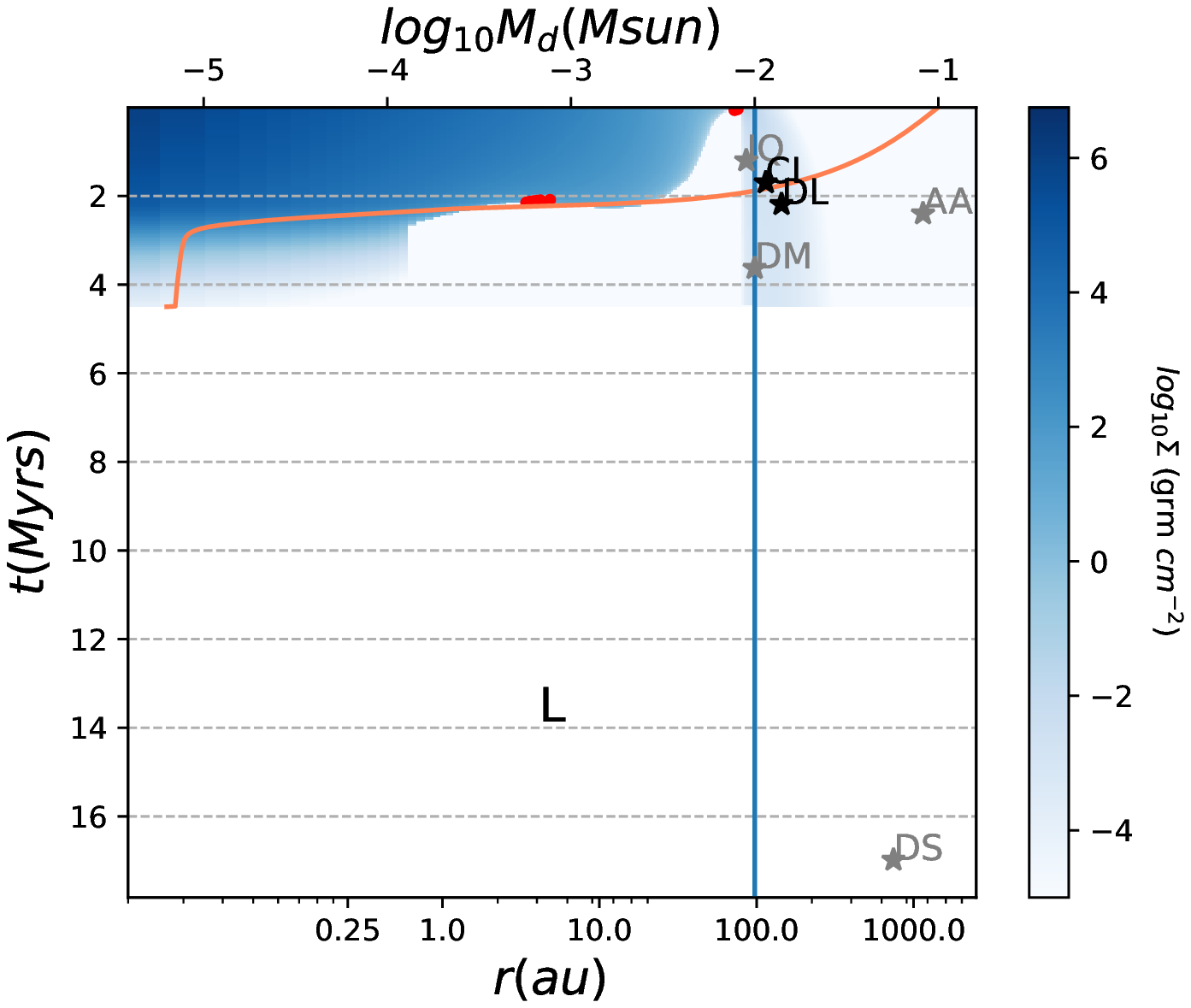} \\
&\includegraphics[width=0.3\linewidth]{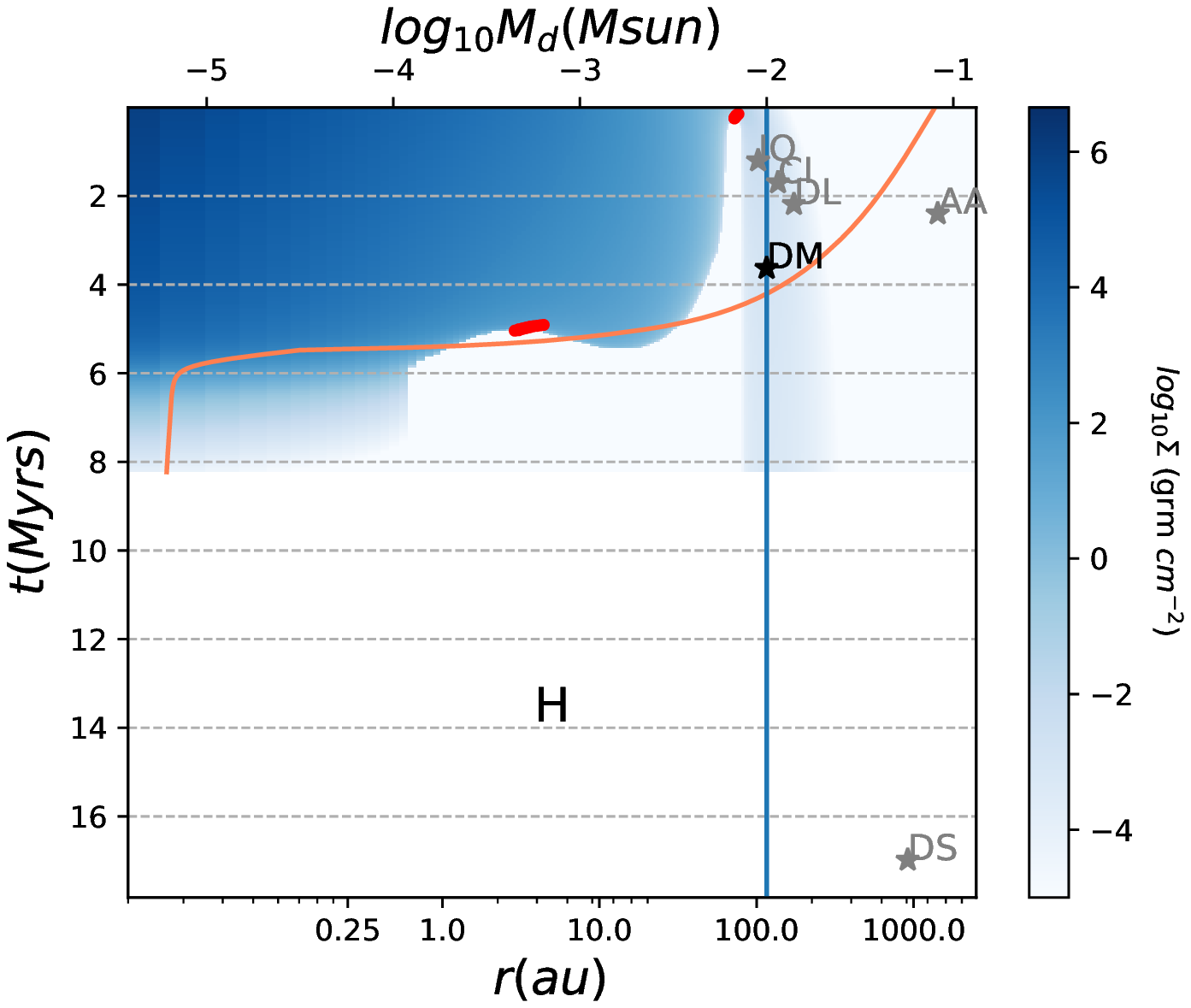} &
\includegraphics[width=0.3\linewidth]{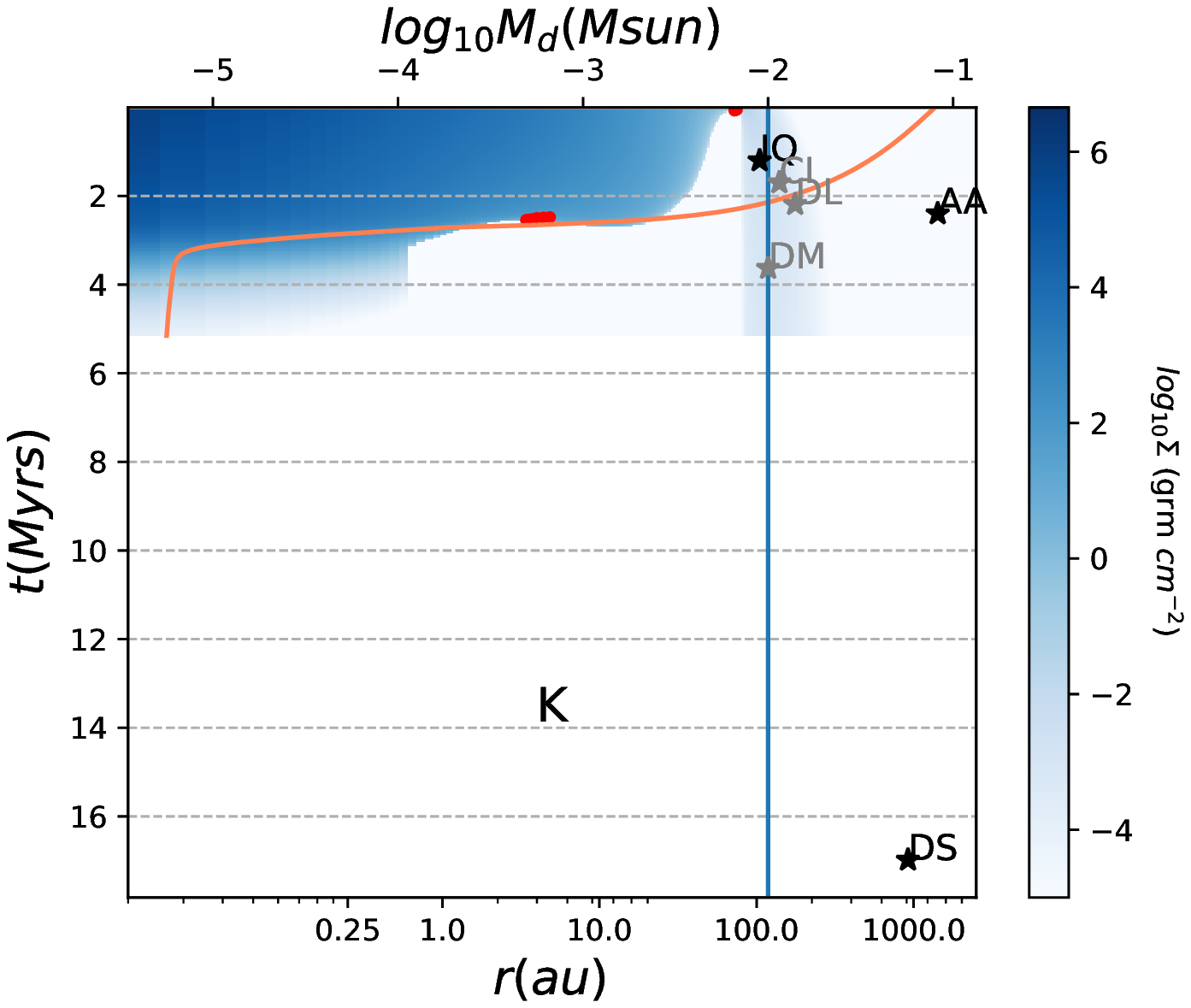} &
\includegraphics[width=0.3\linewidth]{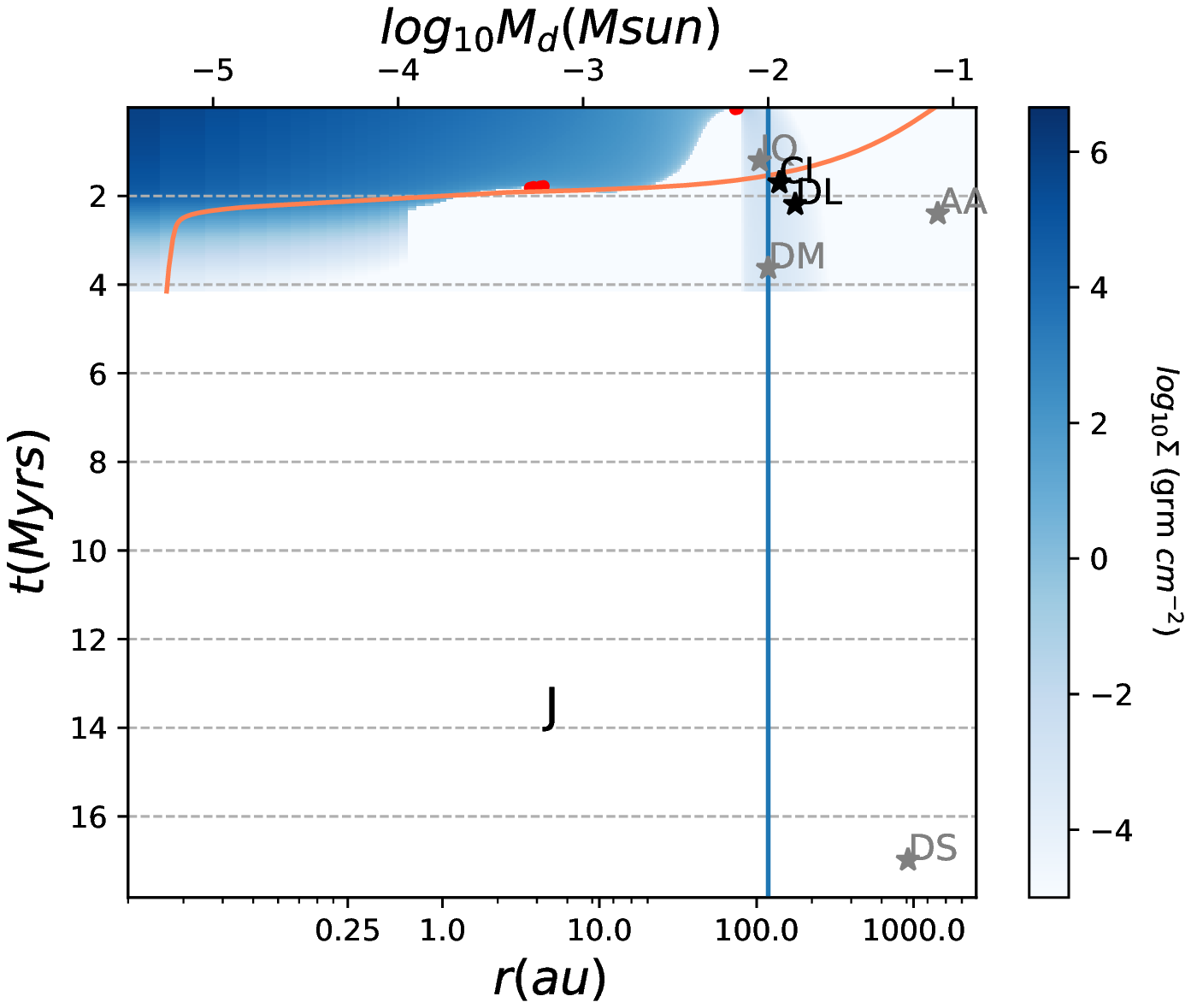} \\
&\includegraphics[width=0.3\linewidth]{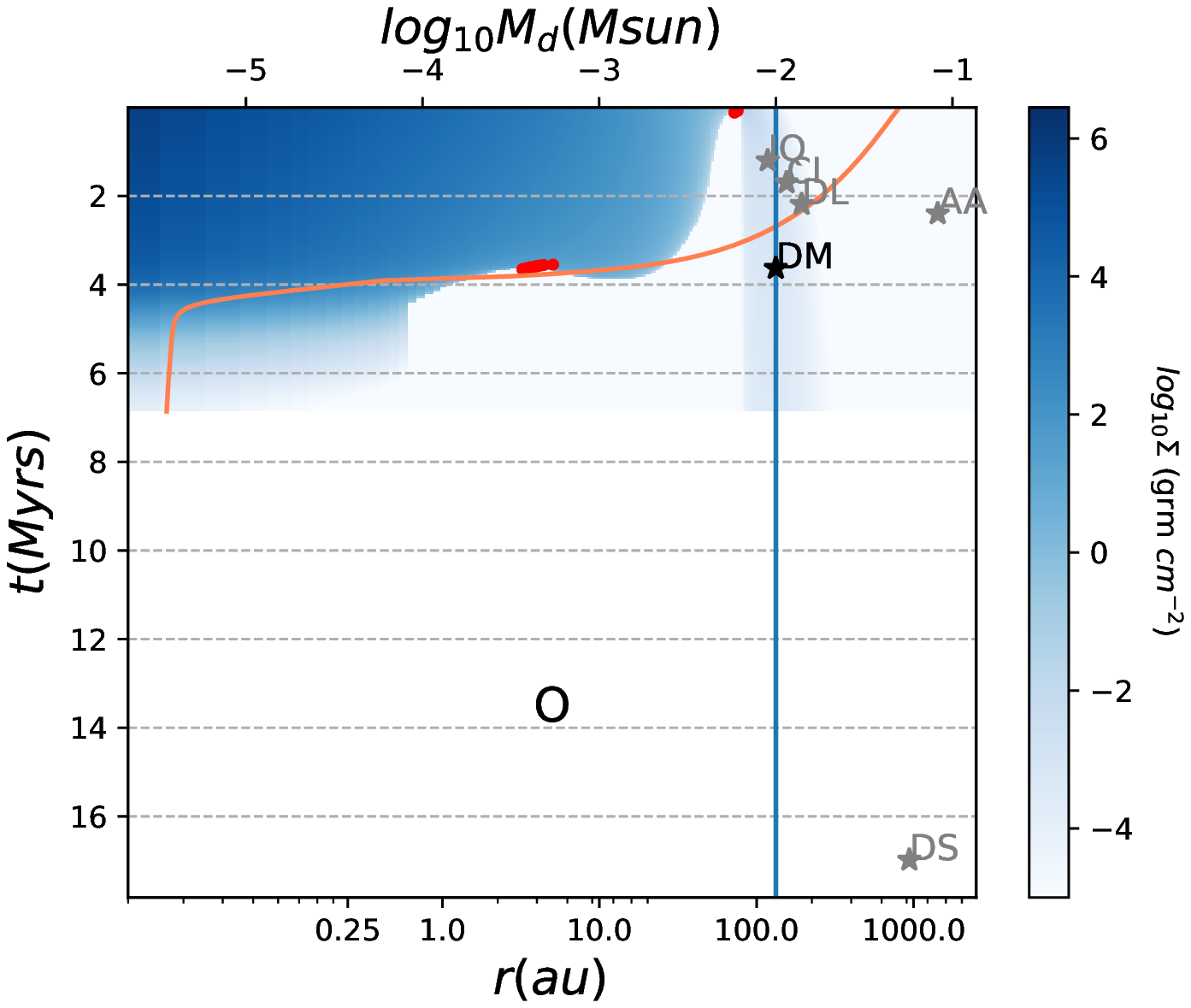} &
\includegraphics[width=0.3\linewidth]{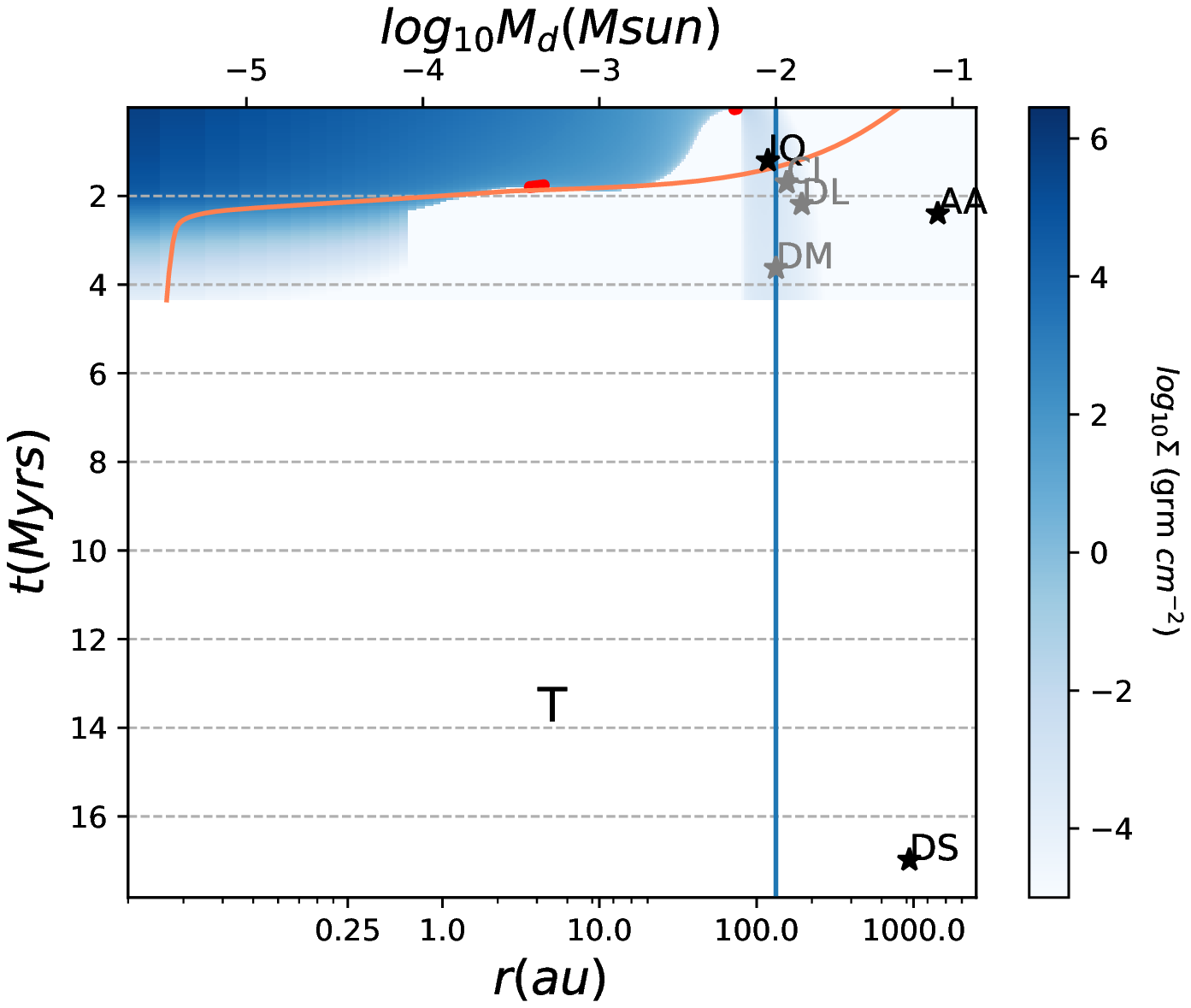} &
\includegraphics[width=0.3\linewidth]{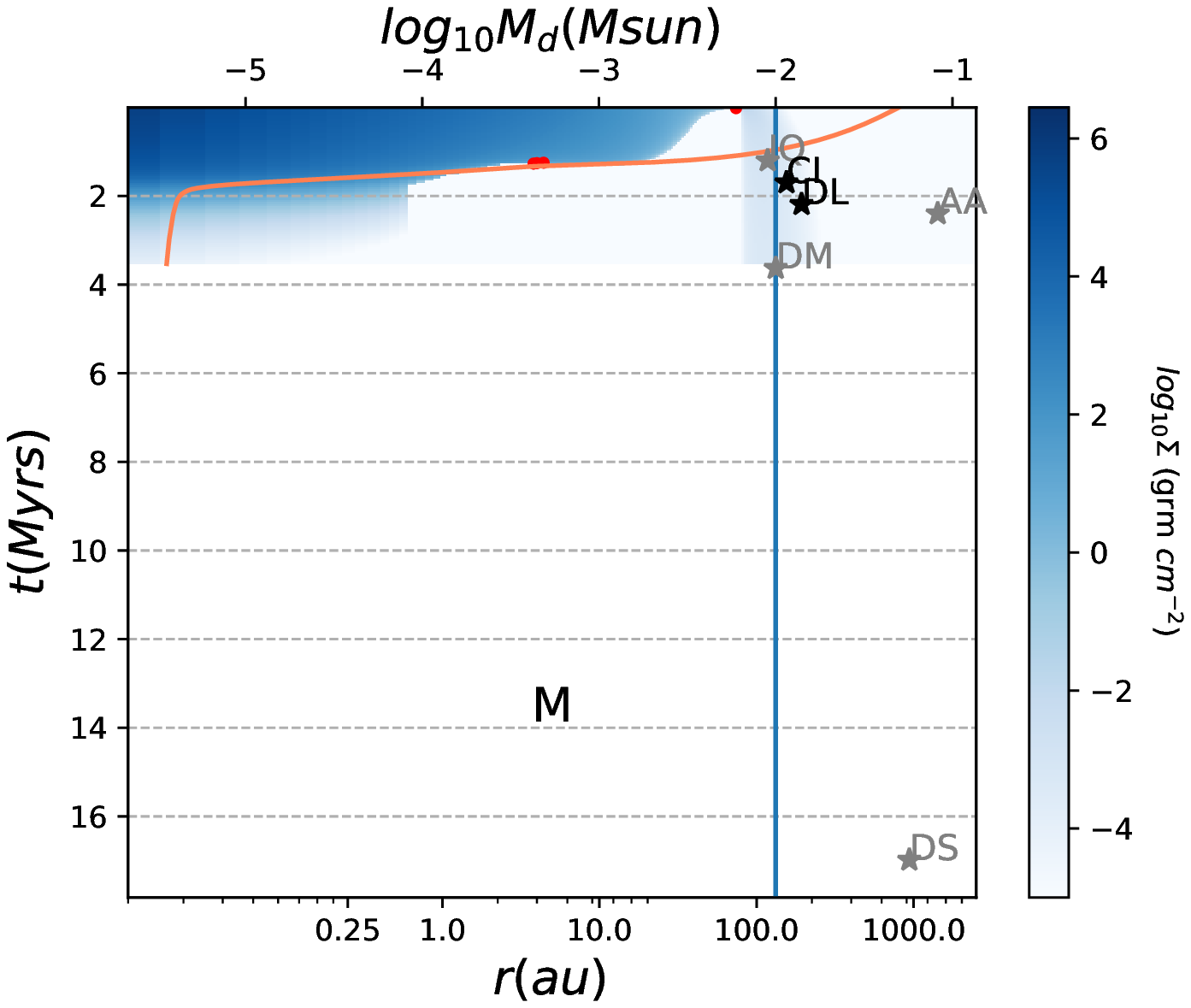} \\
&\multicolumn{3}{c}{  $\xrightarrow{\makebox[6cm]{$M_{*}$}}$  } \\
\end{tabular}
\end{center}
\caption{Evolution of the grid of models 
for a photoevaporation controlled by X-ray heating,
with a wind with efficiency parameter $e_x=5.0$
(meaning the X-ray wind mass losses to be five times the fiducial mass rate),
a viscosity value of $\alpha=10^{-4}$
and a variable flaring profile $P52$.
The surface density in blue gets lighter as the disc is eroded.
The red points indicate local minima in the density profiles. These
are produced much earlier than when $e_x=1.0$, and the disc lifetimes
are much shorter too.}
\label{rx_3b450}
\end{figure*}

The Figure~\ref{rx_3b450} shows the results got when the previous models with viscosity
$\alpha=10^{-4}$ and the flattening profile P52 (where
$\upkappa=0.05$ changes to $\upkappa=0.02$) evolve subject to winds with an increased efficiency factor $e_x=5.0$. 

The eroding of the disc is now stronger and the discs lifetimes
are shorter. The most interesting fact is that 
the second dent and ring-like features produced in the outer part of the 
discs are now produced in almost all models of the grid, and much earlier than in the previous cases,
at times when the disc masses $M_d$ are closer to the observed values.
However, one must also note that the ring-like features produced when two dents or gaps exist
at the same time are transient, and they may have very short lifetimes. 

The disc masses can be obtained from CO measurements \citep{williams14,miotello16,rosotti17} 
or from the dust IR continuum flux, using an adequate dust-to-gas ratio \citep{lodato17}. 
Therefore, the observed disc masses may have systematic uncertainties because the CO 
ratio abundances estimation in the first case or the
selection of the dust-to-gas ratio and the considered dust opacity in the second
case. In any case, we take them as a reference point, and
when we overlay the values corresponding to the Taurus systems, we see that 
they are best matched by the bottom row of the grid, 
which corresponds to the intermediate initial disc masses $M_d^i$.

When compared Figure~\ref{rx_3b450} with the results from 
Figure~\ref{rx_3a410}, that used the fiducial efficiency
$e_x=1.0$, one can note that with these strong winds the cavities just fully develop at the
very end of the disc lifetime. We can also see that 
the second dent grows and develops quicker than the inner one.
Finally, the third very external dent does not appear because
this fast erosion.

The evolution of these models with winds of increased
efficiency is quicker than the evolution of the models
with less-efficient winds. This makes them to have 
ages more consistent with the observed disc lifetimes. 
We have overlaid the Taurus sample systems over our 
grid of models. The observed ring-like features of the real system
with the smallest host stellar mass, the DM Tau system, 
seem to match in age with the $O,G,H$ models. Moreover, they also roughly 
have the adequate disc mass. However, the internal cavity observed in the real system 
seems to form too late in the synthetic models.
Observational uncertainties in age determination, different gas-to-dust ratios and additional mechanisms are again a key issue.

The proper estimation of the age of the disc is then
an important issue for proper calibration of the model.
This age is an observable quantity, subject to large uncertainties, 
such as the selection of the star evolutionary HR track, 
reddening or blue excess and stellar masss.
See a comparison of ages estimated by various methods in \cite{gomez13} and references therein.

In which concerns the intermediate stellar masses, the IQ Tau system
seems to roughly match the $T$ model when considering the production of
ring-like features. 
Regarding the disc masses of AA Tau and DS Tau systems,
they do not match with any model. However, it may happen
that using a very large initial disc mass, 
even larger than the initial mass of model $I$, the AA Tau system may be reproduced.

Finally, regarding the largest stellar masses, the disc masses of the CI and DL Tau systems seems to be explained by the $J$
and $L$ model. 

The creation of these ring-like features is
interesting. Features below tens of AU around young stars that 
are less than $1$ Myr old would need
an early formation of planets at the locations of the rings, 
and whether gas giant planets can form so close on such short 
timescale is still not clear \citep{helled14}, and 
other mechanisms as a common origin of multi-ring systems,
such as snow lines, have been also under debate \citep{marel19}. 
Remarkably, our photoevaporative simulations may provide some complementary explanation
to inner features below tens of AU around young stars.

\begin{figure*}
\begin{center}
\begin{tabular}{cccc}
\multirow{3}{*}{\rotatebox{90}{$\xrightarrow{\makebox[6cm]{$M_{d}$}}$}} &
\includegraphics[width=0.3\linewidth]{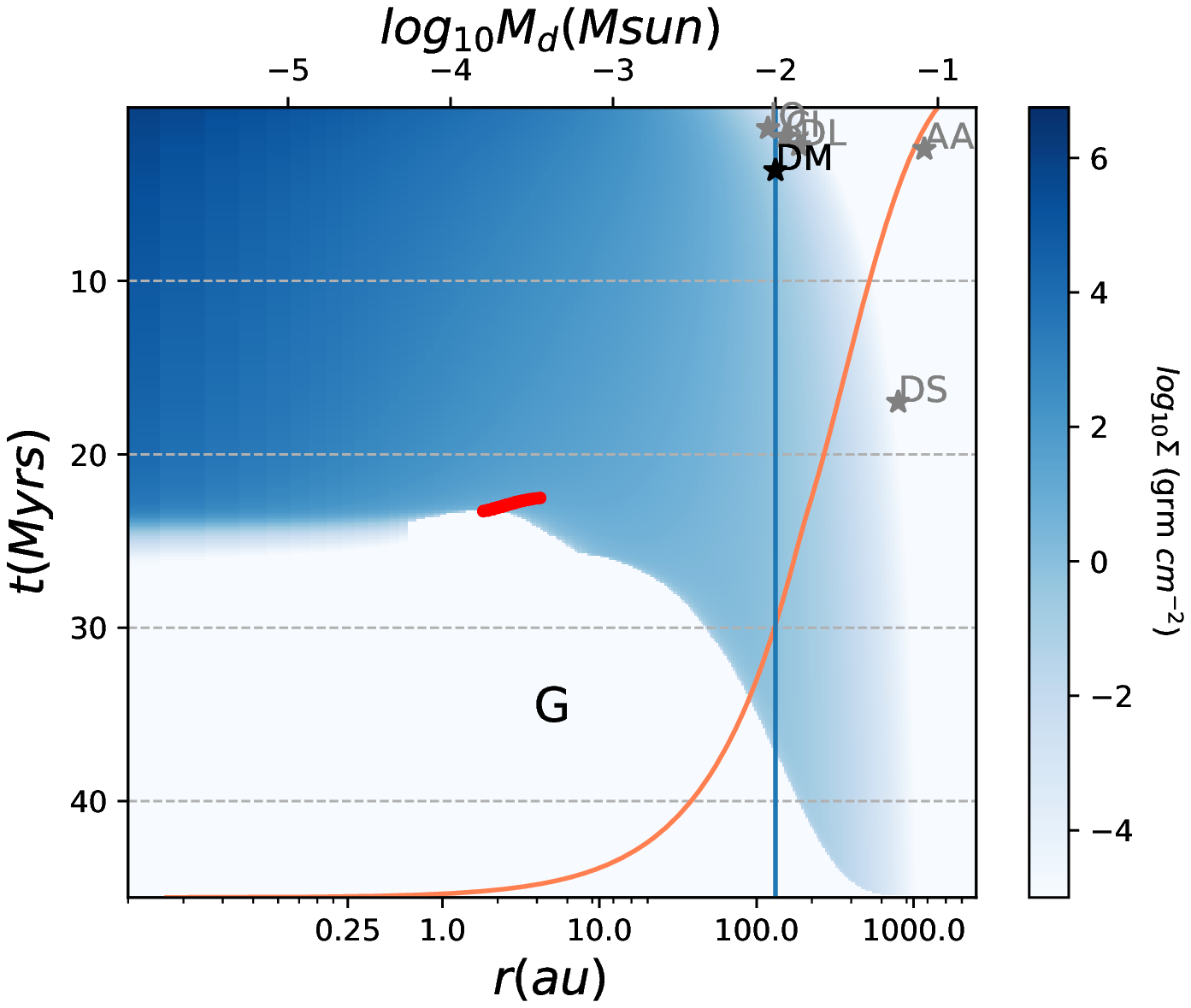} &
\includegraphics[width=0.3\linewidth]{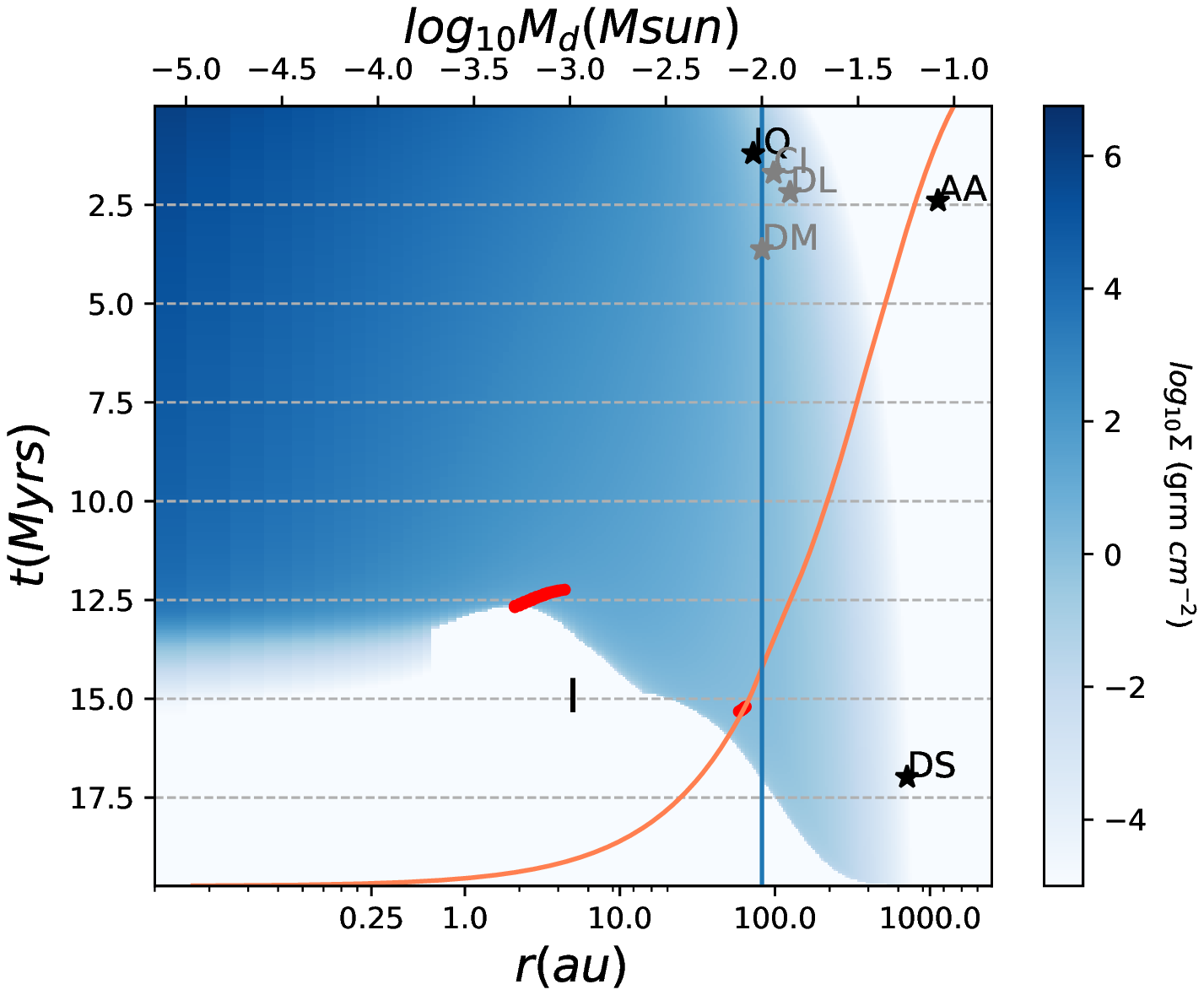} &
\includegraphics[width=0.3\linewidth]{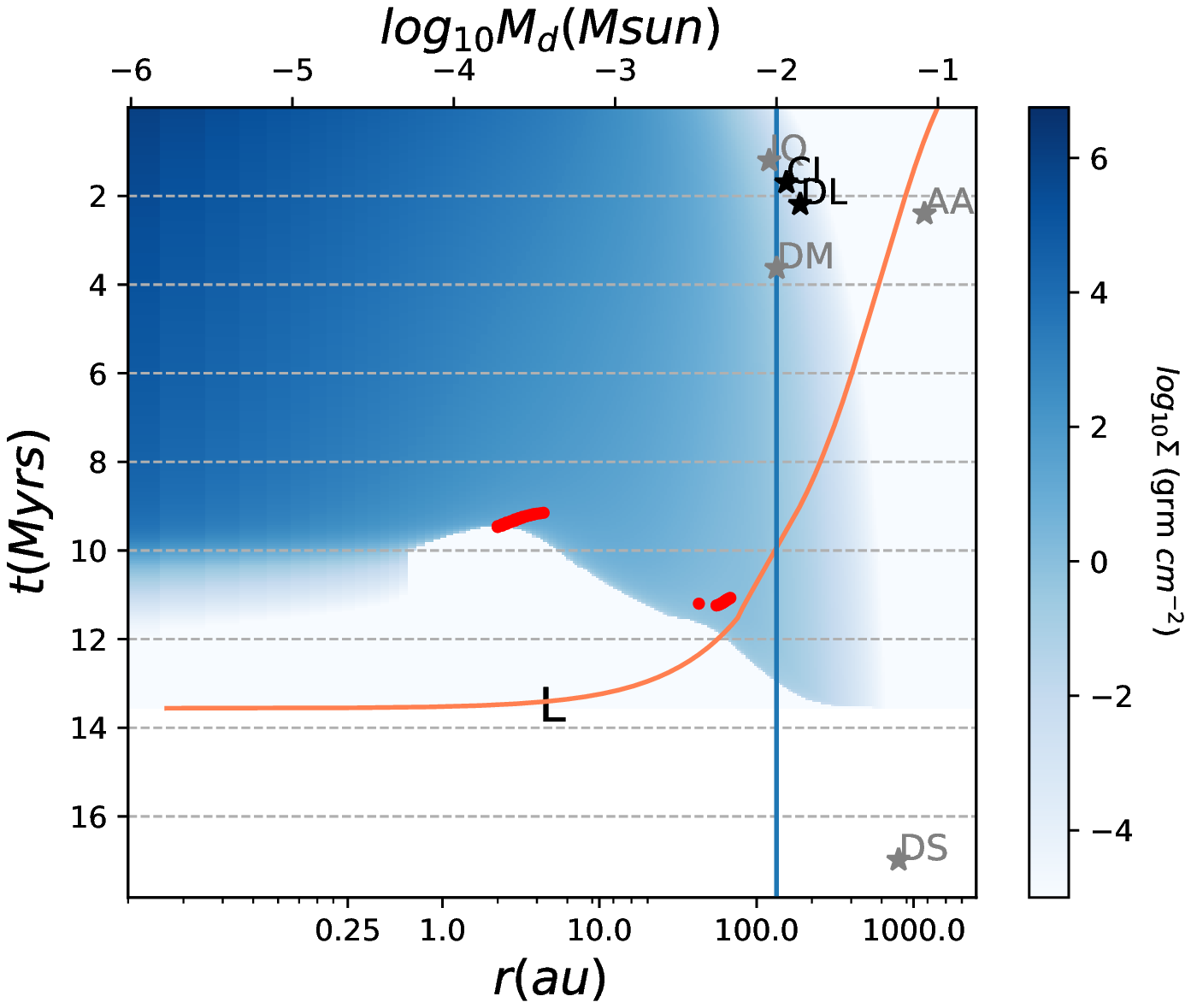} \\
&\includegraphics[width=0.3\linewidth]{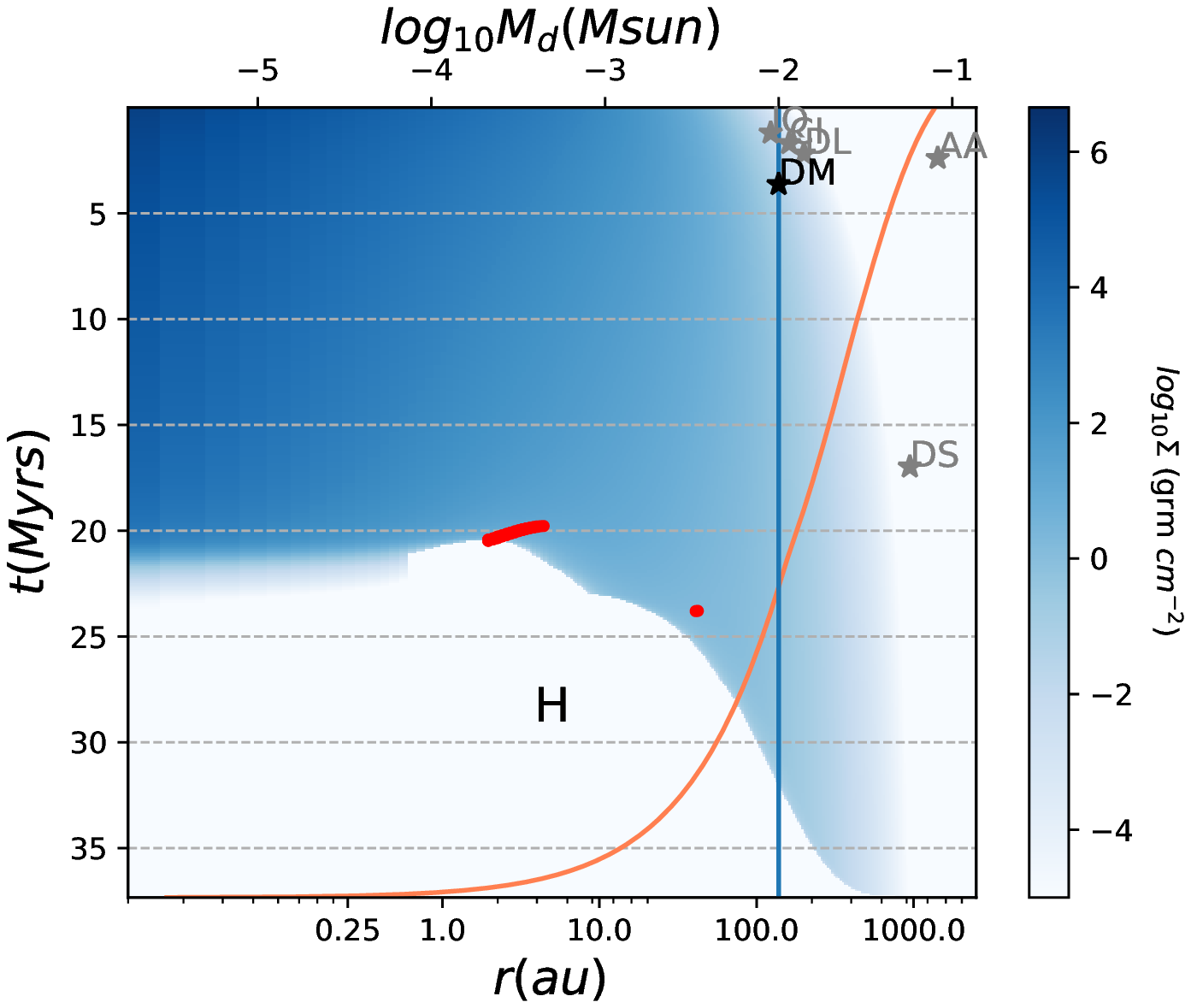} &
\includegraphics[width=0.3\linewidth]{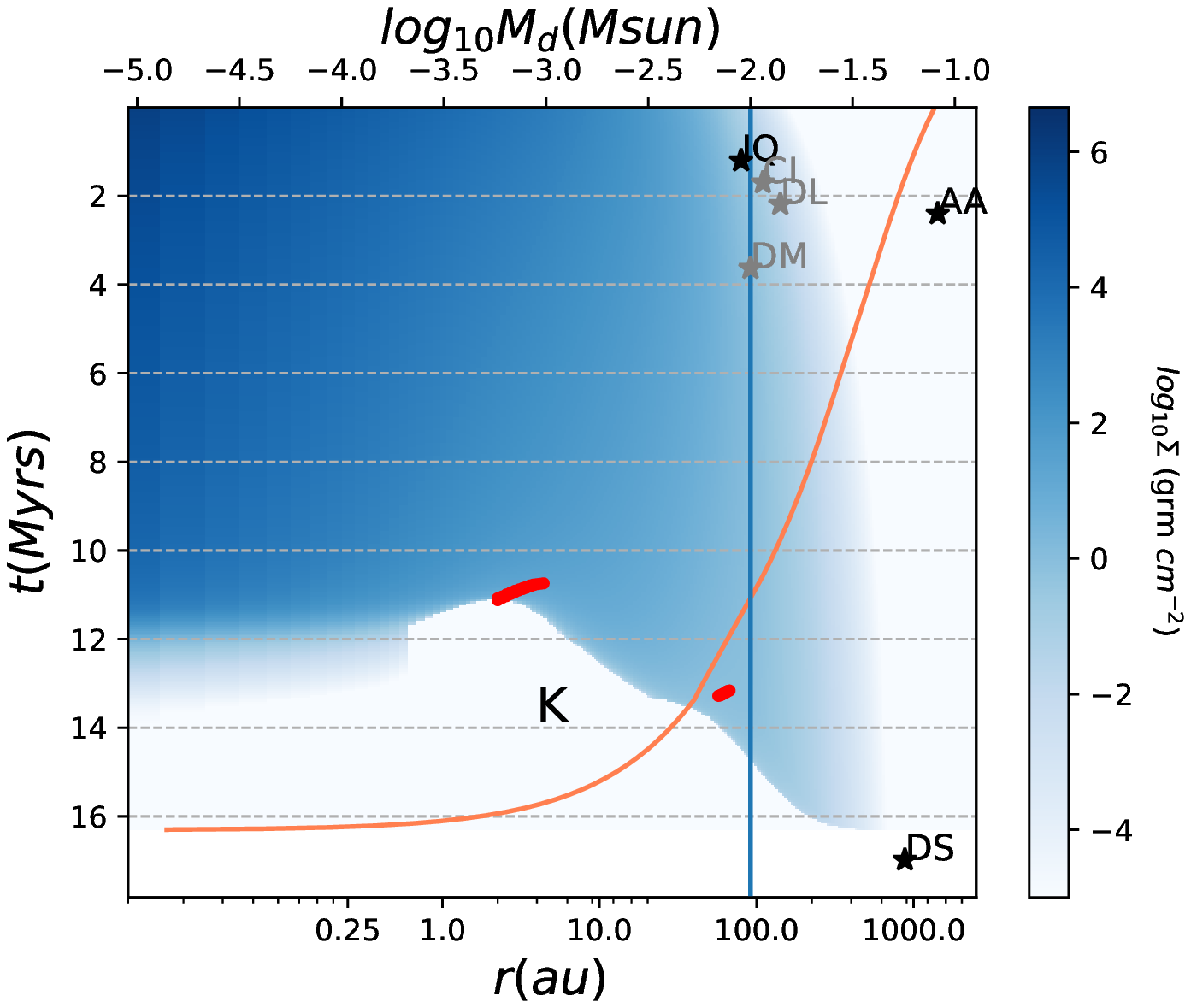} &
\includegraphics[width=0.3\linewidth]{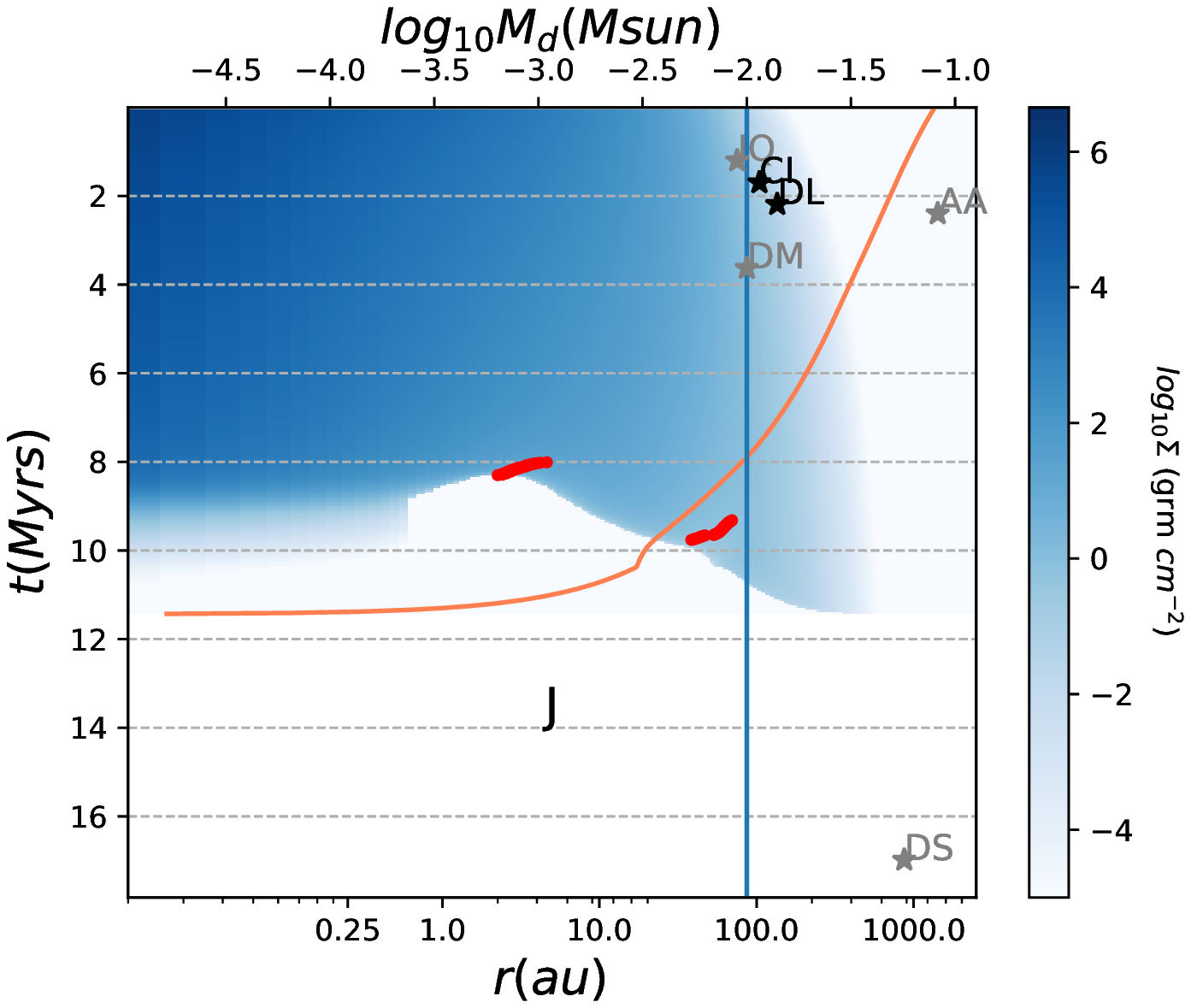} \\
&\includegraphics[width=0.3\linewidth]{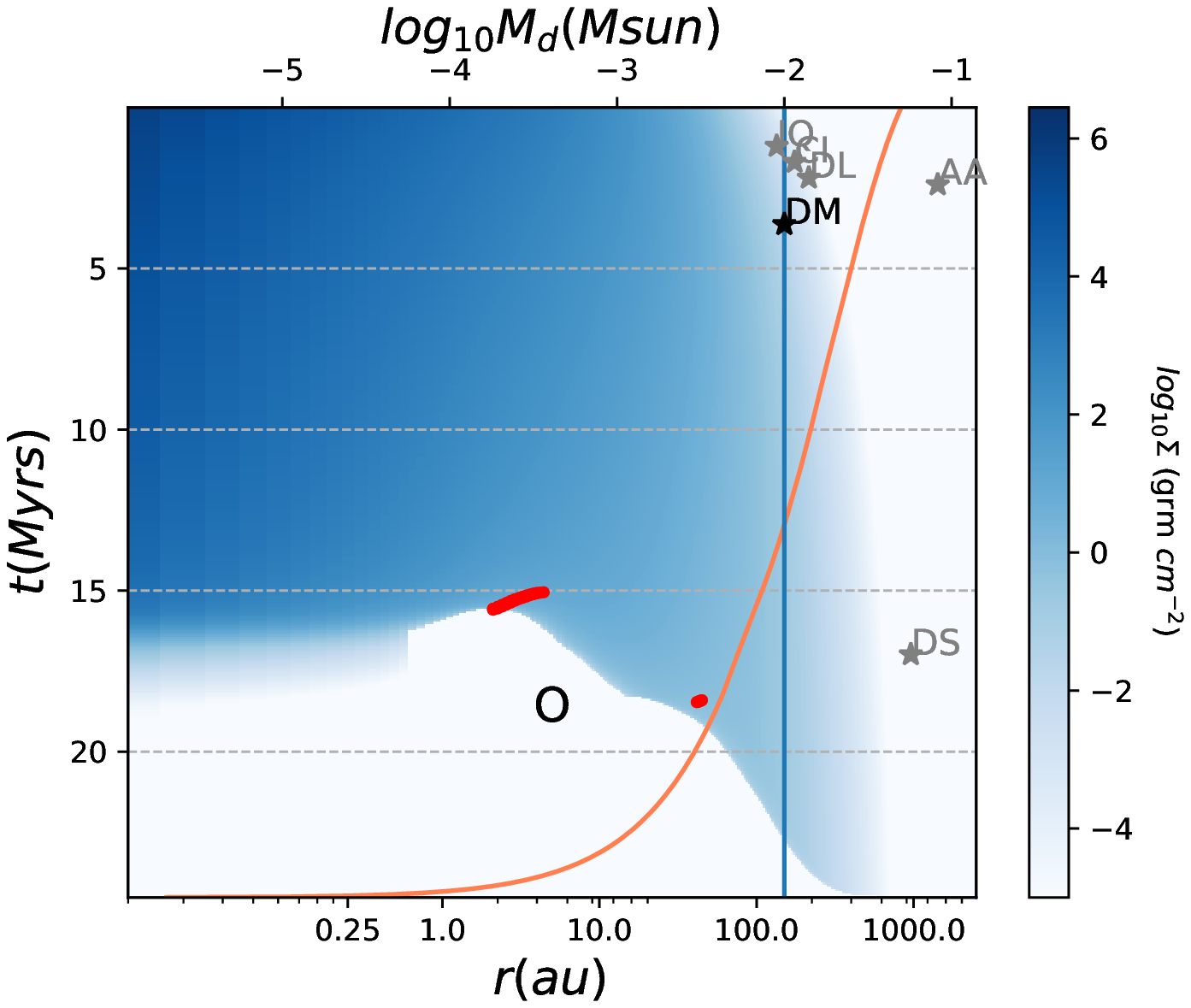} &
\includegraphics[width=0.3\linewidth]{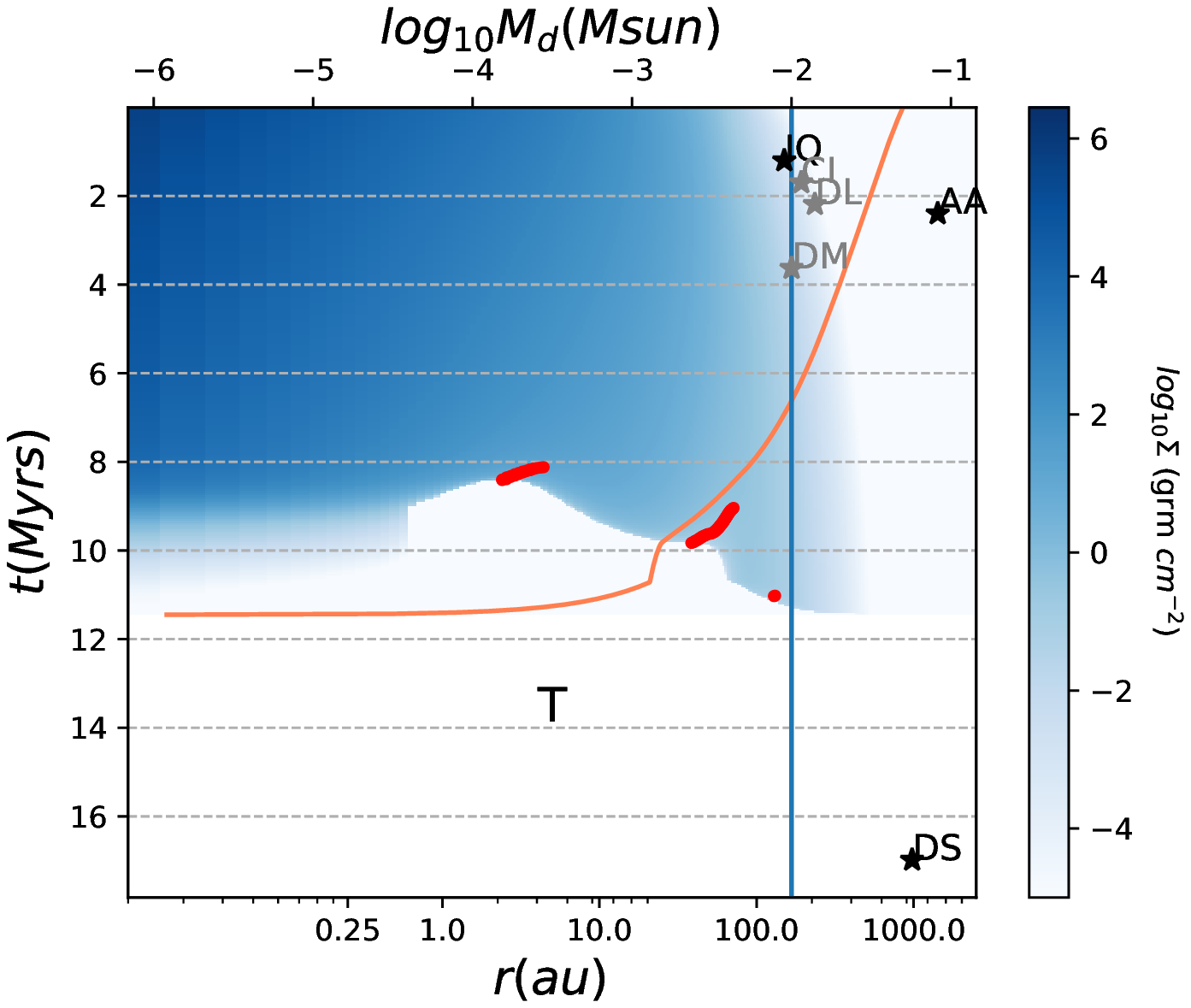} &
\includegraphics[width=0.3\linewidth]{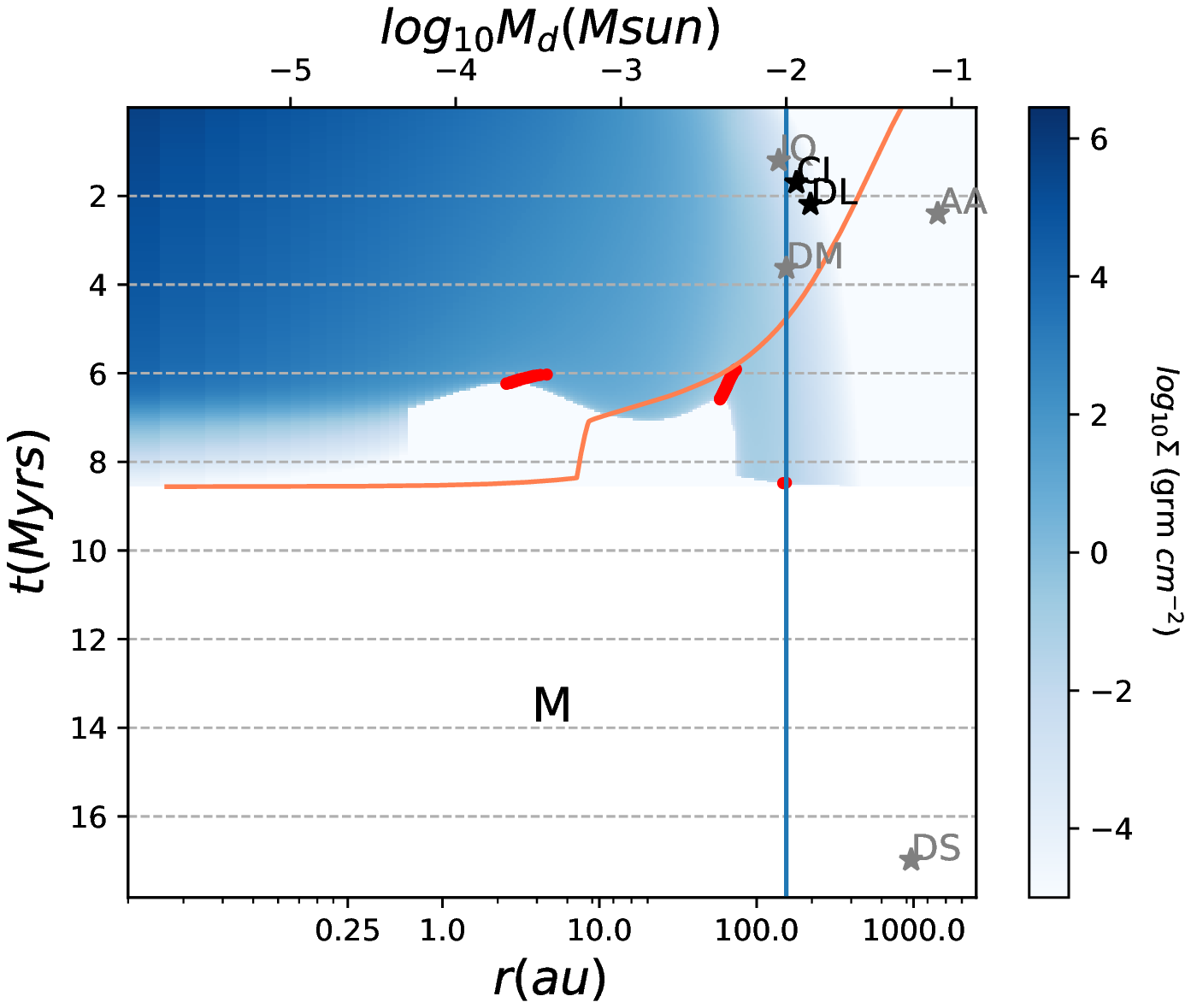} \\
&\multicolumn{3}{c}{  $\xrightarrow{\makebox[6cm]{$M_{*}$}}$  } \\
\end{tabular}
\end{center}
\caption{
Evolution of the grid of models for a 
for a photoevaporation controlled by X-ray heating,
with a wind with efficiency parameter $e_x=0.5$
(meaning the X-ray wind mass losses is half times
smaller than the fiducial value),
a viscosity value of $\alpha=10^{-4}$
and a variable flaring profile $P52$.
The surface density in blue gets lighter as the disc is eroded.
The red points indicate local minima in the density profiles.
These are produced much later than when $e_x=1.0$, and the disc lifetimes
are much longer too.}
\label{rx_3b4X0}
\end{figure*}

The previous results have explored very large 
values of the wind mass loss rates. Now, we will explore the 
opposite case, when the efficiency is lowered instead. 
The Figure~\ref{rx_3b4X0} shows the results got when the previous models with viscosity
$\alpha=10^{-4}$ and the flattening profile $P52$
evolve subject to winds with a decreased efficiency factor $e_x=0.5$,
which corresponds to half of the fiducial mass losses.

The eroding of the disc now slows down and the discs lifetimes
are much longer. This allows the formation of 
cavities with increased lifetimes, mainly in the smallest stellar masses.
The dents and ring-like features produced 
in the outer part of the discs are now generated in the largest stellar mass models of the grid.
These features are best seen for the lowest disc masses, 
where even an additional outer third dent can appear.

These features are generated when the disc masses $M_d$ are too low 
and the ages too large to match the observed values
in the reference sample. However, this low efficiency $e_x$ value might be a way for modelling
systems with very large estimated ages, like the reference system $DS$ Tau.
The initial disc mass should be even larger than the initial
mass of model $I$ in this case. 
One might also think in decreasing the flattening for making the lifetime somehow longer,
slowing the erosion and creating the outer dent at proper age.

\section{Photoevaporation from FUV radiation driven winds}
\label{sec:fuvwinds} 

The importance of internal FUV radiation was first analysed in
\cite{gorti08} assuming large initial disc masses ($\approx 0.1 M_{\odot}$),
though external FUV heating was considered earlier in \cite{johnstone98}
and \cite{adams04}, 
producing high accretion rates. Following these models, 
the X-ray photons do not produce significant photoevaporation by
themselves. But, because they increase the degree
of ionization of the gas in the disc, there is a higher
electron population which in turns reduces the positive charge of the 
dust grains, that helps the FUV-induced grain photoelectric heating of 
the gas.

\begin{figure}
\begin{center}
\includegraphics[width=\columnwidth]{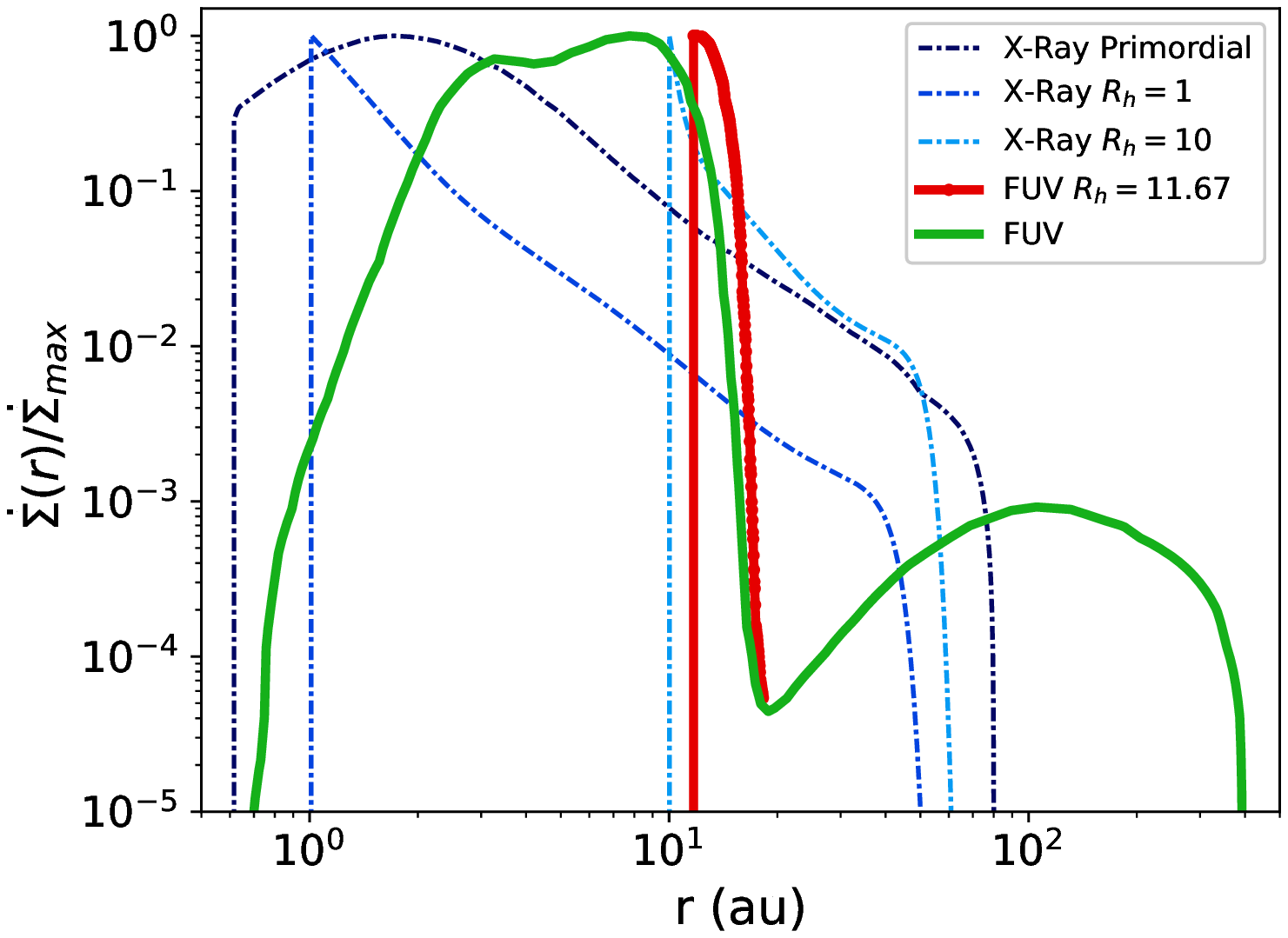} 
\end{center}
\caption{
Normalised photoevaporative wind mass losses corresponding to the
two FUV wind profiles described in \protect\cite{jennings18}.
The FUV profile in the initial phase is a static one,
meanwhile the header of the profile corresponding to the second phase 
is formed by a moving peak (in red) that shifts towards the position of the trough, that remains at fixed radius.
At any time, the integrated mass loss is constant.
The X-ray wind profiles used in Section~\protect\ref{secrxwinds} are overlaid
for comparison.}
\label{fuv_winds}
\end{figure}

We are going to consider here mass losses coming from 
FUV radiation to our semi-analytical $1$D model by adding
the corresponding sink term to the viscous evolution
in the same way we added the X-ray radiation dominated wind.
We add a FUV-wind profile like 
the one described in \cite{jennings18}, that is indeed developed
from the models from \cite{gorti09}. 

This FUV profile, plotted in Figure~\ref{fuv_winds}, splits 
in two phases like the X-ray dominated profile
seen before. There is a primordial profile, considered static, 
and coincident with the profile used in \cite{gorti09}. 
However, the profile switches to another one when the cavity opens. 
Then, the inner peak initially at $7.3$au is shifted towards the 
trough at $18$au, keeping the profile at distances larger than
$18$au unaltered. At any time, the total mass rate loss 
is the same that in the primordial case.

We have seen in previous section that the X-ray dominated profile 
mainly erodes the disc
at given radii, mainly where the X-ray wind peak is located. Once this initial dent 
has grown so much that the floor density is reached, a gap is produced,
and the inner portion of the disc is eroded until it disappears.
The primordial X-ray profile peaks around $2$au, while the primordial FUV profile has a high loss between $3-10$au, peaking at
around $10$au., see Figure~\ref{fuv_winds}. 
Therefore, the FUV profile may produce dents at diferent distances than the X-ray profile. 

Along this work, we will keep the positions of the front and 
trough of the FUV profile fixed and coincident with those values from \cite{jennings18}. 
Evidently, the initial position of the FUV profile might vary depending on the stellar mass and other parameters.
However, in this first approach, we do not want to focus on the 
detailed positions of the gaps. Conversely, we aim to 
understand if a FUV moving front can create ring-like structures at early stages.

Hence, our key control parameter will be the FUV mass rate losses. 
The FUV fluxes can be generated by accretion or chromospherically, and the
mass loss rate will depend on the adopted thermochemistry. As fiducial value, we have 
taken a mass loss rate of $3.0 \cdot 10^{-8} M_{*}$ \citep{owen12}. 
However, this value may be not universal for all systems, and
we will use an efficiency factor $e_{fuv}$, as we did with the 
X-ray dominated winds, to achieve 
the range of different photoevaporative mass rates discussed in \cite{jennings18}.
This $e_{fuv}$ factor will allow to reproduce the huge accretion rates close to $10^{-7}$ 
seen sometimes at early stages \citep{gorti09}, and also 
the lower accretion values produced when the accretion flow
may shade the disc from Lyman-$\alpha$ when it is close to the star. 

\begin{figure*}
\begin{center}
\begin{tabular}{cccc}
\multirow{3}{*}{\rotatebox{90}{$\xrightarrow{\makebox[6cm]{$M_{d}$}}$}} &
\includegraphics[width=0.3\linewidth]{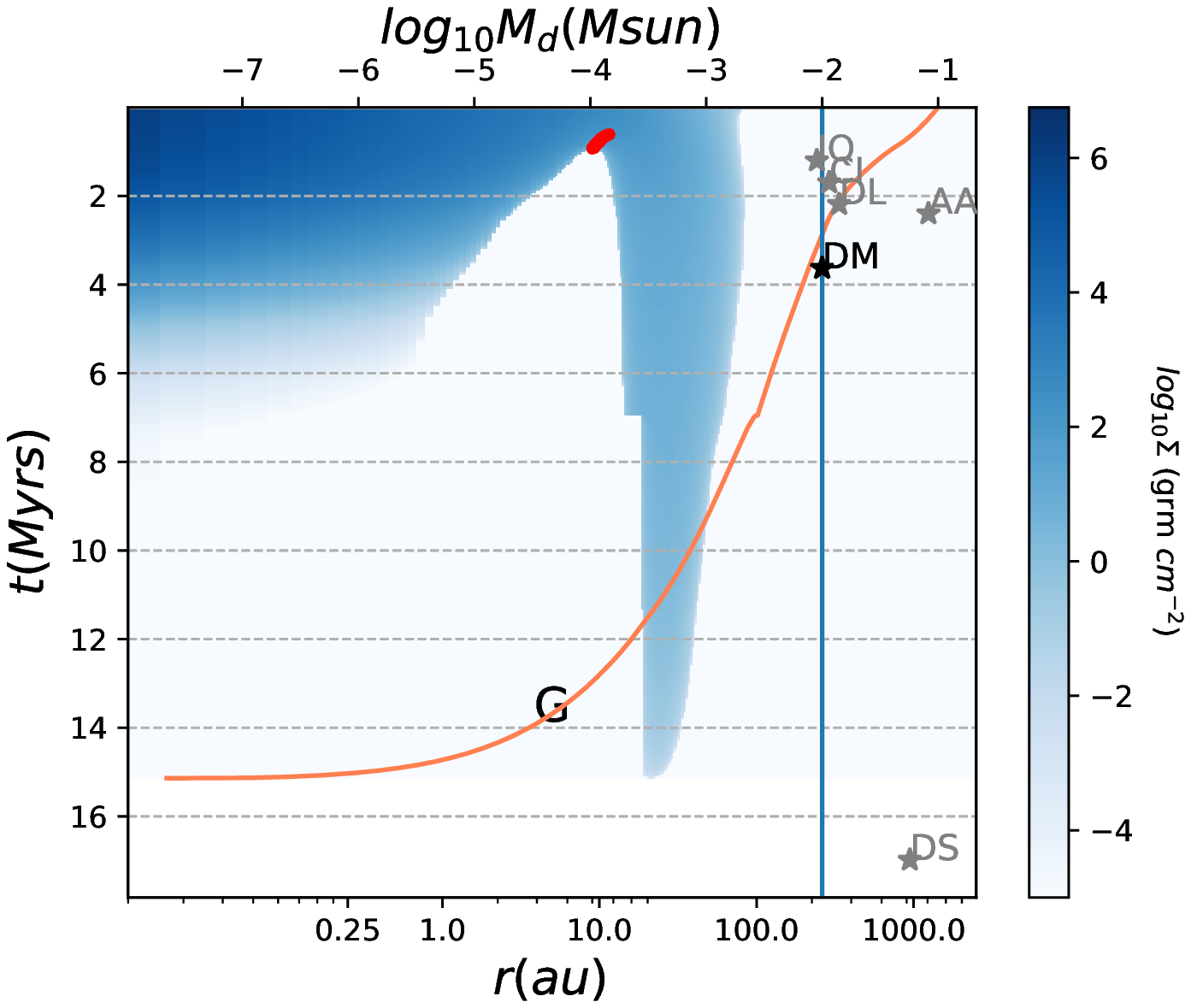} &
\includegraphics[width=0.3\linewidth]{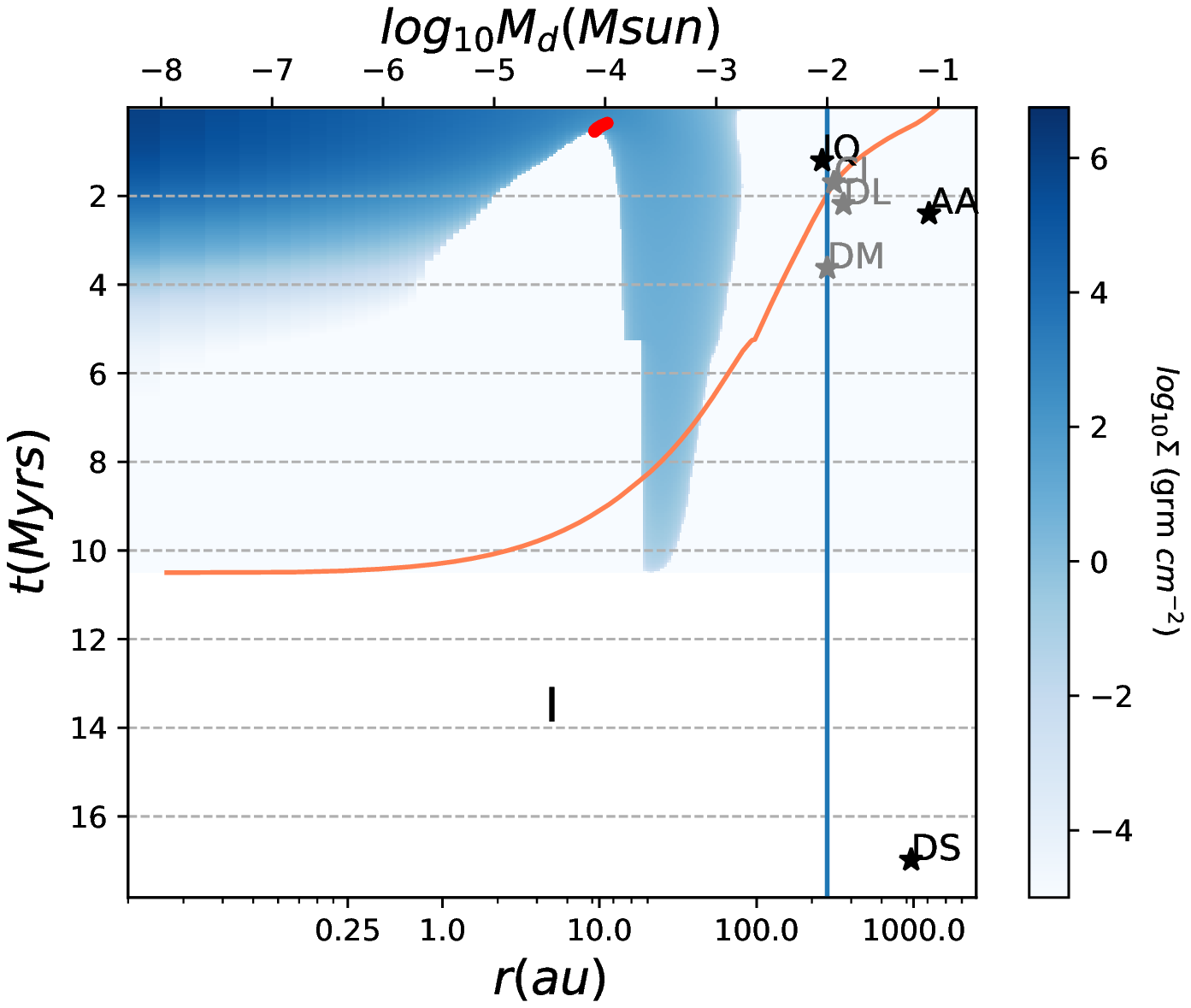} &
\includegraphics[width=0.3\linewidth]{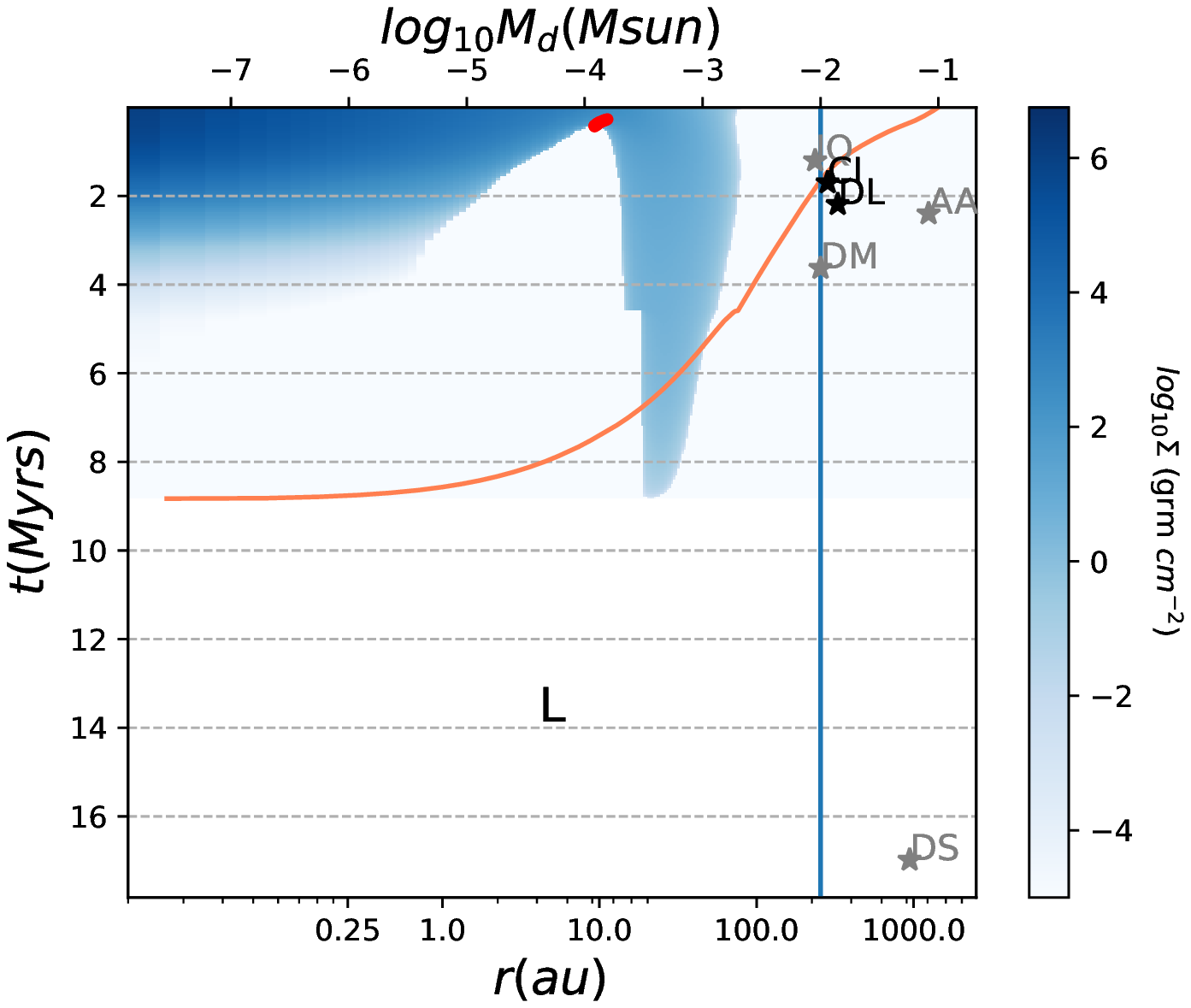} \\
&\includegraphics[width=0.3\linewidth]{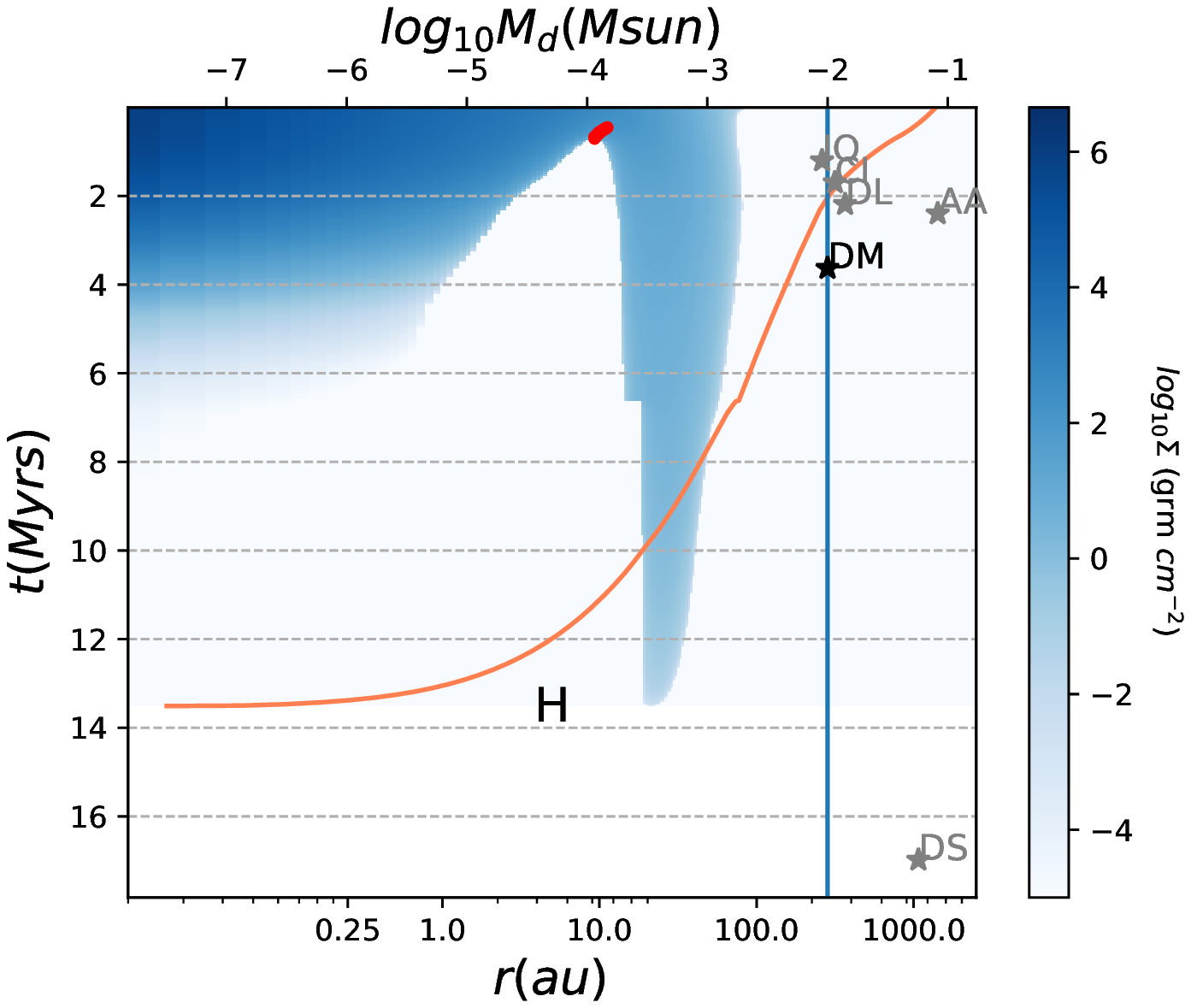} &
\includegraphics[width=0.3\linewidth]{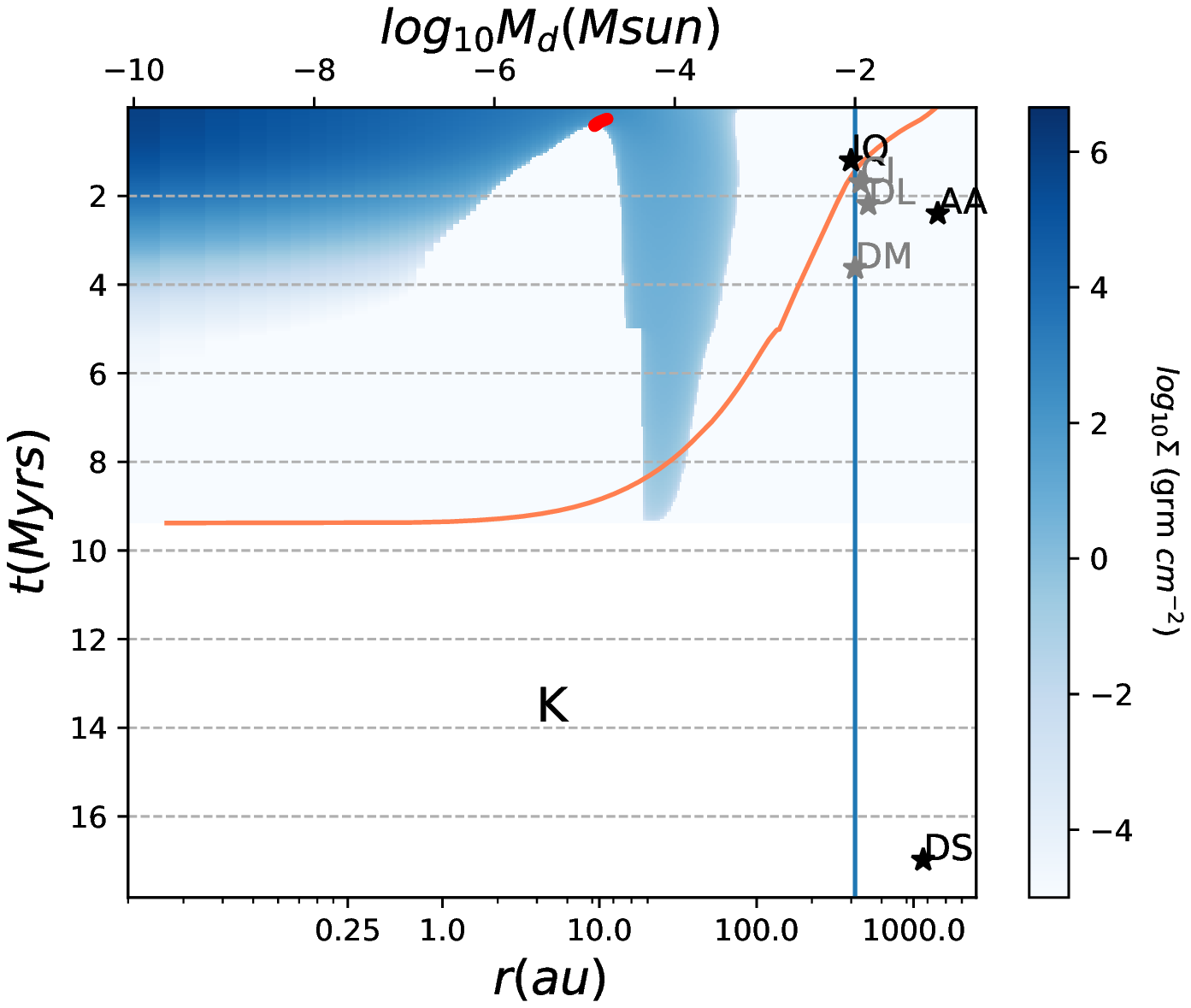} &
\includegraphics[width=0.3\linewidth]{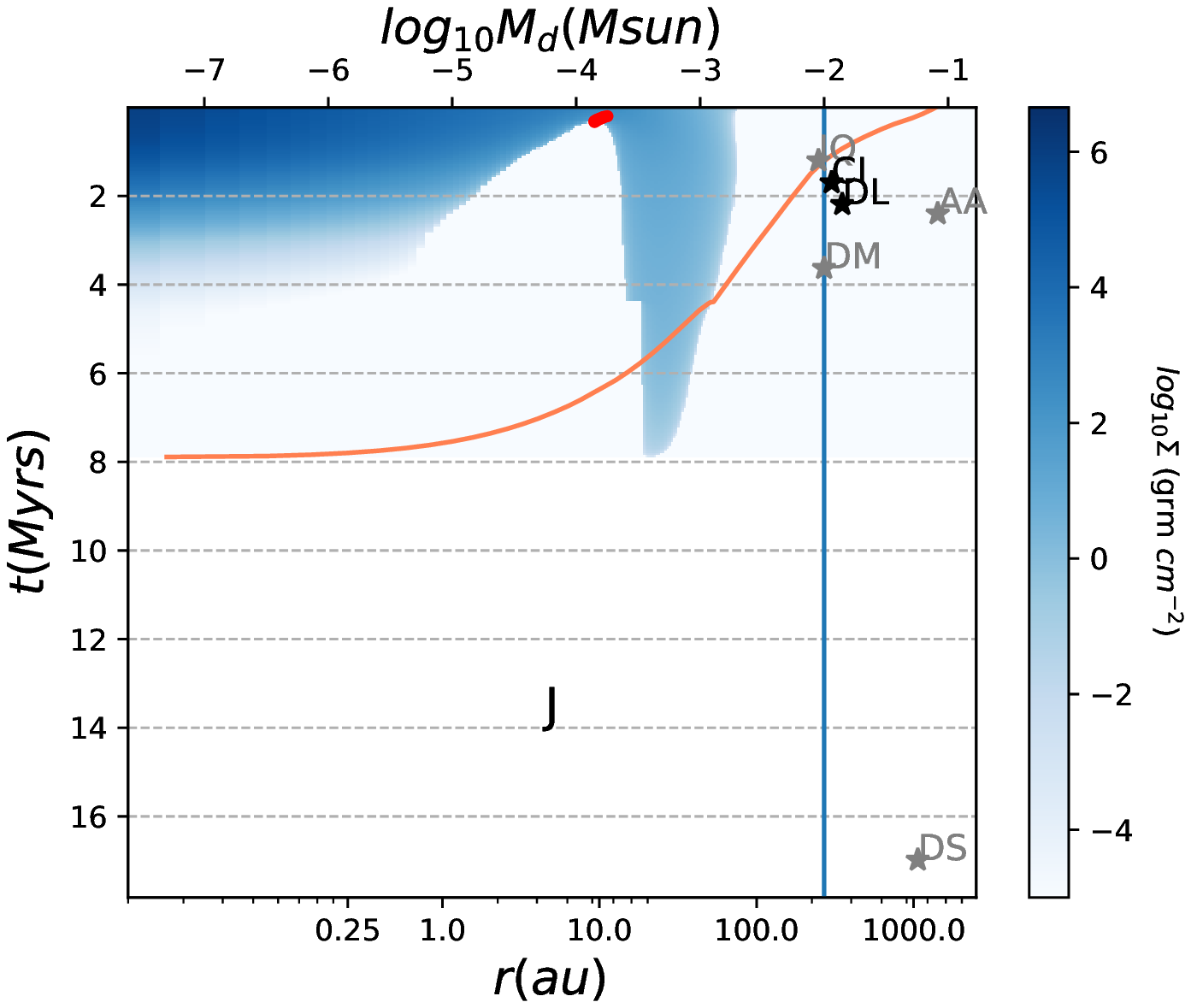} \\
&\includegraphics[width=0.3\linewidth]{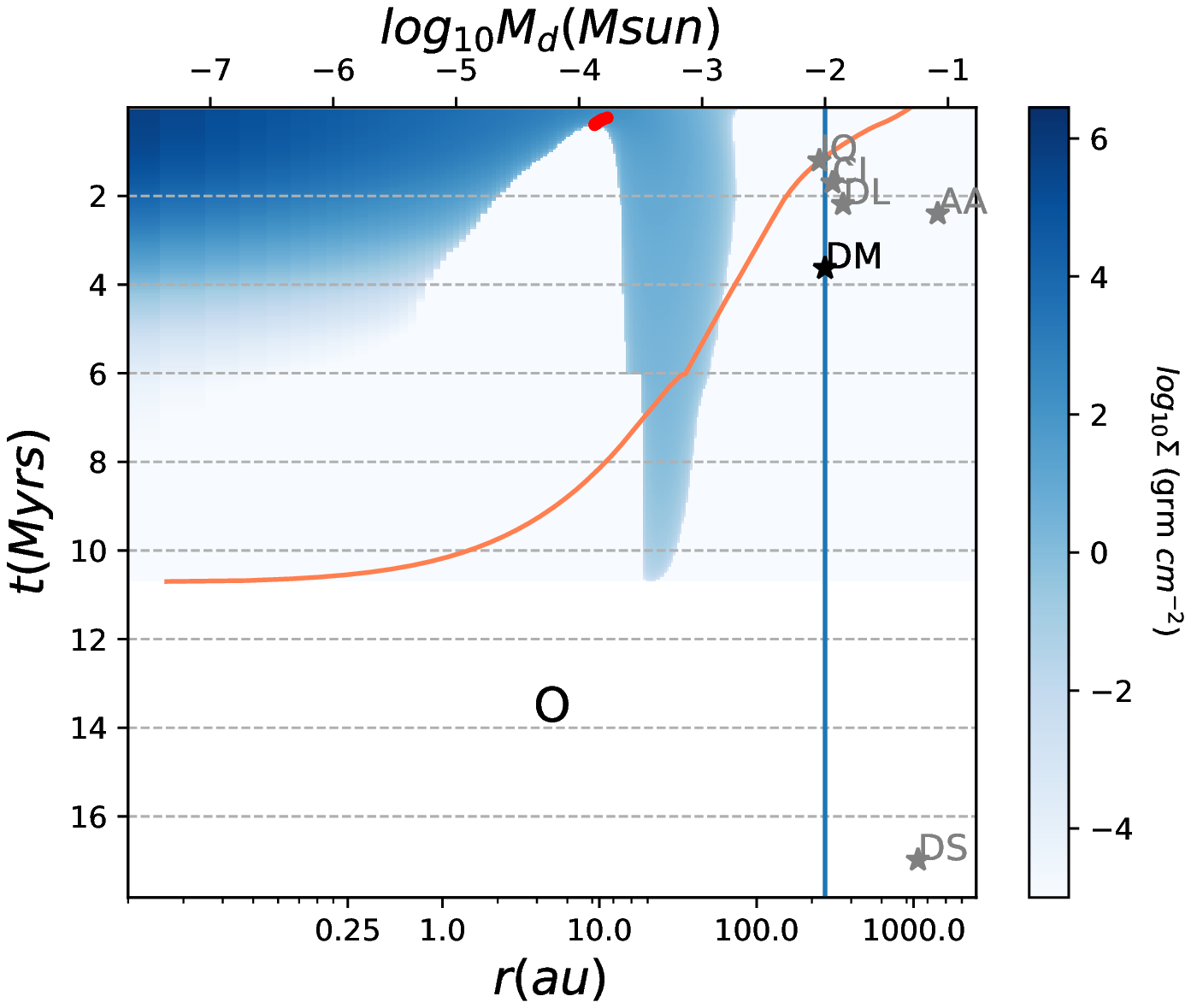} &
\includegraphics[width=0.3\linewidth]{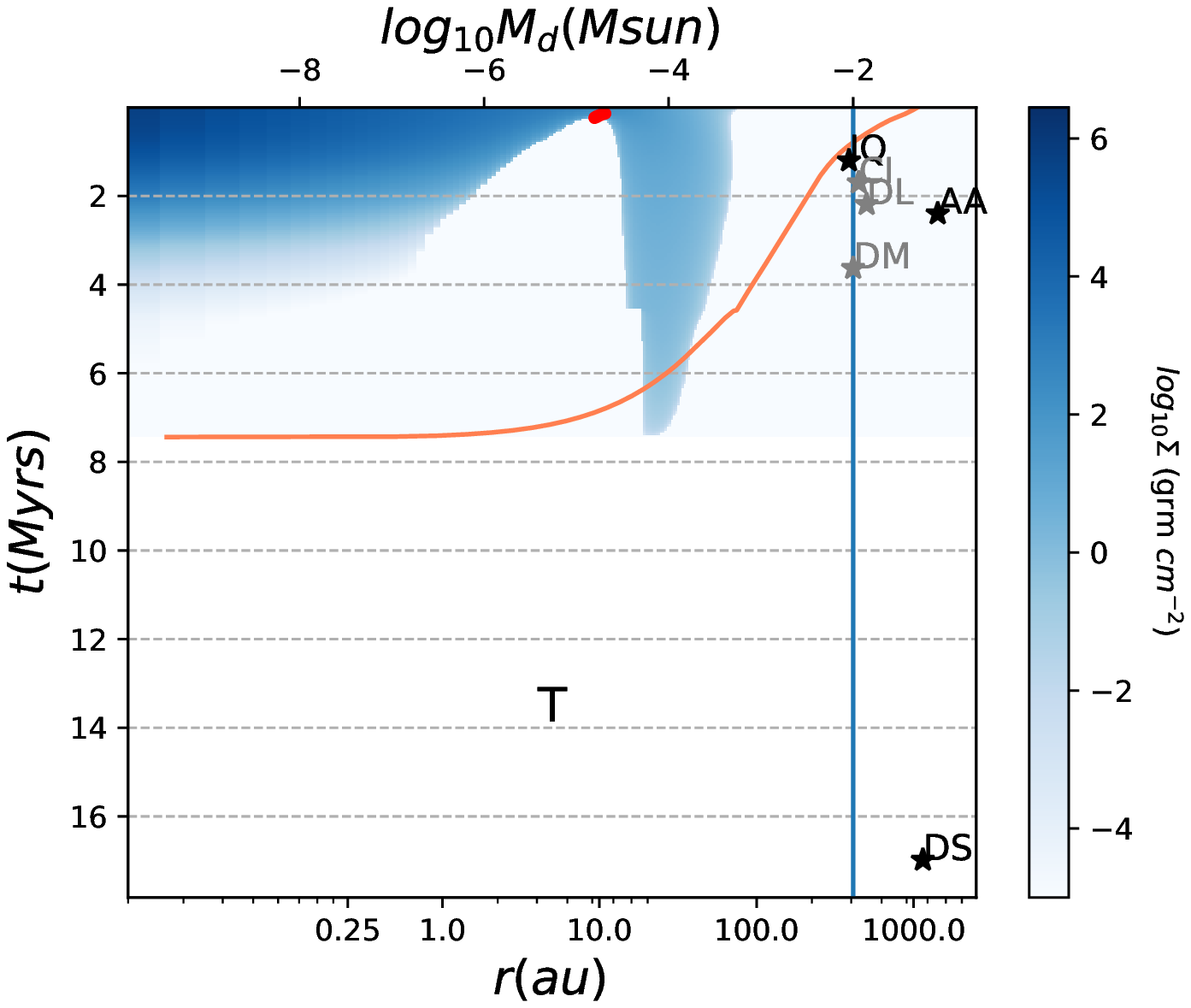} &
\includegraphics[width=0.3\linewidth]{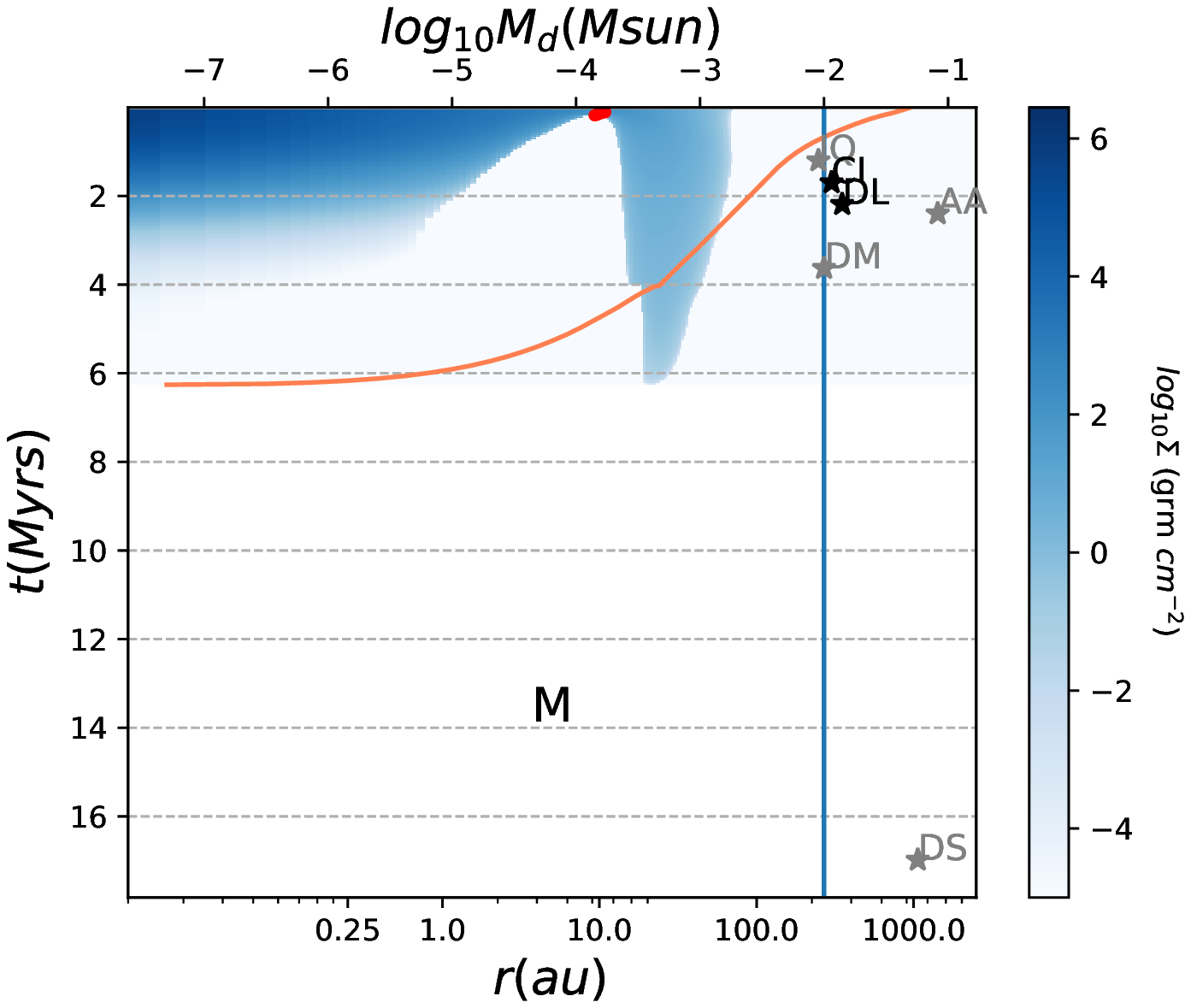} \\
&\multicolumn{3}{c}{  $\xrightarrow{\makebox[6cm]{$M_{*}$}}$  } \\
\end{tabular}
\end{center}
\caption{Evolution of the grid of models subject to
FUV-dominated winds,
with efficiency parameter $e_{fuv}=5.0$
(five times the fiducial values),
a viscosity value of $\alpha=10^{-4}$
and a variable flaring profile $P52$.
The surface density in blue gets lighter as the disc is eroded.
The red points indicate local minima in the density profiles.
The FUV wind seems to produce very good matches 
of the disc masses for the largest initial disc mass
models. However, it creates only one inner gap that will convert into a
long lasting cavity.}
\label{fuv_05}
\end{figure*}

We first computed some simulations using this FUV-dominated wind profile using the fiducial efficiency value
$e_{fuv}=1.0$. These simulations seem to create a single dent in the inner parts
of the disc, around the position of the $18$au peak, which then grows and creates
the inner cavity. No further dents appear, and only systems with inner cavities 
but no external ring-like features might be explained with these FUV-dominated models.
These cavities begin to form just after a pair Myrs, but, notably, they are very long lived,
lasting up to $20$Myrs in average.

Hence, the Figure~\ref{fuv_05} shows the results for simulations run with increased efficiencies $e_{fuv}=5.0$.
These synthetic models open the internal cavities at earlier times. The discs lifetimes
can span up to around $10$Myr for the smaller stellar masses, but makes them shorter than when $e_{fuv}=1.0$.

We have also considered lower than fiducial efficiencies, by running some simulations
with $e_{fuv}=0.5$. These synthetic models reflect a much slower erosion rate, leading
to really very large disc lifetimes.

\section{Accretion rates}

The integrated mass loss rate along all radii can be measured 
using the excess emission in the Balmer 
continuum. Even when this is a measurement that is also
subject to unavoidable uncertainties, 
it is a fundamental parameter when studying the evolution of discs. 
In our models, we can compute it by integrating the surface density as follows, 

\begin{equation}
\label{massrate}
\dot {M}  = 2 \pi \int_{ 0 }^{ \infty } r \dot{\Sigma} (r) dr.
\end{equation}


\begin{figure*}
\begin{center}
\begin{tabular}{cccc}
\multirow{3}{*}{\rotatebox{90}{$\xrightarrow{\makebox[6cm]{$M_{d}$}}$}} &
\includegraphics[width=0.3\linewidth]{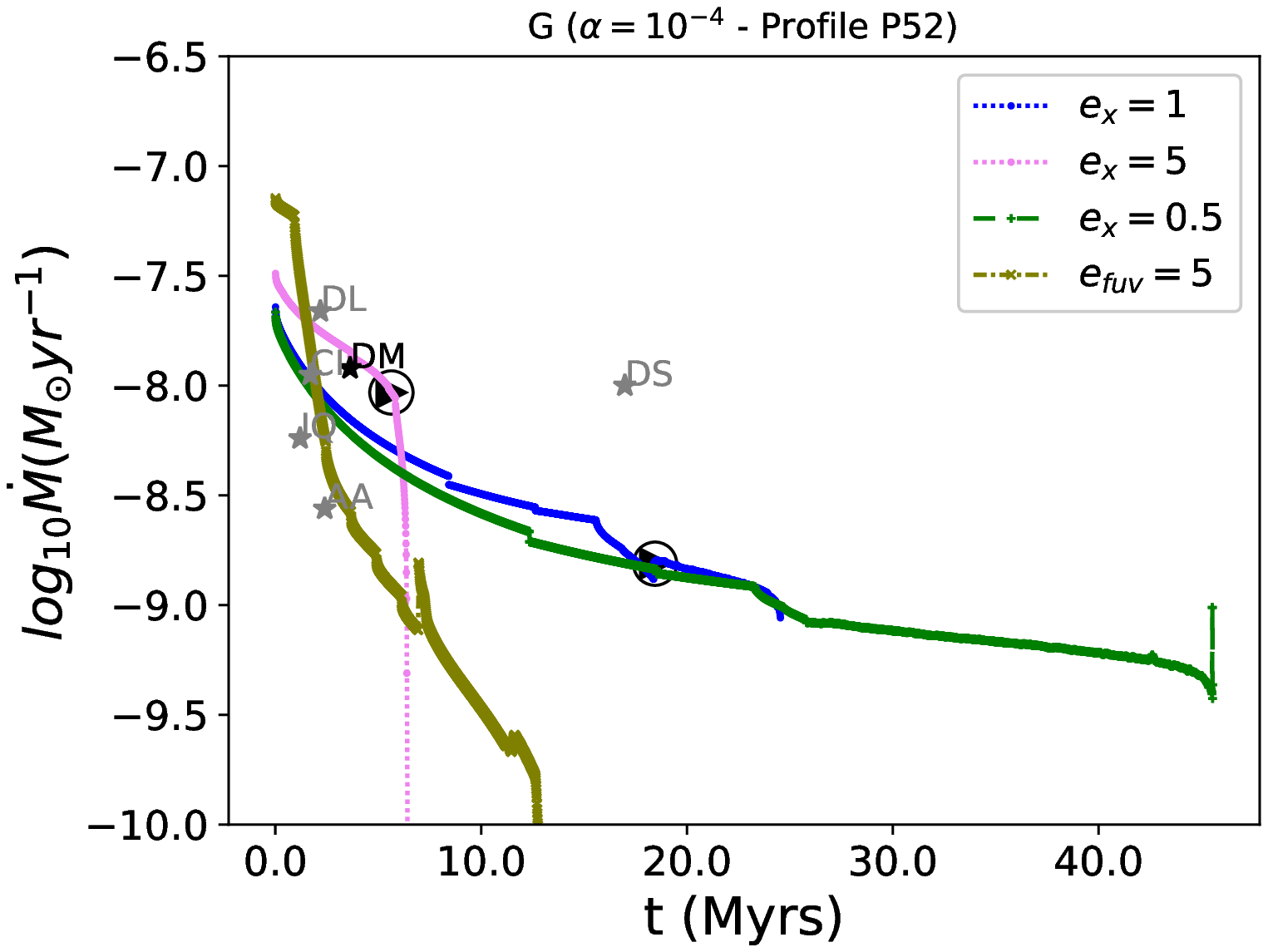} &
\includegraphics[width=0.3\linewidth]{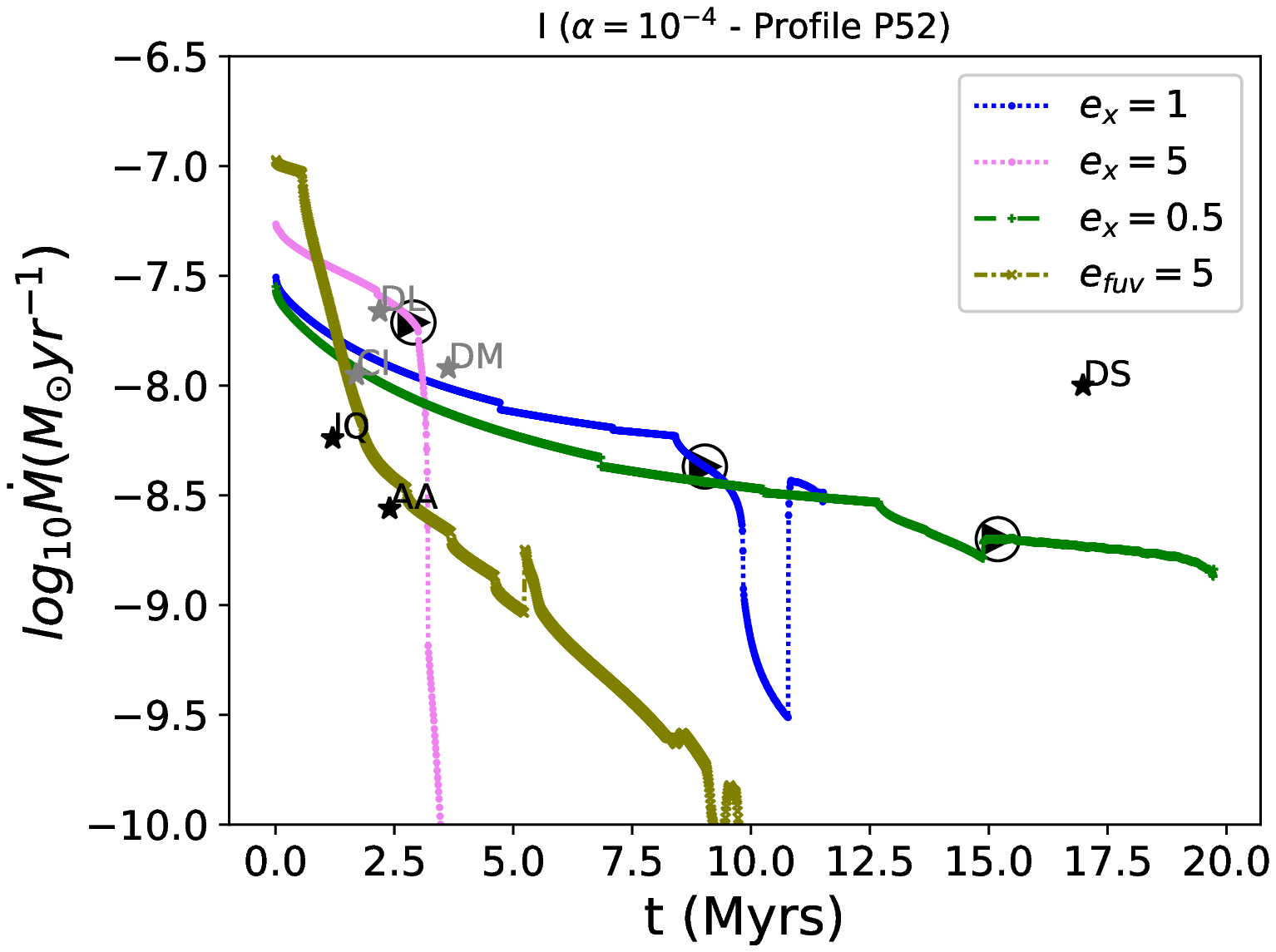} &
\includegraphics[width=0.3\linewidth]{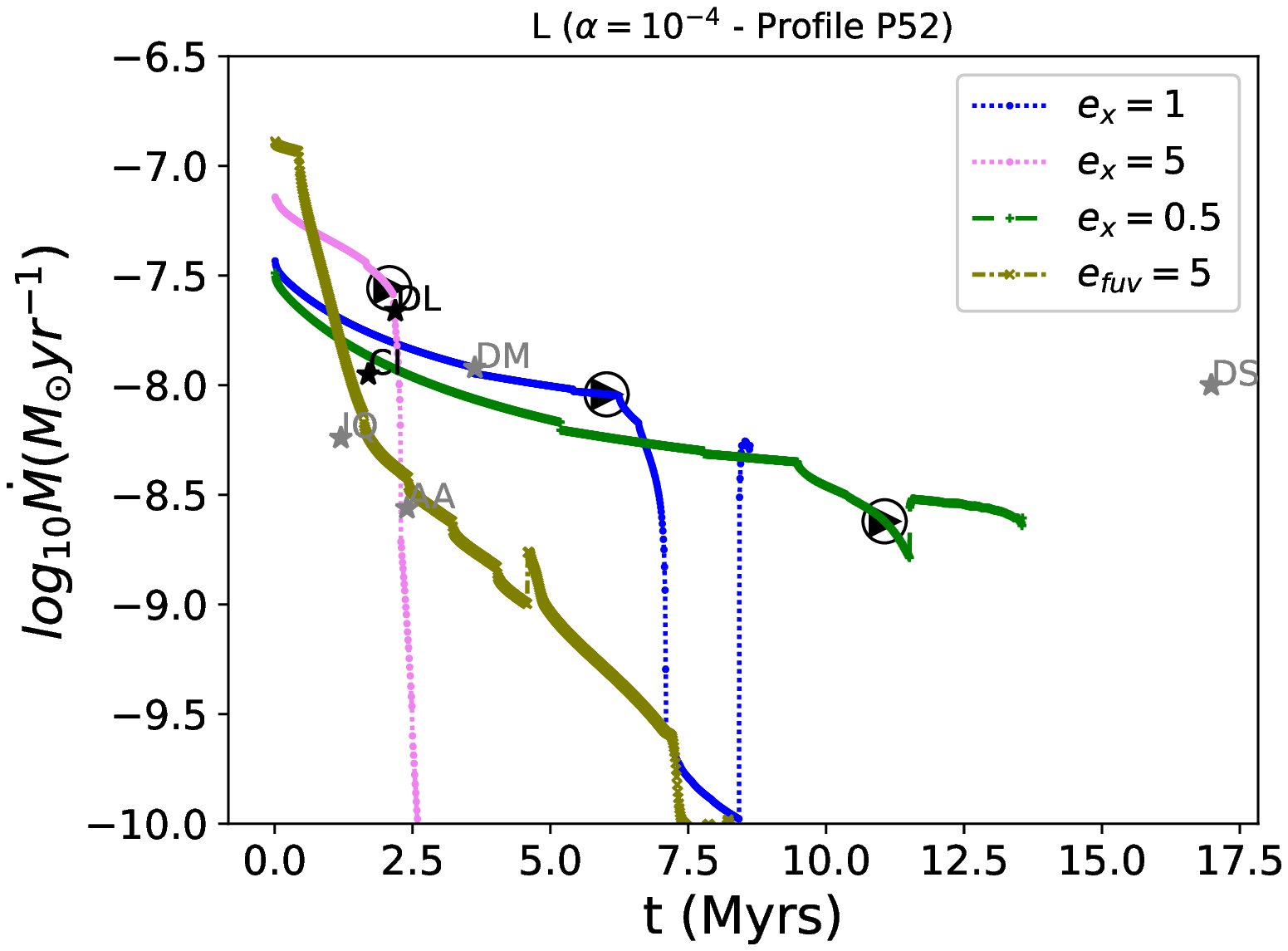} \\
&\includegraphics[width=0.3\linewidth]{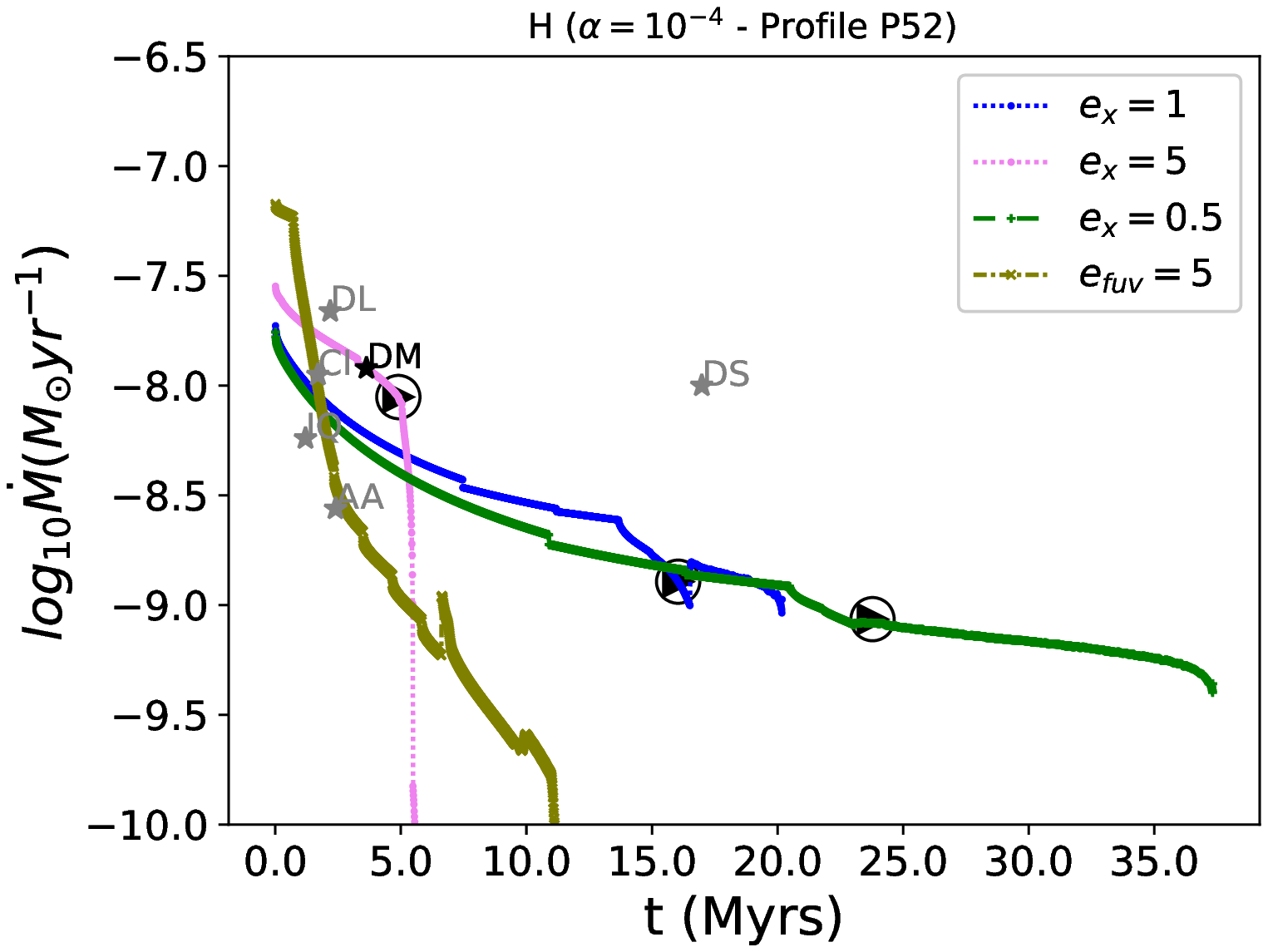} &
\includegraphics[width=0.3\linewidth]{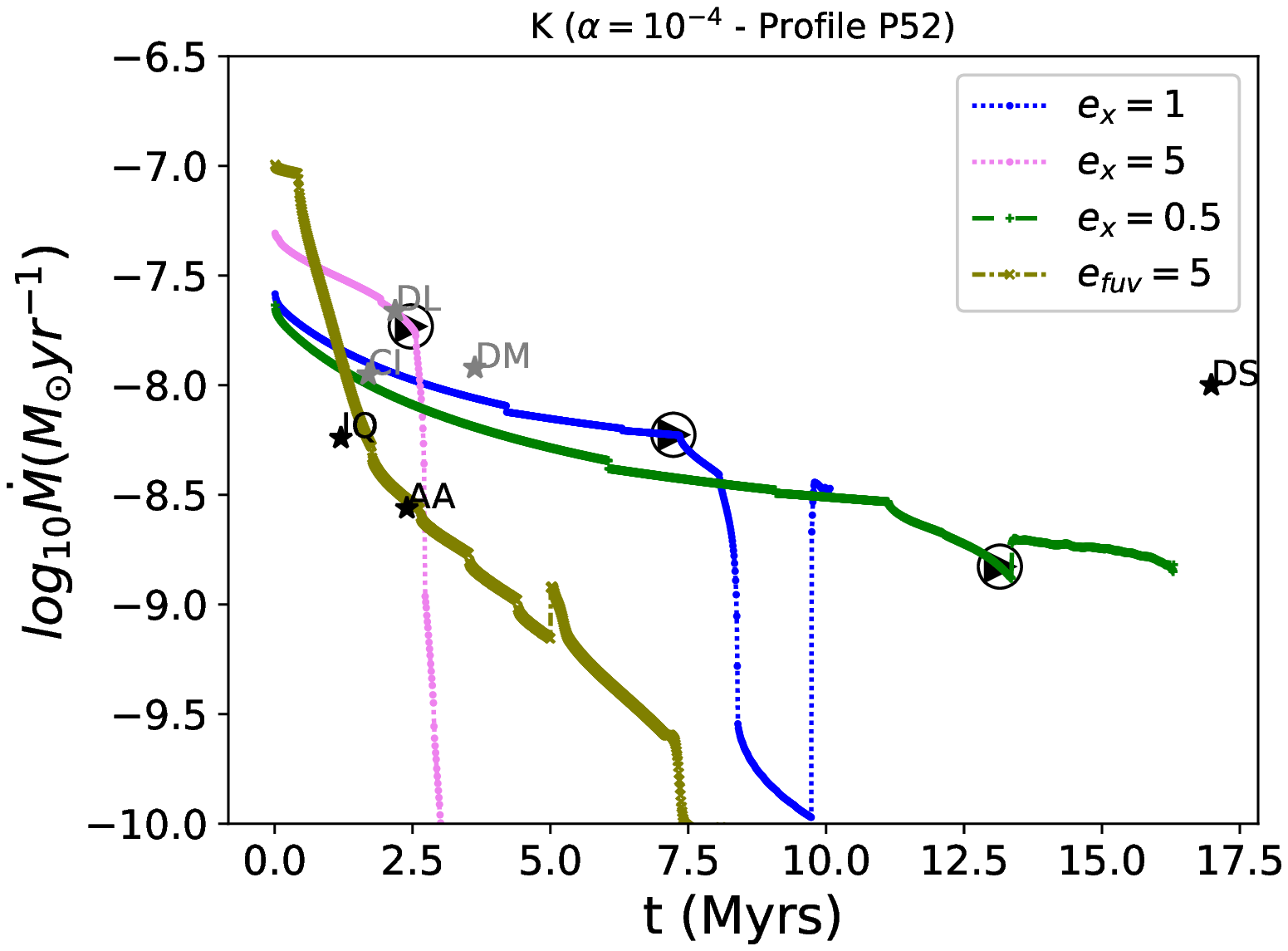} &
\includegraphics[width=0.3\linewidth]{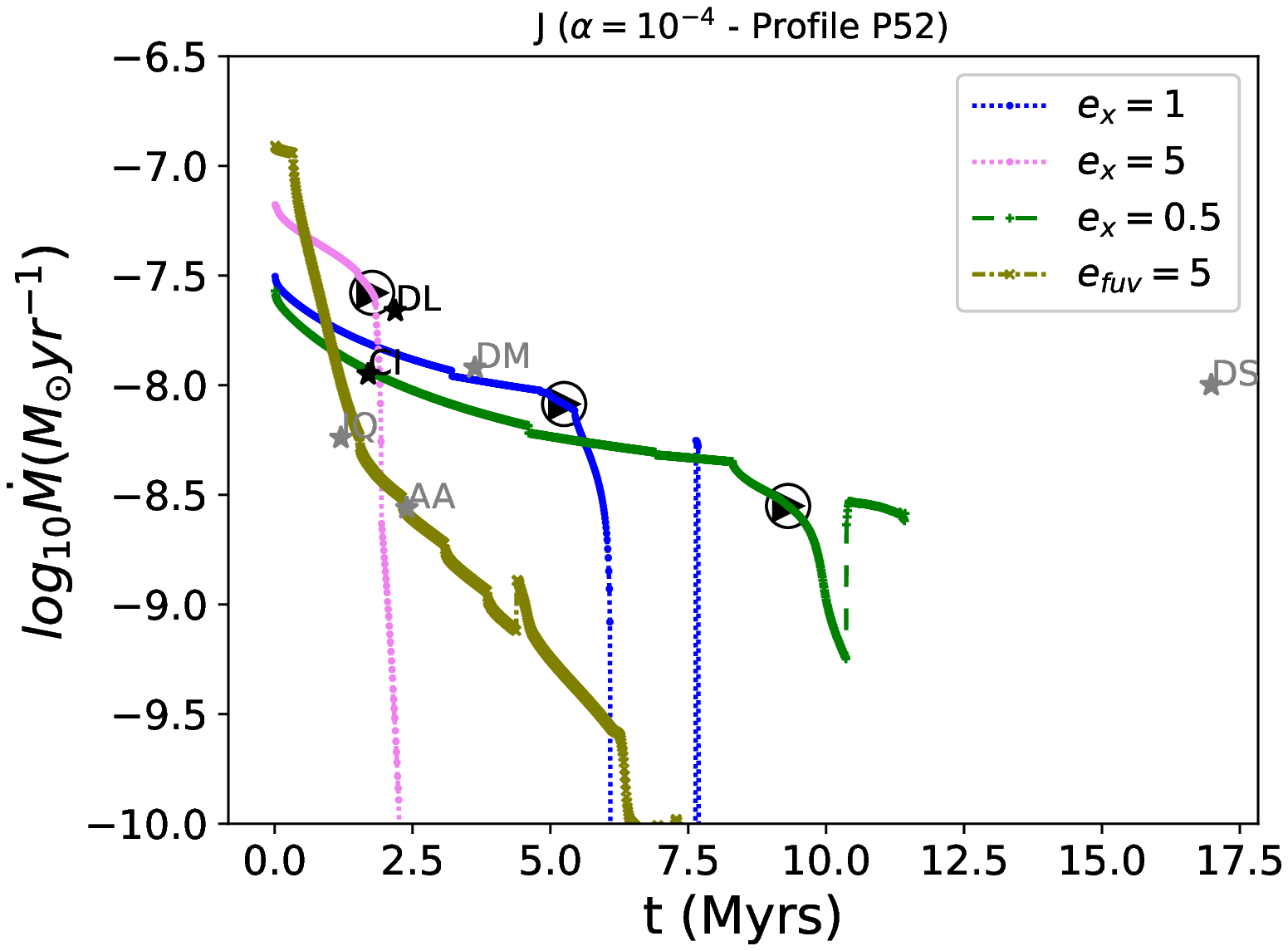} \\
&\includegraphics[width=0.3\linewidth]{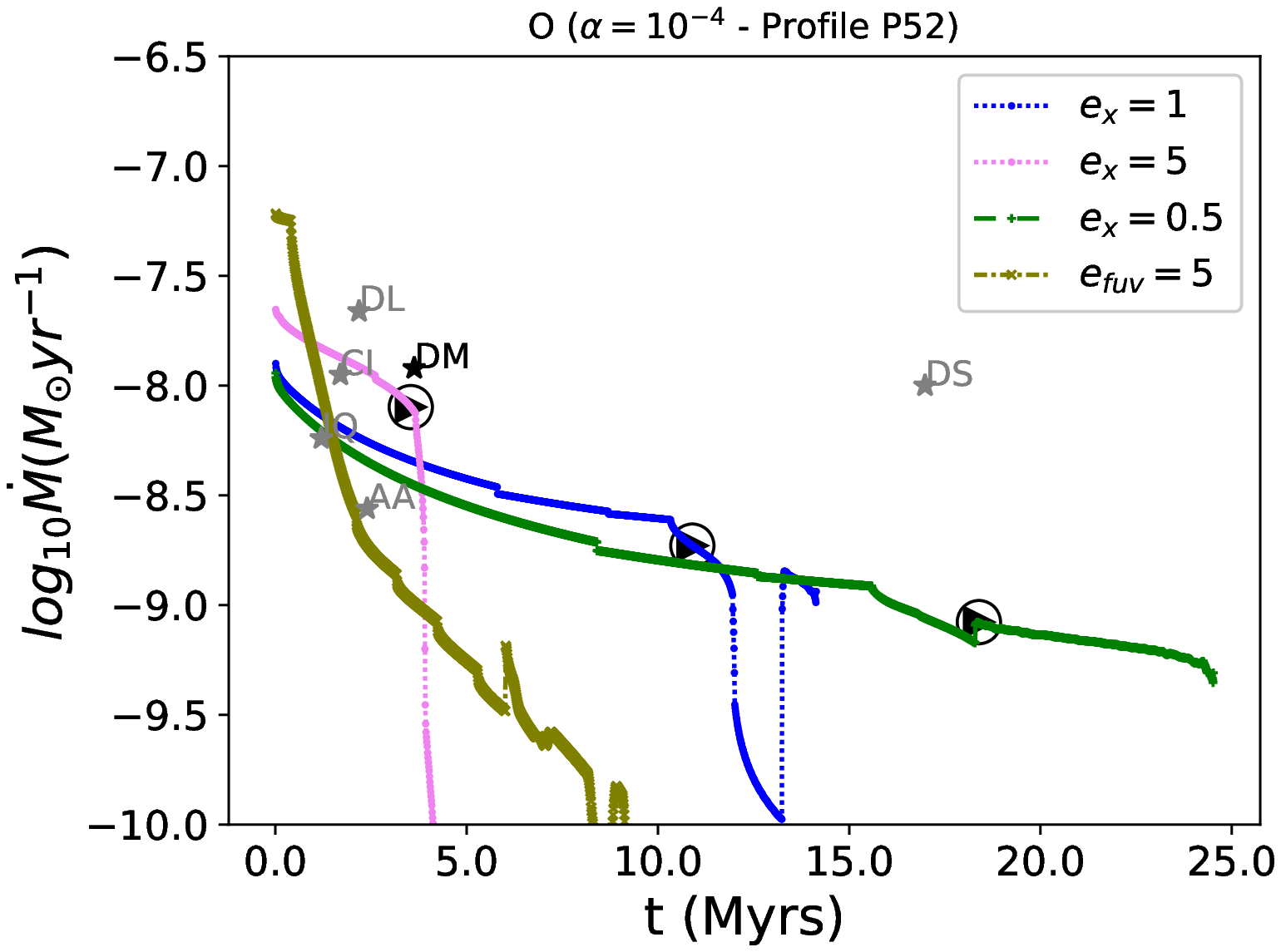} &
\includegraphics[width=0.3\linewidth]{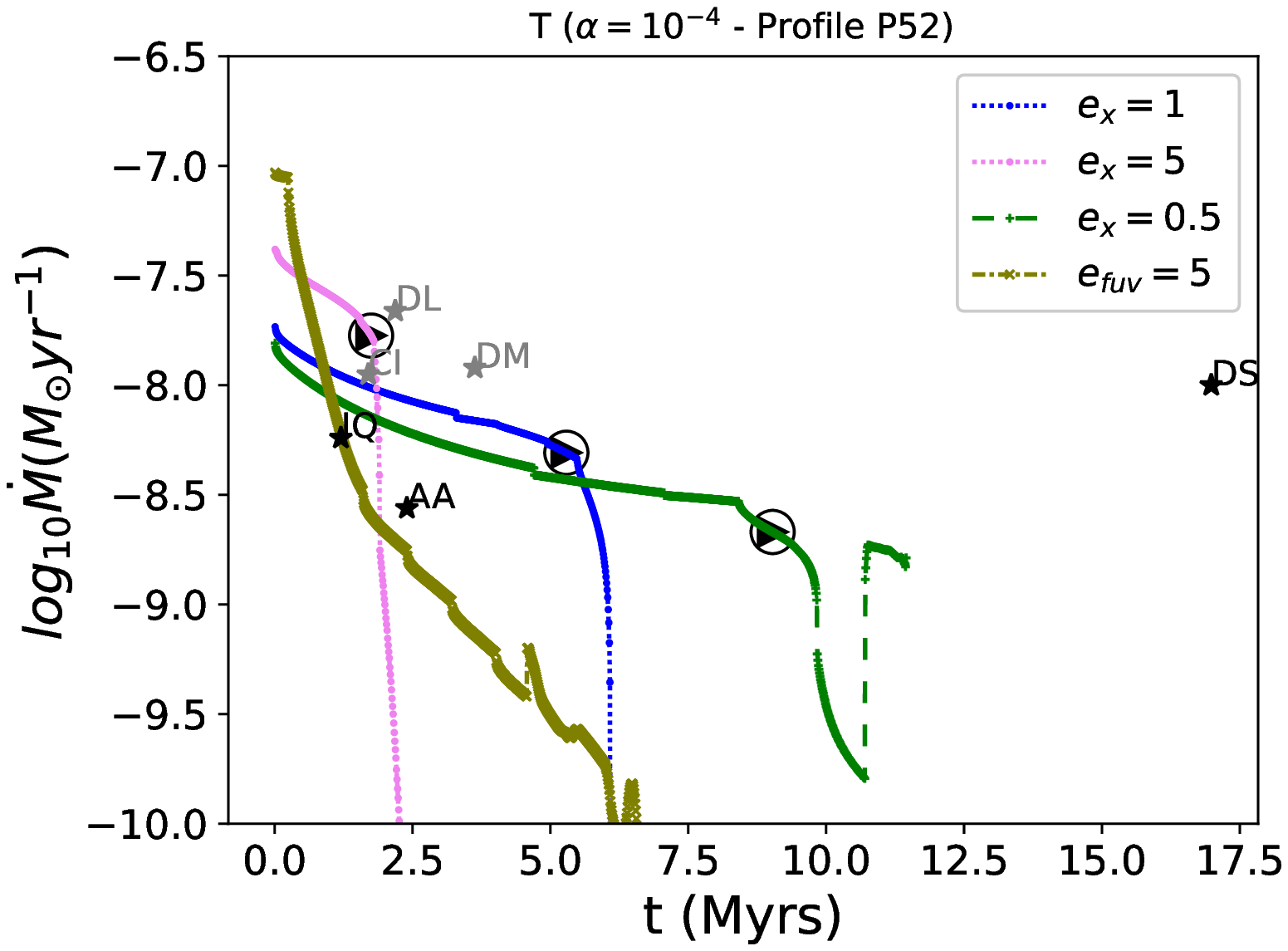} &
\includegraphics[width=0.3\linewidth]{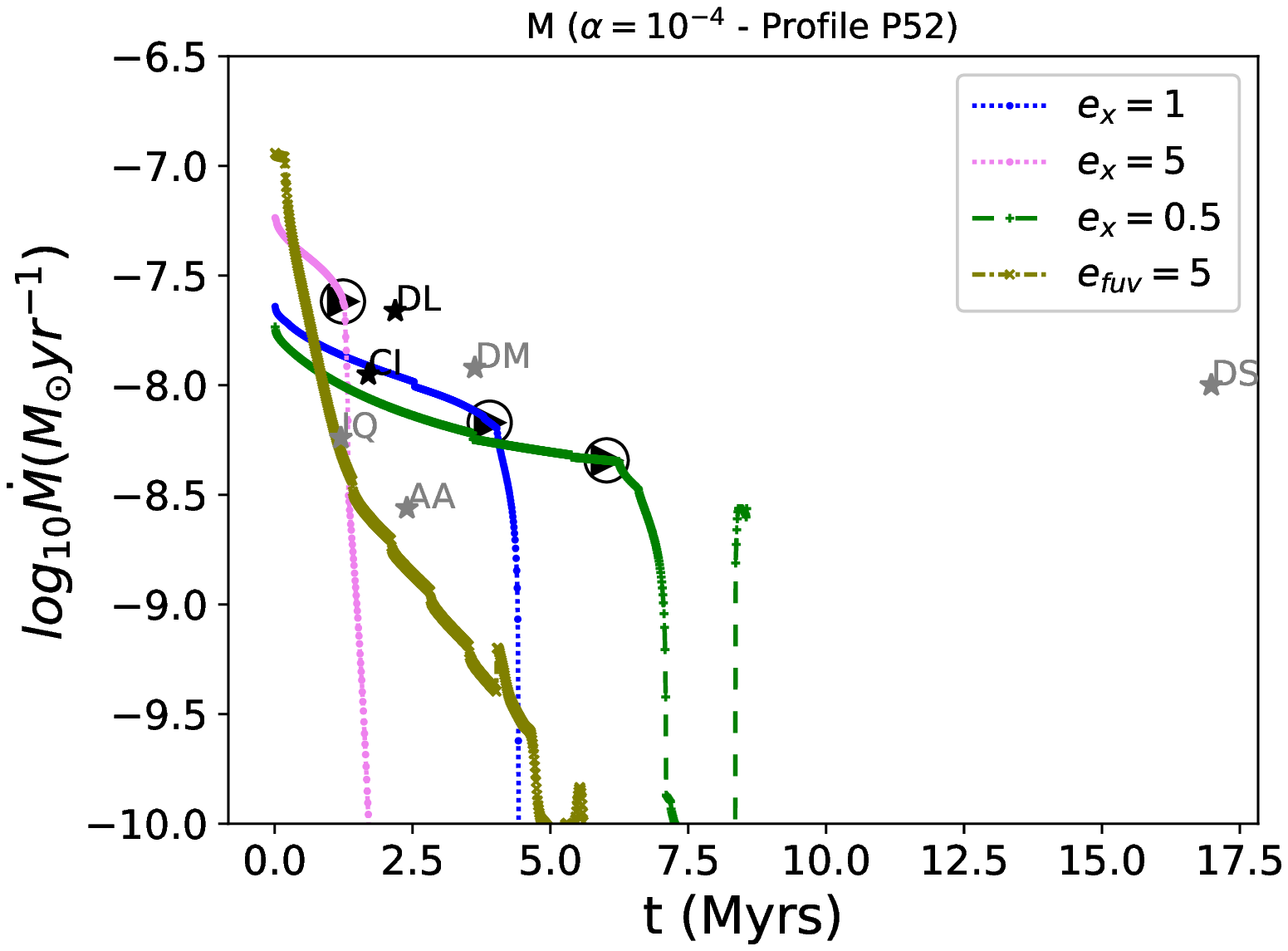} \\
&\multicolumn{3}{c}{  $\xrightarrow{\makebox[6cm]{$M_{*}$}}$  } \\
\end{tabular}
\end{center}
\caption{Accretion rates for X-ray winds and FUV winds, 
$\alpha=0.0001$ and a variable flaring profile changing 
from $0.05$ to $0.02$. The blue curve corresponds to
the fiducial X-ray radiation dominated wind model
$e_x=1$. The pink curve means $e_x=5$, and the green curve $e_x=0.5$. 
The disc subject to FUV-dominated winds, $e_{fuv}=5.0$, appears as olive curve.
The black circles with a triangle 
within it mark when dents are present and ring-like features can be produced.}
\label{rx_rates}
\end{figure*}

The Figure~\ref{rx_rates} shows the accretion rates 
in discs subject to photoevaporation from X-rays dominated or FUV-dominated winds.
All curves correspond to $\alpha=10^{-4}$ and a progressive
flattening $P52$.  
The blue curve means X-rays radiation dominated $e_{x}=1.0$, 
the pink curve means $e_{x}=5.0$ and the green curve means $e_{x}=0.5$.
The disc subject to FUV-dominated winds, $e_{fuv}=5.0$, appears as olive curve.
The black circles with a triangle 
within it marks when a ring-like feature is produced.

This figure shows how the initial value of the accretion
rate depend on the disc mass $M_d$. The larger the initial disc mass, the higher
the initial accretion rate. However, the later evolution of this
initial value will depend on remaining control parameters. 

Obviously, when comparing the evolution of discs with high and low
values of the viscosity, the decay is faster and the dependency 
on the remaining control parameters is stronger in the case of low viscosity
$\alpha$, in coincidence with the EUV-only case \citep{vallejo18}.


If we consider the variable flaring profile $P52$, 
with a primordial phase starting with a higher temperature than the 
standard $\upkappa=0.033$,
the mass rates are higher. The stronger the wind, the larger 
the accreted mass, with larger differences for the 
larger stellar masses.


In general, the mass rate decreases
continuously until the first (inner) dent in the disc opens a gap. 
Then, the decrease in mass rate is faster meanwhile the gap widens 
until the inner disc is fully drained. 
When the inner cavity opens and the direct flux phase
starts, there is in some cases an increase in the mass accretion rate.
This can be explained
because the photoevaporated mass rate is lower in direct flux 
phase than in the primordial case. 
Hence, the  mass loss due to photoevaporation 
decreases, and, in turn, the  accreted mass rate 
that remains after subtracting the wind losses is larger. 
However, one must also note that our models are still very simple, and
further refinements when reaching the direct phase in an under flattening disc.

We have already seen in previous sections that when disc mass is very
small in the direct flux phase, 
the front of the wind can carve an outer dent, 
and even an outer gap on it.
These outer dents are marked by circles in Figure~\ref{rx_rates}.
The stronger the wind and the lower
of the disc mass, the higher 
the likelihood of having these dents and gaps.

The models with a high viscosity $\alpha$ predict a very fast 
reduction in the mass rates. Conversely, the mass rates when one decreases
$\alpha$ are closer to the observed values in the Taurus systems
\citep{vallejo18}. 
The Figure~\ref{rx_rates} shows that even taking $\alpha=10^{-4}$,
the synthetic models produce mass rates slightly lower than the
Taurus systems, but IQ and AA Tau.


The DM Tau seems to roughly match 
with several curves corresponding to different initial disc masses.
The pink curve ($e_x=5.0$) in the $H$ model is the closest to this system.
However, when we also consider when ring-like features are created, 
flagged by the black circle with an arrow inside, the DM Tau 
makes a better match in age, disc mass and accretion rate
with the synthetic model with the smallest initial disc mass 
($O$). 

When considered FUV-dominated fluxes, the accretion rate of DM Tau is closer to the
mass rate curves in $G$ model. But, when one takes into account the 
creation of dents, the FUV-dominated models seems to produce just the 
internal cavity, and no any other ring-like feature. Our models
are very simple, and this may point again to further explore the parametric space.

Regarding the Taurus systems with intermediate stellar mass 
(panels in the central column of the figure), the 
IQ and AA Tau systems seem to be far away from
the mass rates coming from our X-ray dominated models,
and closer to those produced by FUV-dominated winds.
The AA Tau system might be modelled by a synthetic system 
with an initial mass larger than the $I$ model.
Notably, the DS Tau system seems to be far away from all plotted curves because its large age.
However, we note again that we do not aim to explain all Taurus systems. Conversely, 
we just aim to analyse the role of the diverse model parameters in lifetimes, mass rates
and ring-like features.

Notably, the FUV-dominated curve does the best match in these intermediate stellar mass systems.
while predicting the proper cavity ages. However, these models, constrained
by using a FUV wind with fixed peak and trough,  
did not predict the creation of ring-like features at any age.

Finally, in which concerns the largest stellar masses, the CI and DL Tau
systems seem to match with almost any X-ray dominated model. Considering the 
disc mass and observed features at time, the $J$ model have a good match
when considering the highest efficiency of the wind.

The curves describing the mass rate evolution also provide a view 
of the lifetimes of the discs. We see 
how these disc lifetimes depend on the stellar mass, 
the smaller lifetimes corresponding to the largest stellar masses.
As the stellar mass increases, the accreted mass rate grows, as expected.

Notably, some of our photoevaporative models may explain objects 
with large inner cavities and strong accretion rates, 
while much simpler models fail to explain them \citep{ercolano18}.
These transition discs with large cavities and strong accreting 
rates have sometimes be linked to (multiple) giant planet 
formation systems 
\citep{dodson11}.
Alternatively, following \cite{ercolano18}, these features
could result from X-ray photoevaporation in metal (C and O) depleted discs,
and our models might support this photoevaporative approach.

However, we must stress again the simplicity of our models, mainly the FUV-dominated wind 
with fixed geometrical parameters. The best matches of the ages of our models with those from real Taurus 
systems are produced with very large $e_{fuv}$ values. 
The cavities seem to be produced in all cases, but at somehow late ages. Unfortunately, no ring-like features
seems to be present in these initial analyses and further
explorations of the parameter space are needed.

\section{Conclusions}
\label{sec:conclusions} 

We have analysed the impact of a progressive flattening 
in protoplanetary flared discs combined with different photoevaporative winds in 
the creation of ring-like structures, using simple semi-analytical 
$1D$ $\alpha$-disc.

Previous works that used combined X-ray and FUV winds
\citep{gorti09} predicted a significant
population of discs with relatively massive discs
 and low accretion rates with large inner holes. 
By adding variable flaring profiles and different wind efficiencies, 
other combinations of accretion rates and masses are possible,
as seen when we compare our results with measured parameters in
real systems from Taurus.

\begin{figure}
\begin{center}
\begin{tabular}{c}
\includegraphics[width=\columnwidth]{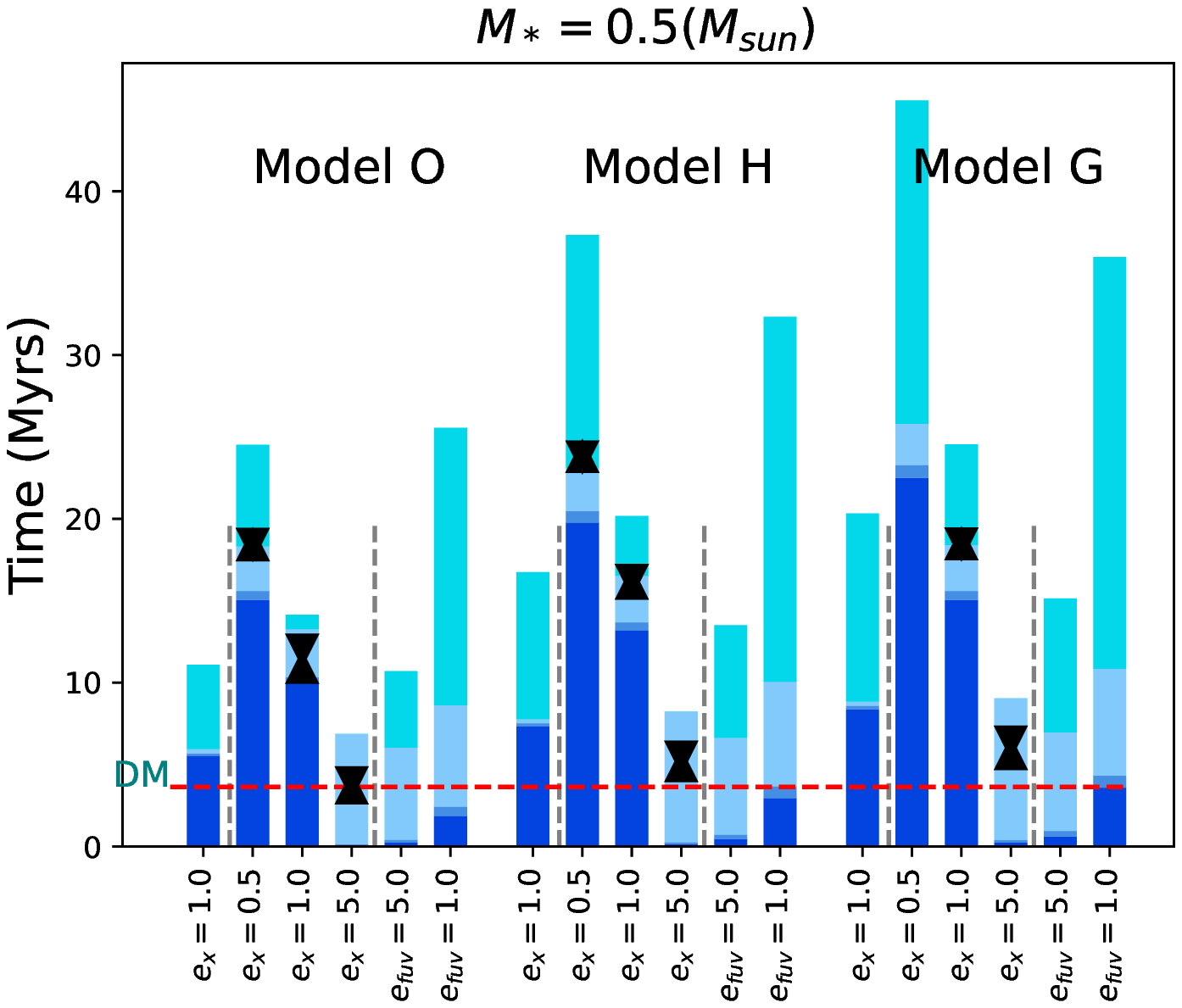} \\
\includegraphics[width=\columnwidth]{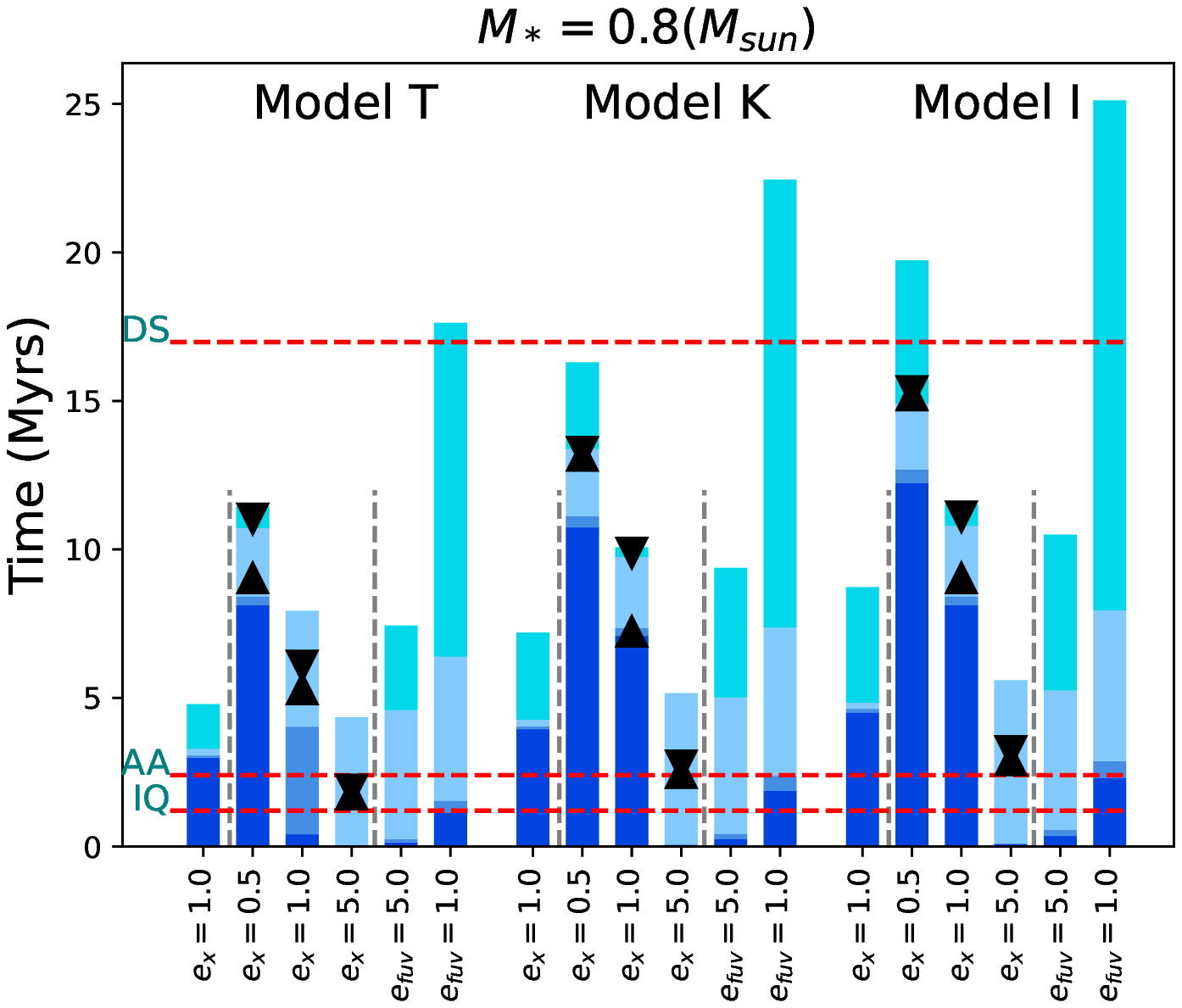} \\
\includegraphics[width=\columnwidth]{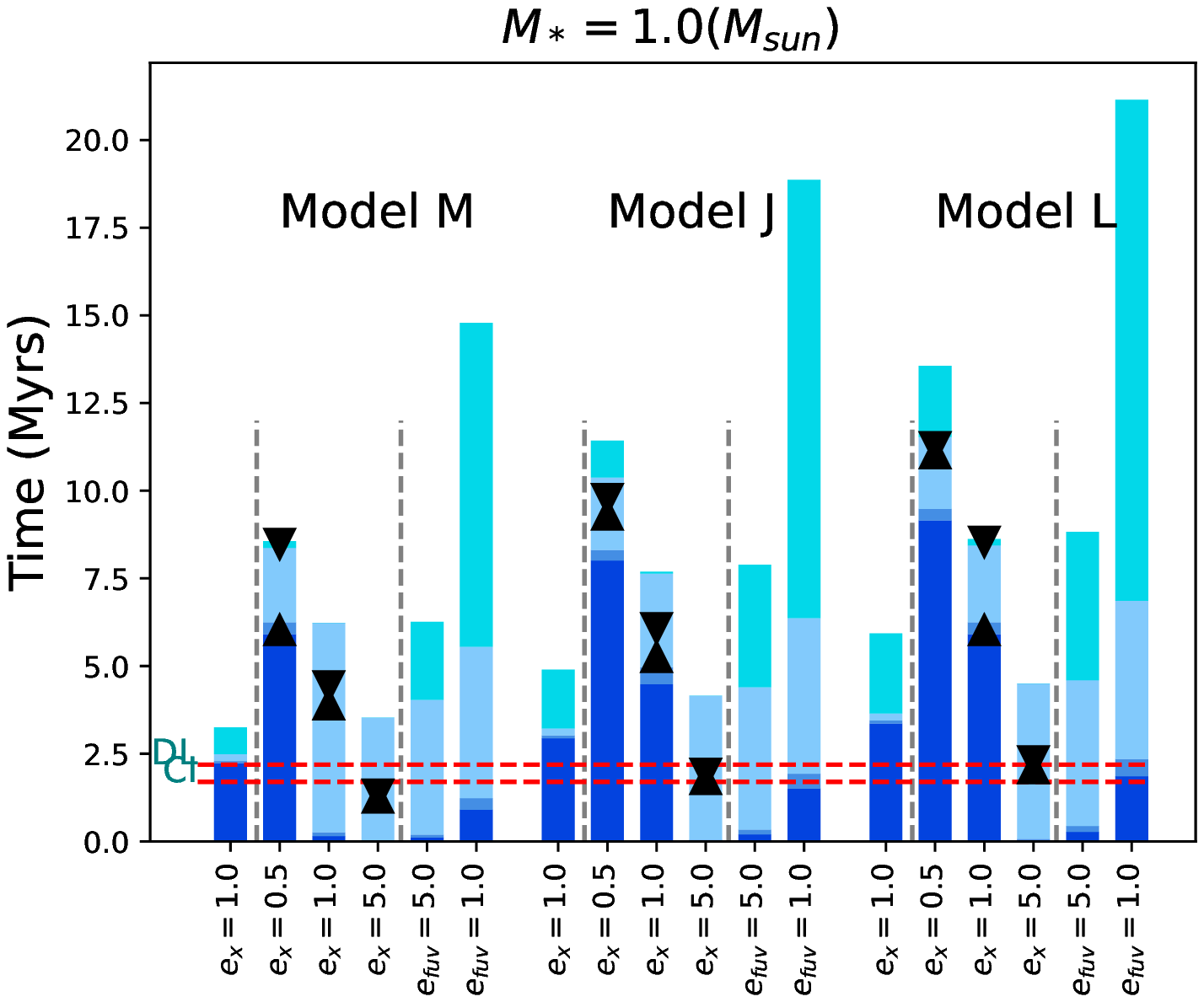} 
\end{tabular}
\end{center}
\caption{
Summary of the evolution and lifetimes of synthetic discs. 
The first bar corresponds to a fiducial X-dominated wind with $\alpha=10^{-3}$, the remaining 
groups of bars correspond to $\alpha=10^{-4}$, the first group
being an X-ray dominated wind, the second group being a 
FUV wind with secondary X-ray emission. 
The horizontal dashed lines mark the ages of the Taurus reference systems.}
\label{rx_evo}
\end{figure}

As summary of previous sections, the creation of ring-like structures 
in photoevaporative discs are facilitated 
in discs with low viscosity values 
and a flared geometry that flattens as the disc evolves.
Several ring-like features can be produced with a variety 
of efficiencies. Up to three dents can be obtained
with fiducial or weak winds, meanwhile only two dents are got with 
strong winds. However, the later case is the one that does not
require long lifetimes until the ring-like features fully develop.
Regarding the generation of long-lived cavities, these are best seen in the FUV-dominated
models. However, they are also present in the X-ray dominated cases
despite being very short-lived.

The Figure~\ref{rx_evo} summarises these results. 
Every phase appears in a different blue tone.
When the inner cavity starts to develop
is dark blue, when it opens is turquoise, when the direct
flux starts is light blue, and the final outer disc evolution is plotted in cyan.
The first bar corresponds to a fiducial X-dominated wind with $\alpha=10^{-3}$, the remaining 
groups of bars correspond to $\alpha=10^{-4}$, the first group
being an X-ray dominated wind, the second group being a 
FUV wind with secondary X-ray emission. 
The time when the first dent appears is marked with an upper arrow. The last time when any dent (it might
have or not evolved to be a gap) is present is marked with a downward arrow.

The Figure~\ref{rx_evo} also provides the different disc lifetimes of the grid models.
When the wind efficiencies increase, the lifetimes are shortened.
In this figure one can also observe than these lifetimes decrease with the host star mass,
and when we fix this parameter, lifetimes grow with initial disc mass,
as expected.

The age when the gas is depleted 
constrains the potential ending of the formation of gas giants planets, 
and migration processes that involve exchanges of 
angular momentum could also stop at these early ages \citep{alexander09}.
We have seen that one can achieve 
very short disc lifetimes when using winds with high efficiency factors.


A FUV-dominated
photoevaporative wind seems to 
reproduce well the accretion rates found in the intermediate
stellar mass cases of our limited sample, but fails
in reproducing the ring-like
features at inner and outer distances observed in real systems, 
and only produce cavities of slow evolution. Conversely, with X-ray dominated winds, one gets
gaps in the range of tens of astronomical units in discs as young as
$1$ Myr old. 

Therefore, a progressive flattening of the disc seems to support the production of ring-like
features at the observed ages.
These features can be produced in our synthetic models when two dents or gaps exist
at the same time. 
Hence, our ring-like features are transient features, typically lasting 
less than $1$Myr. Notably, a third dent can appear, more likely in the higher 
stellar masses and higher disc masses models. However, 
it does not coincide in time with the previously generated dents.

All these features are transient, and they are 
roughly coincident with some of the annular features seen 
in a few of the Taurus disc of our sample  (see Table~\ref{listofstars})
when using somehow too strong efficiencies. 

The major part of these ring-like features are only created 
before the internal cavity is fully developed.
The timescales for the creation
of internal cavities do not match well and further explorations
of the involved parameters are still needed. 
One improvement should be to modify some
of the geometrical baseline parameters of our wind profiles,
such as the peak and trough of the FUV profile at fixed positions
used in this work. Another important enhancement should be to
further refine the flattening process.

Moreover, some additional mechanism may be needed for carving faster the
dents and create real gaps.
For instance, our dents may convert onto dust traps, accelerating
the process of gaps creation in a similar fashion that 
the gaps creation mechanisms is attributed to planets. 
Despite of that, our simple photoevaporative dents still
may produce valid predictions, considering that the dents may trigger
this mechanism and lead to the observed structures.

We note that we have just considered the evolution of the gas, and its evolution
can be somehow different than the evolution of the dust, even when the initial dents created by photoevaporative 
processes may indicate dents into the dust. 
One also may take into account observational
uncertainties. The gas, mainly $H_2$ ($90$\%) and He ($10$\%), 
is difficult to observe, and typically one uses CO (~$0.01$\%) 
as a proxy
\citep{miotello16,molyarova17}. But CO can have a complex distribution due to the disk
structure hence making the interpretation of the results very
difficult, leading to the analysis of other tracers that are 
even more scarce \citep{williams14}.

Many other observations are based on dust continuum, where 
discs present rings and gaps, in addition to the internal cavity.
The gas surface density is assumed to follow the 
dust density evolution, driven by a given dust-to-gas ratio
\citep{birnstiel10}. 
Certainly, this may not be true, and the dust-to-gas mass ratio be
 a free parameter. Anyhow, if gas might be much abundant than dust,
one may find that meanwhile gas erosion keeps as a small
 dent, dust may have already formed a gap.
If this ratio changes with time or is not constant along the disc 
(see, for instance \cite{birnstiel10}), it might be higher in the outer 
parts of the disc than in the inner side, making the outer features in 
the dust to appear before than in the gas.

\begin{figure}
\begin{center}
\includegraphics[width=\columnwidth]{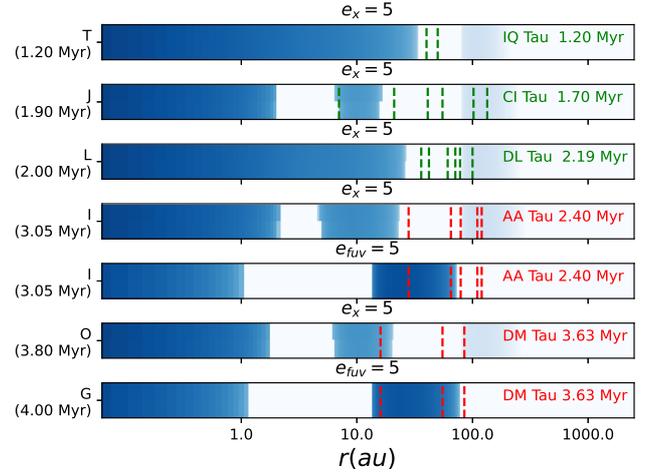} 
\end{center}
\caption{
A schematised view of the position of the ring-like features, inner cavities and gaps, 
of different widths produced by the grid models. The first three models produce
features resembling those observed in the 
 Taurus reference sample at the proper age, with a disc mass and accretion rate similar to the observed values.
These Taurus features appear as dashed lines, red color indicating the internal cavity is present.
The last three models produce a gap at proper ages, but the cavities are fully developed much later than when observed
in real systems.
Note that the colorscale denotes the model surface density.}
\label{models_virtual}
\end{figure}

We have seen that the disc mass evolution curves, the
accretion rates curves and the mass density features predicted by our models can sometimes resemble 
those found in the reference Taurus systems.
This supports the usefulness of using $1D$ models because their 
simplicity in interpreting their results. 
These results can be summarised in Figure~\ref{models_virtual}. 
Here, we can see a schematised view of the position of the ring-like features, inner cavities and gaps, 
of different widths produced by the grid models. Some of the model-generated features resemble those
observed in the Taurus reference sample (denoted by red dashed lines) at proper ages.

The best matches are got with IQ Tau, CI Tau and DL Tau.
Their ages, disc masses and
accretion rates can be reproduced, while 
one can see structures resembling to those seen in some of those real systems.
However, very high mass losses are needed, and some additional gaps should be carved for having a better match. 

Obviously, there are limitations in such simplifications.
Our simple models do not reproduce other reference real 
systems.
The synthetic models closer to DM Tau and AA Tau are still eroding the internal cavity, while
the real DM Tau and AA Tau have already developed this cavity. By other hand, DS Tau seems to remain out of any
of our model predictions, even when DS Tau may be matched
with a model with initial mass larger than $I$ model, may be using a very inefficient X-ray 
dominated wind.

Observations of inner ring-like structures very close
to the host star are not straightforward and remain somehow elusive below the astronomical unit.
Our baseline models focused on low viscosity values in order to 
better match with measured disc ages. 
However, the disc composition is not the only source for
shortening discs lifetimes. We have seen the important role played by
the flattening of the disc profile. Therefore, a quite interesting research topic that can extend the results presented here 
is to modify further the geometrical baseline parameters of the flaring profiles
and the flattening process, especially
in the case of FUV-dominated winds. 
These enhancements may change the location and time where cavities, dents and gaps are created.

\subsection*{Data availability}

The data underlying this article will be shared on reasonable request to the corresponding author.

\section*{Acknowledgements}

The authors thank the Spanish Ministry of Economy, Industry and Competitiveness for 
grants ESP2015-68908-R and ESP2017-87813-R.


\bibliographystyle{mnras}








\bsp	
\label{lastpage}
\end{document}